\documentclass[12pt]{article}
\usepackage{amssymb,amscd,array}
\catcode `\@=11
\@addtoreset{equation}{section}

\newtheorem{thm}{Theorem}[section]
\newtheorem{rem}[thm]{Remark}

\def\qed{\blacksquare}
\newcommand{\be}{\begin{equation}}
\newcommand{\ee}{\end{equation}}
\newcommand{\bea}{\begin{eqnarray}}
\newcommand{\eea}{\end{eqnarray}}
\newcommand{\R}{\mathbb{R}}
\newcommand{\N}{\mathbb{N}}
\newcommand{\C}{\mathbb{C}}

\begin{document}
\begin{titlepage}

\begin{center}
{\bf \Large{Perturbative Gravity in the Causal Approach\\}}
\end{center}
\vskip 1.0truecm
\centerline{D. R. Grigore, 
\footnote{e-mail: grigore@theory.nipne.ro}}
\vskip5mm
\centerline{Department of Theoretical Physics, Institute for Physics and Nuclear
Engineering ``Horia Hulubei"}
\centerline{Institute of Atomic Physics}
\centerline{Bucharest-M\u agurele, P. O. Box MG 6, ROM\^ANIA}

\vskip 2cm
\bigskip \nopagebreak
\begin{abstract}
\noindent
Quantum theory of the gravitation in the causal approach is studied up to the second order of perturbation theory. We prove gauge invariance and renormalizability in the second order of perturbation theory for the pure gravity system (massless and massive). Then we investigate the interaction of massless gravity with matter (described by scalars and spinors) and massless Yang-Mills fields. We obtain a difference with respect to the classical field theory due to the fact that in quantum field theory one cannot enforce the divergenceless property on the vector potential and this spoils the divergenceless property of the usual energy-momentum tensor. To correct this one needs a supplementary ghost term in the interaction Lagrangian.
\end{abstract}
\end{titlepage}

\section{Introduction}

The general framework of perturbation theory consists in the construction of 
the chronological products such that Bogoliubov axioms are verified \cite{BS}, \cite{EG}, \cite{DF}, \cite{ano}; for every set of Wick monomials 
$ 
W_{1}(x_{1}),\dots,W_{n}(x_{n}) 
$
acting in some Fock space generated by the free fields of the model
$
{\cal H}
$
one associates the operator
$ 
T_{W_{1},\dots,W_{n}}(x_{1},\dots,x_{n}); 
$  
all these expressions are in fact distribution-valued operators called chronological products. Sometimes it is convenient to use another notation: 
$ 
T(W_{1}(x_{1}),\dots,W_{n}(x_{n})). 
$ 
The construction of the chronological products can be done recursively according to Epstein-Glaser prescription \cite{EG}, \cite{Gl} (which reduces the induction procedure to a distribution splitting of some distributions with causal support) or according to Stora prescription \cite{PS} (which reduces the renormalization procedure to the process of extension of distributions). These products are not uniquely defined but there are some natural limitation on the arbitrariness. If the arbitrariness does not grow with $n$ we have a renormalizable theory. An equivalent point of view uses retarded products \cite{St1}.

Gauge theories describe particles of higher spin. Usually such theories are not renormalizable. However, one can save renormalizability using ghost fields. Such theories are defined in a Fock space
$
{\cal H}
$
with indefinite metric, generated by physical and un-physical fields (called {\it ghost fields}). One selects the physical states assuming the existence of an operator $Q$ called {\it gauge charge} which verifies
$
Q^{2} = 0
$
and such that the {\it physical Hilbert space} is by definition
$
{\cal H}_{\rm phys} \equiv Ker(Q)/Im(Q).
$
The space
$
{\cal H}
$
is endowed with a grading (usually called {\it ghost number}) and by construction the gauge charge is raising the ghost number of a state. Moreover, the space of Wick monomials in
$
{\cal H}
$
is also endowed with a grading which follows by assigning a ghost number to every one of the free fields generating
$
{\cal H}.
$
The graded commutator
$
d_{Q}
$
of the gauge charge with any operator $A$ of fixed ghost number
\be
d_{Q}A = [Q,A]
\ee
is raising the ghost number by a unit. It means that
$
d_{Q}
$
is a co-chain operator in the space of Wick polynomials. From now on
$
[\cdot,\cdot]
$
denotes the graded commutator.
 
A gauge theory assumes also that there exists a Wick polynomial of null ghost number
$
T(x)
$
called {\it the interaction Lagrangian} such that
\be
~[Q, T] = i \partial_{\mu}T^{\mu}
\ee
for some other Wick polynomials
$
T^{\mu}.
$
This relation means that the expression $T$ leaves invariant the physical states, at least in the adiabatic limit. In all known models one finds out that there exists a chain of Wick polynomials
$
T^{\mu},~T^{\mu\nu},~T^{\mu\nu\rho},\dots
$
such that:
\be
~[Q, T] = i \partial_{\mu}T^{\mu}, \quad
[Q, T^{\mu}] = i \partial_{\nu}T^{\mu\nu}, \quad
[Q, T^{\mu\nu}] = i \partial_{\rho}T^{\mu\nu\rho},\dots
\label{descent}
\ee
In all cases
$
T^{\mu\nu},~T^{\mu\nu\rho},\dots
$
are completely antisymmetric in all indices; it follows that the chain of relation stops at the step $4$ (if we work in four dimensions). We can also use a compact notation
$
T^{I}
$
where $I$ is a collection of indices
$
I = [\nu_{1},\dots,\nu_{p}]~(p = 0,1,\dots,)
$
and the brackets emphasize the complete antisymmetry in these indices. All these polynomials have the same canonical dimension
\be
\omega(T^{I}) = \omega_{0},~\forall I
\ee
and because the ghost number of
$
T \equiv T^{\emptyset}
$
is supposed null, then we also have:
\be
gh(T^{I}) = |I|.
\ee
One can write compactly the relations (\ref{descent}) as follows:
\be
d_{Q}T^{I} = i~\partial_{\mu}T^{I\mu}.
\label{descent1}
\ee

For concrete models the equations (\ref{descent}) can stop earlier: for instance in the case of gravity
$
T^{\mu\nu\rho\sigma} = 0.
$

Now we can construct the chronological products
$$
T^{I_{1},\dots,I_{n}}(x_{1},\dots,x_{n}) \equiv T(T^{I_{1}}(x_{1}),\dots,T^{I_{n}}(x_{n}))
$$
according to the recursive procedure. We say that the theory is gauge invariant in all orders of the perturbation theory if the following set of identities generalizing (\ref{descent1}):
\be
d_{Q}T^{I_{1},\dots,I_{n}} = 
i \sum_{l=1}^{n} (-1)^{s_{l}} {\partial\over \partial x^{\mu}_{l}}
T^{I_{1},\dots,I_{l}\mu,\dots,I_{n}}
\label{gauge}
\ee
are true for all 
$n \in \N$
and all
$
I_{1}, \dots, I_{n}.
$
Here we have defined
\be
s_{l} \equiv \sum_{j=1}^{l-1} |I|_{j}
\ee
(see also \cite{DB}). In particular, the case
$
I_{1} = \dots = I_{n} = \emptyset
$
is sufficient for the gauge invariance of the scattering matrix, at least
in the adiabatic limit.

Such identities can be usually broken by {\it anomalies} i.e. expressions of the type
$
A^{I_{1},\dots,I_{n}}
$
which are quasi-local and might appear in the right-hand side of the relation (\ref{gauge}). These expressions verify some consistency conditions - the so-called Wess-Zumino equations. One can use these equations in the attempt to eliminate the anomalies by redefining the chronological products. All these operations can be proved to be of cohomological nature and naturally lead to descent equations of the same type as (\ref{descent1}) but for different ghost number and canonical dimension. 

If one can choose the chronological products such that gauge invariance is true then  there is still some freedom left for redefining them. To be able to decide if the theory is renormalizable one needs the general form of such arbitrariness. Again, one can reduce the study of the arbitrariness to descent equations of the type as (\ref{descent1}).

Such type of cohomology problems have been extensively studied in the more popular approach to quantum gauge theory based on functional methods (following from some path integration method). In this setting the co-chain operator is non-linear and makes sense only for classical field theories. On the contrary, in the causal approach the co-chain operator is linear so the cohomology problem makes sense directly in the Hilbert space of the model. For technical reasons
one needs however a classical field theory machinery to analyze the descent equations more easily.

In this paper we want to give a general description of these methods and we will apply them to gravitation models. We consider the case of massless and massive gravity.

In the next Section we remind the axioms verified by the chronological products (for simplicity we give them only for the second order of perturbation theory) and consider the particular case of gauge models. In Section \ref{WZ} we give some general results about the structure of the anomalies and reduce the proof of (\ref{gauge}) to descent equations. We will use a convenient geometric setting for our problem presented in \cite{cohomology}. In Section \ref{q} we determine the cohomology of the operator 
$
d_{Q}
$
for gravity models. Using this cohomology and the algebraic Poincar\'e lemma we can solve the descent equations in various ghost numbers in Section \ref{relative}. In Section \ref{int} we use these methods to prove the gauge invariance and renormalization properties of the pure gravity model in the second order of perturbation theory. In Section \ref{int1} we determine the interaction between massless gravity, matter and Yang-Mills fields; we consider here only massless Yang-Mills fields and we will treat the massive case elsewere. An interesting fact appears in this case due to the fact that, as it is well known, one cannot enforce the property
\be 
\partial_{\mu}v_{a}^{\mu} = 0
\ee
for quantum massless fields. This means that the usual expression for the energy-momentum tensor
$
T^{\mu\nu}
$
will not be divergenceless in the quantum context and this spoils the gauge invariance property of the interaction. Fortunately one can correct this adding a new ghost term in the interaction Lagrangian. 

In \cite{Sc2} one can find similar results for massless gravity and its interaction with a scalar field but the cohomological methods are not used for the proofs.

We note that the renormalizability of quantum gravity has attracted a lot of attention and we mention \cite{GW} and \cite{Kr}. Here we concentrate only on some technical procedures of cohomological nature which can be used to simplify the understanding of the lower order of perturbation theory.

\section{General Gauge Theories\label{ggt}}
 
We give here the essential ingredients of perturbation theory. For simplicity we emphasize the second order of the perturbation theory. 

\subsection{Bogoliubov Axioms}{\label{bogoliubov}}

Suppose that we have a Fock space
$
{\cal H}
$
generated by some set of free fields and consider a set of Wick monomials
$
W_{j},~j = 1,\dots,n
$ 
acting in this Hilbert space. The chronological products
$ 
T(W_{1}(x_{1}),\dots,W_{n}(x_{n})) \equiv T_{W_{1},\dots,W_{n}}(x_{1},\dots,x_{n})
\quad n = 1,2,\dots
$
are some operator-valued distribution acting in the Fock space and verifying a set of axioms (named {\it Bogoliubov axioms}) explained in detail in \cite{cohomology}. For the case of gravity we will concentrate in this paper on the second order of perturbation theory, so we give the the axioms only for the cases
$
n= 1, 2.
$ 
We postulate the ``initial condition"
\be
T_{W}(x) = W(x)
\ee
and we give the axioms for the chronological products
$ 
T_{W_{1},W_{2}}(x_{1},x_{2});
$
for the particular case 
$
W_{1} = W_{2} = T
$
one denotes
$ 
T_{W_{1},W_{2}}(x_{1},x_{2}) = T_{2}(x_{1},x_{2})
$
and these axioms guarantee that the scattering matrix
\be
S(g) = {\bf I} + \int_{\R^{4n}} dx g(x) T(x) +
\int_{\R^{8n}} dx_{1}~dx_{2}~g(x_{1})~g(x_{2})~T_{2}(x_{1},x_{2}) + \cdots
\ee
is Poincar\'e covariant, unitary and causal. In general we require:
\begin{itemize}
\item
Skew-symmetry:
\be
T_{W_{1},W_{2}}(x_{1},x_{2}) = (-1)^{f_{1} f_{2}}~T_{W_{2},W_{1}}(x_{2},x_{1})
\ee
where
$f_{i}$
is the number of Fermi fields appearing in the Wick monomial
$W_{i}$.
\item
Poincar\'e invariance: we have a natural action of the Poincar\'e group in the space of Wick monomials and we impose that for all 
$(a,A) \in inSL(2,\C)$
we have:
\be
U_{a, A} T_{W_{1},W_{2}}(x_{1},x_{2}) U^{-1}_{a, A} =
T_{A\cdot W_{1},A\cdot W_{2}}(A\cdot x_{1}+a,A\cdot x_{2}+a);
\label{invariance}
\ee

Sometimes it is possible to supplement this axiom by other invariance properties: space and/or time inversion, charge conjugation invariance, global symmetry invariance with respect to some internal symmetry group, supersymmetry, etc.
\item
Causality: if
$
x_{1}
$
succeeds causally
$
x_{2}
$
(which one denotes
$
x_{1} \geq x_{2})
$
then we have:
\be
T_{W_{1},W_{2}}(x_{1},x_{2}) = T_{W_{1}}(x_{1})~~T_{W_{2}}(x_{2}) = W_{1}(x_{1})~W_{2}(x_{2});
\label{causality}
\ee
\item
Unitarity:
\be
T_{W^{\dagger}_{1},W^{\dagger}_{2}}(x_{1},x_{2})^{\dagger} = 
- T_{W_{1},W_{2}}(x_{1},x_{2}) + W_{1}(x_{1})~W_{2}(x_{2}) + W_{2}(x_{2})~W_{1}(x_{1}).
\label{unitarity}
\ee
\end{itemize}

It can be proved that this system of axioms can be supplemented with
\be
T_{W_{1},W_{2}}(x_{1},x_{2}) = \sum \epsilon \quad
<\Omega, T_{W^{\prime}_{1},W^{\prime}_{2}}(x_{1},x_{2})\Omega>
~:W_{1}^{\prime\prime}(x_{1}),W_{2}^{\prime\prime}(x_{2}):
\label{wick-chrono2}
\ee
where
$W^{\prime}_{i}$
and
$W^{\prime\prime}_{i}$
are Wick submonomials of
$W_{i}$
such that
$W_{i} = :W^{\prime}_{i} W^{\prime\prime}_{i}:$
and the sign
$\epsilon$
takes care of the permutation of the Fermi fields. This is called the {\it Wick expansion property}. 

We can also include in the induction hypothesis a limitation on the order of singularity of the vacuum averages of the chronological products associated to arbitrary Wick monomials
$W_{1},W_{2}$;
explicitly:
\be
\omega(<\Omega, T_{W_{1},W_{2}}(x_{1},x_{2})\Omega>) \leq \omega(W_{1}) + \omega(W_{2}) - 4
\label{power}
\ee
where by
$\omega(d)$
we mean the order of singularity of the (numerical) distribution $d$ and by
$\omega(W)$
we mean the canonical dimension of the Wick monomial $W$; in particular this means
that we have
\be
T_{W_{1},W_{2}}(x_{1},x_{2}) = \sum_{g} t_{g}(x_{1} - x_{2})~W_{g}(x_{1},x_{2})
\label{generic}
\ee
where
$W_{g}$
are Wick polynomials of fixed canonical dimension,
$t_{g}$
are distributions in one variable with the order of singularity bounded by the power counting
theorem \cite{EG}:
\be
\omega(t_{g}) + \omega(W_{g}) \leq \omega(W_{1}) + \omega(W_{2}) - 4
\label{power1}
\ee
and the sum over $g$ is essentially a sum over Feynman graphs. We indicate briefly the simplest way to obtain the chronological products \cite{EG}. We compute the commutator
$
[W_{1}(x_{1}), W_{2}(x_{2})]
$
and we first consider only the contributions coming from three graphs; we end up with an expression of the form 
\be
[W_{1}(x_{1}), W_{2}(x_{2})]_{\rm tree} = \sum {\partial \over \partial x^{\mu_{1}}}\cdots
{\partial \over \partial x^{\mu_{k}}}D_{m_{j}}(x_{1} - x_{2})~
W^{\mu_{1}\dots\mu_{k}}_{j}(x_{1},x_{2}) + \cdots
\label{commutator}
\ee 
where
$
D_{m}(x_{1} - x_{2})
$
is the Pauli-Villars causal distribution of mass $m$,
$
W^{\mu_{1}\dots\mu_{k}}_{j}(x_{1},x_{2})
$
are Wick polynomials and by 
$
\cdots
$
we mean similar terms for the Fermi sector where instead of 
$
D_{m}(x_{1} - x_{2})
$
we have the corresponding causal function
$
S_{m}(x_{1} - x_{2})
$
(see \cite{Sc2} for the explicit expressions). Now one defines
\be
T_{W_{1},W_{2}}(x_{1}, x_{2})_{\rm tree} = \sum {\partial \over \partial x^{\mu_{1}}} \cdots
{\partial \over \partial x^{\mu_{k}}}D^{F}_{m_{j}}(x_{1} - x_{2})~
W^{\mu_{1}\dots\mu_{k}}_{j}(x_{1},x_{2}) + \cdots
\label{feynman}
\ee 
obtained from the previous one by replacing the causal distributions by the corresponding Feynman propagators. A similar procedure works for loop graphs also only the procedure of obtaining the Feynman propagators from the corresponding causal distributions is more complicated (however, as for the tree contributions, is based on a standard procedure of distribution splitting). The resulting chronological products do verify all the axioms.

Up to now, we have defined the chronological products only for Wick monomials 
$
W_{1}, W_{2}
$
but we can extend the definition for Wick polynomials by linearity.

One can modify the chronological products without destroying the basic property of causality {\it iff} one can make
\be
T_{W_{1},W_{2}}(x_{1},x_{2}) \rightarrow T_{W_{1},W_{2}}(x_{1},x_{2})
+ R_{W_{1},W_{2}}(x_{1},x_{2})
\label{renorm}
\ee
where $R$ are quasi-local expressions; by a {\it quasi-local expression} we mean in this case an expression of the form
\be
R_{W_{1},W_{2}}(x_{1},x_{2}) = \sum_{g} \left[ P_{g}(\partial)\delta(x_{1} - x_{2})\right]
W_{g}(x_{1},x_{2}) 
\label{renorm1}
\ee
with 
$P_{g}$ 
monomials in the partial derivatives and 
$W_{g}$
are Wick polynomials. Because of the delta function we can consider that   
$P_{g}$
is a monomial only in the derivatives with respect to, say
$
x_{2}.
$
If we want to preserve (\ref{power}) we impose the restriction
\be
deg(P_{g}) + \omega(W_{g}) \leq \omega(W_{1}) + \omega(W_{2}) - 4
\label{power2}
\ee
and some other restrictions are following from the preservation of Lorentz covariance and unitarity.

The redefinitions of the type (\ref{renorm}) are the so-called {\it finite renormalizations}. Let us note that in higher orders of perturbation theory this arbitrariness, described by the number of independent coefficients of the polynomials
$
P_{g}
$
can grow with $n$ and in this case the theory is called {\it non-renormalizable}. This can happen if some of the Wick monomials
$
W_{j}, j = 1,\dots,n
$
have canonical dimension greater than $4$. This seems to be the case for quantum gravity.

It is not hard to prove that any finite renormalization can be rewritten in the form
\be
R(x_{1},x_{2}) =
\delta(x_{1} - x_{2})~W(x_{1}) 
+ {\partial \over \partial x^{\mu}_{2}}\delta(x_{1} - x_{2})~W^{\mu}(x_{1})
\label{renorm2}
\ee
where the expression
$
W, W^{\mu}
$
are Wick polynomials. But it is clear that the second term in the above expression is null in the adiabatic limit so we can postulate that these type of finite renormalizations are {\it trivial}. This means that we can admit that the finite renormalizations have a much simpler form, namely
\be
R(x_{1},x_{2}) = \delta(X)~W(x_{1})
\label{renorm3}
\ee  
where the Wick polynomial $W$ is constrained by
\be
\omega(W) \leq \omega(W_{1}) + \omega(W_{2}) - 4.
\label{power3}
\ee

\subsection{Gauge Theories and Anomalies\label{anomalies}}

From now on we consider that we work in the four-dimensional Minkowski space and we have the Wick polynomials
$
T^{I}
$
such that the descent equations (\ref{descent1}) are true and we also have
\be
T^{I}(x_{1})~T^{J}(x_{2}) = (-1)^{|I||J|}~T^{J}(x_{2})~T^{I}(x_{1}),~~
\forall~x_{1} \sim x_{2}
\label{graded-comm}
\ee
i.e. for 
$
x_{1} - x_{2}
$
space-like these expressions causally commute in the graded sense. 

The equations (\ref{descent1}) are called a {\it relative cohomology} problem. The co-boundaries for this problem are of the type
\be
T^{I} = d_{Q}B^{I} + i~\partial_{\mu}B^{I\mu}.
\label{coboundary}
\ee

In the second order of perturbation theory we construct the associated chronological products
$$
T^{I_{1},I_{2}}(x_{1},x_{2}) = T_{T^{I_{1}},T^{I_{2}}}(x_{1},x_{2}).
$$

We will impose the graded symmetry property:
\be
T^{I_{1},I_{2}}(x_{1},x_{2}) = (-1)^{|I_{1}| |I_{2}|}~T^{I_{2},I_{1}}(x_{2},x_{1}).
\label{symmetryT}
\ee
We also have
\be
gh(T^{I_{1},I_{2}}) = |I_{1}| + |I_{2}|.
\label{ghT}
\ee

In the case of a gauge theory the set of {\it trivial} finite renormalizations is larger; we can also include co-boundaries because they induce the null operator on the physical space:
\be
R^{I_{1},I_{2}}(x_{1},x_{2}) = d_{Q}B^{I_{1},I_{2}}(x_{1}) 
+ i~{\partial \over \partial x^{\mu}_{2}}\delta(x_{1} - x_{2})~B^{I_{1},I_{2};\mu}(x_{1})
\label{renorm4}
\ee

One can write the gauge invariance condition (\ref{gauge}) in a more compact form \cite{cohomology} but for 
$
n = 2
$
this will not be necessary. 

We now determine the obstructions for the gauge invariance relations (\ref{gauge}). These relations are true for $n = 1$ according to (\ref{descent1}). Then one can prove that in order $n = 2$ we must have:
\be
d_{Q}T^{I_{1},I_{2}} = i~{\partial\over \partial x^{\mu}_{1}}T^{I_{1}\mu,I_{2}}
+ i~(-1)^{|I_{1}|} {\partial\over \partial x^{\mu}_{2}}T^{I_{1},I_{2}\mu}
+ A^{I_{1},I_{2}}(x_{1},x_{2})
\label{gauge2}
\ee
where the expressions
$
A^{I_{1},I_{2}}(x_{1},x_{2})
$
are quasi-local operators:
\bea
A^{I_{1},I_{2}}(x_{1},x_{2})
= \sum_{k}~\left[ {\partial \over \partial x_{2}^{\rho_{1}}} \dots 
{\partial\over \partial x_{2}^{\rho_{k}}} \delta(x_{2} - x_{1}) \right]~W^{I_{1},I_{2};\{\rho_{1},\dots,\rho_{k}\}}(x_{1})
\label{genericA}
\eea
and are called {\it anomalies}. In this expression the Wick polynomials
$
W^{I_{1},\dots,I_{n};\{\rho_{1},\dots,\rho_{k}\}}
$
are uniquely defined. From (\ref{power1}) we have
\be
\omega(W^{I_{1},I_{2};\{\rho_{1},\dots,\rho_{k}\}}) \leq 2~\omega_{0} - 3 - k
\label{power4}
\ee
where we remind that
$
\omega_{0} \equiv \omega(T);
$
this gives a bound on $k$ in the previous sum. It is clear that we have from (\ref{symmetryT}) a similar symmetry for the anomalies: namely we have:
\be
A^{I_{1},I_{2}}(x_{1},x_{2}) = (-1)^{|I_{1}| |I_{2}|}~A^{I_{2},I_{1}}(x_{2},x_{1})
\label{symmetryA}
\ee
and we also have
\be
gh(A^{I_{1},I_{2}}) = |I_{1}| + |I_{2}| + 1
\label{ghA}
\ee
and
\be
A^{I_{1},I_{2}} = 0
\quad {\it iff} \quad |I_{1}| + |I_{2}| > 2~\omega_{0} - 4.
\label{limit}
\ee

We also have some consistency conditions verified by the anomalies. If one applies the operator
$d_{Q}$
to (\ref{gauge2}) one obtains the so-called {\it Wess-Zumino consistency conditions} for the cases
$
n =2
$:
\be
d_{Q}A^{I_{1},I_{2}} = - i~{\partial\over \partial x^{\mu}_{1}}A^{I_{1}\mu,I_{2}}
- i~(-1)^{|I_{1}|} {\partial\over \partial x^{\mu}_{2}}A^{I_{1},I_{2}\mu}.
\label{wz}
\ee

Let us note that we can suppose, as for the finite renormalizations ( see (\ref{renorm3})) that all anomalies which are total divergences are trivial because they spoil gauge invariance by terms which can be made as small as one wishes (in the adiabatic limit), i.e. we can take the form: 
\be
A^{I_{1},I_{2}}(x_{1},x_{2}) =\delta(x_{1} - x_{2})~W^{I_{1},I_{2}}(x_{1}).
\label{ano}
\ee
In the case of quantum gravity it is not necessary to postulate this relation: one can prove it if one makes convenient finite renormalizations! For Yang-Mills models one can prove even more: such type of relations can be implemented in an arbitrary order of perturbation theory.

Suppose now that we have fixed the gauge invariance (\ref{gauge}) (for $n = 2$) and we investigate the renormalizability issue i.e. we make the redefinitions
\be
T^{I_{1},I_{2}} \rightarrow T^{I_{1},I_{2}} + R^{I_{1},I_{2}}
\label{renorm5}
\ee
where $R$ are  quasi-local expressions. As before we have
\be
R^{I_{1},I_{2}}(x_{1},x_{2}) = (-1)^{|I_{1}| |I_{2}|}~R^{I_{2},I_{1}}(x_{2},x_{1}).
\label{symmetryR}
\ee
We also have
\be
gh(R^{I_{1},I_{2}}) = |I_{1}| + |I_{2}|
\label{ghR}
\ee
and
\be
R^{I_{1},I_{2}} = 0~~{\it iff}~~
\quad
|I_{1}| + |I_{1}| > 2~\omega_{0} - 4.
\label{limit1}
\ee

If we want to preserve (\ref{gauge}) it is clear that the quasi-local operators
$
R^{I_{1},I_{2}}
$
should also verify 
\be
d_{Q}R^{I_{1},I_{2}} = i~{\partial\over \partial x^{\mu}_{1}}R^{I_{1}\mu,I_{2}}
- i~(-1)^{|I_{1}|} {\partial\over \partial x^{\mu}_{2}}R^{I_{1},I_{2}\mu}
\label{wz3}
\ee
i.e. equations of the type (\ref{wz}). In this case we note that we have more structure;
according to the previous discussion we can impose the structure (\ref{renorm3}):
\be
R^{I_{1},I_{2}}(x_{1},x_{2}) = \delta(x_{1} - x_{2})~W^{I_{1},I_{2}}(x_{1})
\ee
and we obviously have:
\be
gh(W^{I_{1},I_{2}}) = |I_{1}| + |I_{2}|
\label{ghR1}
\ee
and
\be
W^{I_{1},I_{2}} = 0~~{\it iff}~~
\quad
|I_{1}| + |I_{2}| > 2~\omega_{0} -4.
\label{limit2}
\ee

From (\ref{wz3}) we obtain after some computations that there are Wick polynomials
$
R^{I}
$
such that
\be
W^{I_{1},I_{2}} = (-1)^{|I_{1}| |I_{2}|}~R^{I_{1} \cup I_{2}}.
\label{gauge6}
\ee

Moreover, we have
\be
gh(R^{I}) = |I|
\label{ghR2}
\ee
and
\be
R^{I} = 0~~{\it iff}~~
\quad
|I| > 2~\omega_{0} -4.
\label{limit3}
\ee

Finally, the following descent equations are true:
\be
d_{Q}R^{I} = i~\partial_{\mu}R^{I\mu}
\label{gauge5}
\ee
and we have obtained another relative cohomology problem similar to the one from the Introduction.
\section{Wess-Zumino Consistency Conditions \label{WZ}}

In this Section we consider a particular form of (\ref{gauge2}) and (\ref{wz}) namely the case when all polynomials
$
T^{I}
$
have canonical dimension
$
\omega_{0} = 5
$
and
$
T^{\mu\nu\rho\sigma} = 0.
$
In this case (\ref{limit}) becomes:
\be
A^{I_{1},I_{2}}(X) = 0
\quad {\it iff} \quad |I_{1}| + |I_{2}| > 6.
\label{limit4}
\ee

It is convenient to define 
\bea
A_{1} \equiv A^{\emptyset,\emptyset},~
A_{2}^{\mu} \equiv A^{[\mu],\emptyset},~
A_{3}^{[\mu\nu]} \equiv A^{[\mu\nu],\emptyset},
A_{4}^{\mu;\nu} \equiv A^{[\mu],[\nu]},
\nonumber \\
A_{5}^{[\mu\nu];\rho} \equiv A^{[\mu\nu],\rho},~
A_{6}^{[\mu\nu];[\rho\sigma]} \equiv A^{[\mu\nu],[\rho\sigma]},
A_{7}^{\mu\nu\rho} \equiv A^{[\mu\nu\rho],\emptyset},
\nonumber \\
A_{8}^{[\mu\nu\rho];\sigma} \equiv A^{[\mu\nu\rho],[\sigma]},~
A_{9}^{[\mu\nu\rho];[\sigma\lambda]} \equiv A^{[\mu\nu\rho],[\sigma\lambda]},~
A_{10}^{[\mu\nu\rho];[\sigma\lambda\omega]} \equiv A^{[\mu\nu\rho],[\sigma\lambda\omega]}
\eea
where we have emphasized the antisymmetry properties with brackets. We have from (\ref{gauge2}) the following anomalous gauge equations:
\bea
d_{Q} T(T(x_{1}),T(x_{2})) =
\nonumber \\
i {\partial\over \partial x^{\mu}_{1}}T(T(x_{1}),T^{\mu}(x_{2}))
+ i {\partial\over \partial x^{\mu}_{2}}T(T(x_{1}),T^{\mu}(x_{2}))
+ A_{1}(x_{1},x_{2})
\label{g1}
\eea
\bea
d_{Q} T(T^{\mu}(x_{1}),T(x_{2})) =
\nonumber \\
i {\partial\over \partial x^{\mu}_{1}}T(T^{\mu\nu}(x_{1}),T(x_{2}))
-i {\partial\over \partial x^{\nu}_{2}} T(T^{\mu}(x_{1}),T^{\nu}(x_{2}))
+ A^{\mu}_{2}(x_{1},x_{2})
\label{g2}
\eea
\bea
d_{Q} T(T^{\mu\nu}(x_{1}),T(x_{2})) =
\nonumber \\
i {\partial\over \partial x^{\rho}_{1}}T(T^{\mu\nu\rho}(x_{1}),T(x_{2}))
+ i {\partial\over \partial x^{\rho}_{2}} 
T(T^{\mu\nu}(x_{1}),T^{\rho}(x_{2}))
+ A^{[\mu\nu]}_{3}(x_{1},x_{2})
\label{g3}
\eea
\bea
d_{Q} T(T^{\mu}(x_{1}),T^{\nu}(x_{2})) =
\nonumber \\
i {\partial\over \partial x^{\rho}_{1}}T(T^{\mu\rho}(x_{1}),T^{\nu}(x_{2}))
- i {\partial\over \partial x^{\rho}_{2}}T(T^{\mu}(x_{1}),T^{\nu\rho}(x_{2}))
+ A^{\mu;\nu}_{4}(x_{1},x_{2})
\label{g4}
\eea
\bea
d_{Q} T(T^{\mu\nu}(x_{1}),T^{\rho}(x_{2})) =
\nonumber \\
i {\partial\over \partial x^{\sigma}_{1}}T(T^{\mu\nu\sigma}(x_{1}),T^{\rho}(x_{2}))
+ i {\partial\over \partial x^{\sigma}_{2}}T(T^{\mu\nu}(x_{1}),T^{\rho\sigma}(x_{2}))
+ A^{[\mu\nu];\rho}_{5}(x_{1},x_{2})
\label{g5}
\eea
\bea
d_{Q} T(T^{\mu\nu}(x_{1}),T^{\rho\sigma}(x_{2})) =
\nonumber \\
i {\partial\over \partial x^{\lambda}_{1}}T(T^{\mu\nu\lambda}(x_{1}),T^{\rho\sigma}(x_{2}))
+ i {\partial\over \partial x^{\lambda}_{2}}T(T^{\mu\nu}(x_{1}),T^{\rho\sigma\lambda}(x_{2}))
+ A^{[\mu\nu];[\rho\sigma]}_{6}(x_{1},x_{2})
\label{g6}
\eea
\bea
d_{Q} T(T^{\mu\nu\rho}(x_{1}),T(x_{2})) =
- i {\partial\over \partial x^{\sigma}_{2}}T(T^{\mu\nu\rho}(x_{1}),T^{\sigma}(x_{2}))
+ A^{[\mu\nu\rho]}_{7}(x_{1},x_{2})
\label{g7}
\eea
\bea
d_{Q} T(T^{\mu\nu\rho}(x_{1}),T^{\sigma}(x_{2})) =
- i {\partial\over \partial x^{\lambda}_{2}}T(T^{\mu\nu\rho}(x_{1}),T^{\sigma\lambda}(x_{2}))
+ A^{[\mu\nu\rho];\sigma}_{8}(x_{1},x_{2})
\label{g8}
\eea
\bea
d_{Q} T(T^{\mu\nu\rho}(x_{1}),T^{\sigma\lambda}(x_{2})) =
- i {\partial\over \partial x^{\omega}_{2}} 
T(T^{\mu\nu\rho}(x_{1}),T^{\sigma\lambda\omega}(x_{2}))
+ A^{[\mu\nu\rho];[\sigma\lambda]}_{9}(x_{1},x_{2})
\label{g9}
\eea
\bea
d_{Q} T(T^{\mu\nu\rho}(x_{1}),T^{\sigma\lambda\omega}(x_{2})) = 0.
\label{g10}
\eea

From (\ref{symmetryA}) we get the following symmetry properties:
\be
A_{1}(x_{1},x_{2}) = A_{1}(x_{2},x_{1})
\label{sA1}
\ee
and we also have:
\be
A_{4}^{\mu;\nu}(x_{1},x_{2}) = - A_{4}^{\nu;\mu}(x_{2},x_{1}),
\label{s4'}
\ee
\be
A_{6}^{[\mu\nu];[\rho\sigma]}(x_{1},x_{2}) = 
A_{6}^{[\rho\sigma];[\mu\nu]}(x_{2},x_{1}),
\label{s6'}
\ee
and
\be
A_{10}^{[\mu\nu\rho];[\sigma\lambda\omega]}(x_{1},x_{2}) 
= - A_{10}^{[\sigma\lambda\omega];[\mu\nu\rho]}(x_{2},x_{1}).
\label{s7'}
\ee
 
The Wess-Zumino consistency conditions are in this case:
\be
d_{Q} A_{1}(x_{1},x_{2}) 
= - i {\partial\over \partial x^{\mu}_{1}}A^{\mu}_{2}(x_{1},x_{2})
- i {\partial\over \partial x^{\mu}_{2}}A^{\mu}_{2}(x_{2},x_{1})
\label{WZ1}
\ee
\be
d_{Q} A^{\mu}_{2}(x_{1},x_{2})
= - i {\partial\over \partial x^{\nu}_{1}}A^{[\mu\nu]}_{3}(x_{1},x_{2})
+ i {\partial\over \partial x^{\nu}_{2}}A^{\mu;\nu}_{4}(x_{1},x_{2})
\label{WZ2}
\ee
\be
d_{Q} A^{[\mu\nu]}_{3}(x_{1},x_{2})
= - i {\partial\over \partial x^{\rho}_{1}}A^{[\mu\nu\rho]}_{7}(x_{1},x_{2})
- i {\partial\over \partial x^{\rho}_{2}}A^{[\mu\nu];\rho}_{5}(x_{1},x_{2})
\label{WZ3}
\ee
\be
d_{Q} A^{\mu;\nu}_{4}(x_{1},x_{2})
= - i {\partial\over \partial x^{\rho}_{1}}A^{[\mu\rho];\nu}_{5}(x_{1},x_{2})
+ i {\partial\over \partial x^{\rho}_{2}}A^{[\nu\rho];\mu}_{5}(x_{2},x_{1})
\label{WZ4}
\ee
\be
d_{Q} A^{[\mu\nu];\rho}_{5}(x_{1},x_{2})
= - i {\partial\over \partial x^{\sigma}_{1}}A^{[\mu\nu\sigma];\rho}_{8}(x_{1},x_{2})
- i {\partial\over \partial x^{\sigma}_{2}}A^{[\mu\nu];[\rho\sigma]}_{6}(x_{1},x_{2})
\label{WZ5}
\ee
\be
d_{Q} A^{[\mu\nu];[\rho\sigma]}_{6}(x_{1},x_{2}) = 
- i {\partial\over \partial x^{\lambda}_{1}}A^{[\mu\nu\lambda];[\rho\sigma]}_{9}(x_{1},x_{2})
- i 
{\partial\over \partial x^{\lambda}_{2}}A^{[\rho\sigma\lambda];[\mu\nu]}_{9}(x_{2},x_{1});
\label{WZ6}
\ee
\be
d_{Q} A^{[\mu\nu\rho]}_{7}(x_{1},x_{2})
= i {\partial\over \partial x^{\sigma}_{2}}A^{[\mu\nu\rho];\sigma}_{8}(x_{1},x_{2});
\label{WZ7}
\ee
\be
d_{Q} A^{[\mu\nu\rho];\sigma}_{8}(x_{1},x_{2}) 
= i 
{\partial\over \partial x^{\lambda}_{2}}A^{[\mu\nu\rho];[\sigma\lambda]}_{9}(x_{1},x_{2});
\label{WZ8}
\ee
\be
d_{Q} A^{[\mu\nu\rho];[\sigma\lambda]}_{9}(x_{1},x_{2}) 
= i {\partial\over \partial x^{\omega}_{2}}A^{[\mu\nu\rho];[\sigma\lambda\omega]}_{10}(x_{1},x_{2});
\label{WZ9}
\ee
\be
d_{Q} A^{[\mu\nu\rho];[\sigma\lambda\omega]}_{10}(x_{1},x_{2}) = 0. 
\label{WZ10}
\ee

We suppose from now on that we work in a $4$-dimensional Minkowski space-time and we have the following result:
\begin{thm}

One can redefine the chronological products such that
\bea
A_{1}(x_{1},x_{2}) = \delta(x_{1} - x_{2})~W(x_{1}), \qquad
A^{\mu}_{2}(x_{1},x_{2}) = \delta(x_{1} - x_{2})~W^{\mu}(x_{1})
\nonumber \\
A^{[\mu\nu]}_{3}(x_{1},x_{2}) = \delta(x_{1} - x_{2})~W^{[\mu\nu]}(x_{1}), \qquad
A^{\mu;\nu}_{4}(x_{1},x_{2}) = - \delta(x_{1} - x_{2})~W^{[\mu\nu]}(x_{1}),
\nonumber \\
A^{[\mu\nu];\rho}_{5}(x_{1},x_{2}) = \delta(x_{1} - x_{2})~W^{[\mu\nu\rho]}(x_{1}), \qquad 
A^{[\mu\nu];[\rho\sigma]}_{7}(x_{1},x_{2}) = - \delta(x_{1} - x_{2})~W^{[\mu\nu\rho]}(x_{1})
\eea
and
$
A_{j} = 0,~j = 6, 8, 9, 10.
$
Moreover one has the following descent equations:
\be
d_{Q}W = - i~\partial_{\mu}W^{\mu},\qquad
d_{Q}W^{\mu} = i~\partial_{\nu}W^{[\mu\nu]},\qquad
d_{Q}W^{[\mu\nu]} = - i\partial_{\rho}W^{[\mu\nu\rho]},\qquad
d_{Q}W^{[\mu\nu\rho]} = 0.
\label{W}
\ee

The expressions 
$
W,~W^{\mu}
$ 
and
$
W^{[\mu\nu]}
$
are relative co-cycles and are determined up to relative co-boundaries. The expression
$
W^{[\mu\nu\rho]}
$
is a co-cycle and it is determined up to a co-boundary.
\label{ano-2}
\end{thm}

{\bf Proof:} The symmetry properties and the Wess-Zumino equations of consistency will be enough to obtain the result from the statement. We will rely on some computations done in \cite{cohomology}. We will use (\ref{genericA}) together with the restriction (\ref{power4}). Because we also have
\be
gh(W^{I_{1},I_{2};\{\rho_{1},\dots,\rho_{k}\}}) = |I_{1}| + |I_{2}| + 1
\label{gh2}
\ee
the sum goes in fact up to
$
k = 6.
$
If we get rid of the top terms (i.e. corresponding to
$
k = 5, 6
$)
from the preceding sum then we are, at least for
$
|I_{1}|, |I_{2}| \leq 2
$,
in the case studied in \cite{cohomology}.

We divide the proof in a number of steps.

(i) From (\ref{genericA}) we have:
\be
A_{1}(x_{1},x_{2})
= \sum_{k \leq 6} 
\partial_{\mu_{1}} \dots \partial_{\mu_{k}} \delta(x_{2} - x_{1})
W^{\{\mu_{1},\dots,\mu_{k}\}}_{1}(x_{1})
\label{A1-2}
\ee
and we have the restrictions 
\be
\omega(W^{\{\mu_{1},\dots,\mu_{k}\}}_{1}) \leq 7 - k, \qquad
gh(W^{\{\mu_{1},\dots,\mu_{k}\}}_{1}) = 1
\ee
for all 
$
k = 0, \dots, 6.
$
We perform the finite renormalization:
\be
T(T^{\mu_{1}}(x_{1}),T(x_{2})) \rightarrow T(T^{\mu_{1}}(x_{1}),T(x_{2})) 
+ \partial_{\mu_{2}}\cdots\partial_{\mu_{6}}~\delta(x_{2} - x_{1})
U^{\mu_{1};\{\mu_{2},\dots,\mu_{6}\}}_{2}(x_{1})
\label{R2}
\ee
and it is easy to see that if we choose
$
U^{\mu_{1};\{\mu_{2},\dots,\mu_{6}\}}_{2} = 
- {i\over 2}~W^{\{\mu_{1},\dots,\mu_{6}\}}_{1}
$
then we obtain a new expression (\ref{A1-2}) for the anomaly 
$
A_{1}
$
where the sum goes only up to 
$
k = 5.
$
(Although the monomials
$
W^{\{\mu_{1},\dots,\mu_{k}\}}_{1}
$
will be changed after this finite renormalization we keep the same notation.) Now we impose the symmetry property (\ref{sA1}) and consider only the terms with five derivatives on 
$\delta$;
it easily follows that
$
W^{\{\mu_{1},\dots,\mu_{5}\}}_{1} = 0
$
i.e. in the expression (\ref{A1-2}) for the anomaly 
$
A_{1}
$
the sum goes only up to 
$
k = 4.
$
Now we have the expression (3.53) from \cite{cohomology} and we can perform the succession of finite renormalizations from there. In the end the expression (\ref{A1-2}) will have the form from the statement.

(ii) From (\ref{genericA}) we have:
\be
A^{\mu}_{2}(x_{1},x_{2})
= \sum_{k \leq 5} 
\partial_{\rho_{1}} \dots \partial_{\rho_{k}} \delta(x_{2} - x_{1})
W^{\mu;\{\rho_{1},\dots,\rho_{k}\}}_{2}(x_{1})
\label{A2-2}
\ee
and we have the restrictions 
\be
\omega(W^{\mu;\{\rho_{1},\dots,\rho_{k}\}}_{2}) \leq 7 - k, \quad
gh(W^{\mu;\{\rho_{1},\dots,\rho_{k}\}}_{2}) = 2
\ee
for all 
$
k = 0, \dots, 5.
$
We use Wess-Zumino consistency condition (\ref{WZ1}); if we consider only the terms with six derivatives on 
$\delta$
we obtain that the completely symmetric part of
$
W^{\mu_{1};\mu_{2},\dots,\mu_{6}}_{2}
$
is null:
$
W^{\{\mu_{1};\mu_{2},\dots,\mu_{6}\}}_{2} = 0.
$
In this case it is easy to prove that one can write 
$
W^{\mu_{1};\mu_{2},\dots,\mu_{6}}_{2}
$
in the following form:
\be
W^{\mu_{1};\mu_{2},\dots,\mu_{6}}_{2} = 
{1 \over 5}~\sum_{j=2}^{6}~
\tilde{W}^{[\mu_{1}\mu_{j}];\{\mu_{2},\dots\hat{\mu_{j}},\dots,\mu_{6}\}}_{2}
\ee 
with 
\be
\tilde{W}^{[\mu_{1}\mu_{2}];\{\mu_{3},\dots,\mu_{6}\}}_{2} \equiv 
{5 \over 4}~W^{\mu_{1};\mu_{2},\dots,\mu_{6}}_{2} - (\mu_{1} \leftrightarrow \mu_{2}).
\ee

We perform the finite renormalization
\be
T(T^{[\mu_{1}\mu_{2}]}(x_{1}),T(x_{2})) \rightarrow T(T^{[\mu_{1}\mu_{2}]}(x_{1}),T(x_{2})) 
+ \partial_{\mu_{3}}\cdots\partial_{\mu_{6}}\delta(x_{2} - x_{1})
~U_{3}^{[\mu_{1}\mu_{2}];\{\mu_{3},\dots,\mu_{6}\}}(x_{1})
\label{R3a}
\ee
with
$
U_{3}^{[\mu_{1}\mu_{2}];\{\mu_{3},\dots,\mu_{6}\}} = - i~\tilde{W}^{[\mu_{1}\mu_{2}];\{\mu_{3},\dots,\mu_{6}\}}_{2}
$
and we eliminate the contributions corresponding to 
$k = 5$
from (\ref{A2-2}). We use again the Wess-Zumino consistency condition (\ref{WZ1}); if we consider only the terms with five derivatives on 
$\delta$
we obtain that the completely symmetric part of
$
W^{\mu_{1};\mu_{2},\dots,\mu_{5}}_{2}
$
is null
$
W^{\{\mu_{1};\mu_{2},\dots,\mu_{5}\}}_{2} = 0
$
and write:
\be
W_{2}^{\mu_{1};\mu_{2},\dots,\mu_{5}} = 
{1\over 4}~\sum_{j=2}^{5}~
\tilde{W}^{[\mu_{1}\mu_{j}];\{\mu_{2},\dots\hat{\mu_{j}},\dots,\mu_{5}\}}_{2}
\ee
with 
\be
\tilde{W}_{2}^{[\mu_{1}\mu_{2}];\{\mu_{3},\mu_{4},\mu_{5}\}} = 
{4\over 5}~W_{2}^{\mu_{1};\mu_{2},\dots,\mu_{5}} 
- (\mu_{1} \leftrightarrow \mu_{2}).
\ee

Now we consider the finite renormalization
\be
T(T^{[\mu\nu]}(x_{1}),T(x_{2})) \rightarrow T(T^{[\mu\nu]}(x_{1}),T(x_{2})) 
+ \partial_{\rho_{1}}\partial_{\rho_{2}}\partial_{\rho_{3}}
\delta(x_{2} - x_{1})~U_{3}^{[\mu\nu];\{\rho_{1}\rho_{2}\rho_{3}\}}(x_{1})
\label{R3b-2}
\ee
with 
$
U_{3}^{[\mu\nu];\rho_{1}\rho_{2}\rho_{3}} = i~\tilde{W}_{2}^{[\mu\nu];\rho_{1}\rho_{2}\rho_{3}} 
$
and we get a new expressions (\ref{A2-2}) for which
$
W_{2}^{\mu_{1};\{\mu_{2},\dots,\mu_{5}\}} = 0,
$
i.e. the summation in (\ref{A2-2}) goes only up to
$
k = 4.
$
As a result we have the expression (3.57) from \cite{cohomology} and we can perform the succession of finite renormalizations from there. In the end the expression (\ref{A2-2}) will have the form from the statement.

It is easy to prove that the Wess-Zumino equation (\ref{WZ1}) is now equivalent to:
\be
d_{Q}W_{1} = - i~\partial_{\mu}W^{\mu}_{2}.
\label{WZ1'}
\ee

(iii) From (\ref{genericA}) we have:
\be
A^{[\mu\nu]}_{3}(x_{1},x_{2})
= \sum_{k \leq 4} 
\partial_{\rho_{1}} \dots \partial_{\rho_{k}} \delta(x_{2} - x_{1})
W^{[\mu\nu];\{\rho_{1},\dots,\rho_{k}\}}_{3}(x_{1})
\label{A3-2}
\ee
and we have the restrictions 
\be
\omega(W^{[\mu\nu];\{\rho_{1},\dots,\rho_{k}\}}_{3}) \leq 7 - k,\quad
gh(W^{[\mu\nu];\{\rho_{1},\dots,\rho_{k}\}}_{3}) = 3
\ee
for all 
$
k = 0,\dots,4.
$

We perform the finite renormalization
\be
T(T^{[\mu\nu]}(x_{1}),T^{\rho}(x_{2})) \rightarrow T(T^{[\mu\nu]}(x_{1}),T^{\rho}(x_{2})) 
+ \partial_{\sigma_{1}}\partial_{\sigma_{2}}\partial_{\sigma_{3}}
\delta(x_{2} - x_{1})~U_{5}^{[\mu\nu];\rho;\{\sigma_{1}\sigma_{2}\sigma_{3}\}}(x_{1})
\label{R5a}
\ee
with
$
U_{5}^{[\mu\nu];\rho_{1};\{\rho_{2}\rho_{3}\rho_{4}\}} = 
i~W^{[\mu\nu];\{\rho_{1},\dots,\rho_{4}\}}_{3}
$
and we eliminate the contributions corresponding to 
$k = 4$
from (\ref{A3-2}). Now we consider the finite renormalization
\be
T(T^{[\mu\nu]}(x_{1}),T^{\rho}(x_{2})) \rightarrow T(T^{[\mu\nu]}(x_{1}),T^{\rho}(x_{2})) 
+ \partial_{\sigma_{1}}\partial_{\sigma_{2}}
\delta(x_{2} - x_{1})~U_{3}^{[\mu\nu];\rho;\{\sigma_{1}\sigma_{2}\}}(x_{1})
\label{R5b}
\ee
with
$
U_{5}^{[\mu\nu];\rho_{1};\{\rho_{2}\rho_{3}\}} = 
i~W_{3}^{[\mu\nu];\{\rho_{1}\rho_{2}\rho_{3}\}} 
$
and we get a new expressions (\ref{A3-2}) with
$
k \leq 2
$.
As a result we have the expression (3.67) from \cite{cohomology} and we can perform the succession of finite renormalizations from there. In the end the expression (\ref{A3-2}) will have the form from the statement.

(iv) From (\ref{genericA}) we have:
\be
A^{\mu;\nu}_{4}(x_{1},x_{2})
= \sum_{k \leq 4} 
\partial_{\rho_{1}} \dots \partial_{\rho_{k}} \delta(x_{2} - x_{1})
W^{\mu;\nu;\{\rho_{1},\dots,\rho_{k}\}}_{4}(x_{1})
\label{A4-2}
\ee
and we have the restrictions 
\be
\omega(W^{\mu;\nu;\{\rho_{1},\dots,\rho_{k}\}}_{4}) \leq 7 - k, \quad
gh(W^{\mu;\nu;\{\rho_{1},\dots,\rho_{k}\}}_{4}) = 3
\ee
for all 
$
k = 0,\dots,4.
$

We will have to consider the (anti)symmetry (\ref{s4'}). From the terms with four derivatives on delta we obtain that
$
W^{\mu;\nu;\{\rho_{1},\dots,\rho_{4}\}}_{4}
$
is antisymmetric in the first two indices i.e. we have the writing
$
W^{\mu;\nu;\{\rho_{1},\dots,\rho_{4}\}}_{4} = W^{[\mu\nu];\{\rho_{1},\dots,\rho_{4}\}}_{4}.
$

Next we consider the Wess-Zumino consistency condition (\ref{WZ2}). From the terms with five derivatives on delta we obtain
\be
{\cal S}_{\nu,\rho_{1},\dots,\rho_{4}}~W^{[\mu\nu];\{\rho_{1},\dots,\rho_{4}\}}_{4} = 0
\ee
where 
$
{\cal S}
$
denotes symmetrization in the corresponding indices. We note now that in the finite renormalization (\ref{R5a}) we have used only the expression
$
U_{5}^{[\mu\nu];\{\rho_{1};\rho_{2}\rho_{3}\rho_{4}\}}
$
i.e. we still can use 
$
U_{5}^{[\mu\nu];\rho_{1};\rho_{2}\rho_{3}\rho_{4}}
$
with
$
U_{5}^{[\mu\nu];\{\rho_{1};\rho_{2}\rho_{3}\rho_{4}\}} = 0.
$
It is not so complicated to prove (using the preceding relation) that the choice:
$
U_{5}^{[\mu\nu];\rho_{1};\rho_{2}\rho_{3}\rho_{4}} = c~
(W^{\mu;\rho_{1};\{\nu\rho_{2}\rho_{3}\rho_{4}\}}_{4} 
+ {1\over 4}~ W^{\mu;\nu;\{\rho_{1},\dots,\rho_{4}\}}_{4}) 
$
is possible i.e. it verifies the preceding relation; moreover if we take 
$
c = {8 i \over 15}
$
we get a new expression (\ref{A4-2}) for which
$
k \leq 3
$. 
We use again the (anti)symmetry property (\ref{s4'}); from the terms with three derivatives on $\delta$ we obtain:
\bea
W^{\mu;\nu;\{\rho_{1}\rho_{2}\rho_{3}\}}_{4} = W^{\nu;\mu;\{\rho_{1}\rho_{2}\rho_{3}\}}_{4}
\eea
i.e. we have the writing
$
W^{\mu;\nu;\{\rho_{1}\rho_{2}\rho_{3}\}}_{4} = 
W^{\{\mu\nu\};\{\rho_{1}\rho_{2}\rho_{3}\}}_{4}.
$
We consider again the Wess-Zumino consistency condition (\ref{WZ2}); from the terms with 
four derivatives on $\delta$ we obtain: 
\be
{\cal S}_{\nu,\rho_{1}\rho_{2}\rho_{3}}~W^{\{\mu\nu\};\{\rho_{1}\rho_{2}\rho_{3}\}}_{4} = 0.
\ee
As before we note now that in the finite renormalization (\ref{R5b}) we have used only the expression
$
U_{5}^{[\mu\nu];\{\rho_{1};\rho_{2}\rho_{3}\}}
$
i.e. we still can use 
$
U_{5}^{[\mu\nu];\rho_{1};\rho_{2}\rho_{3}}
$
with
$
U_{5}^{[\mu\nu];\{\rho_{1};\rho_{2}\rho_{3}\}} = 0.
$
A possible choice is:
$
U_{5}^{[\mu\nu];\rho_{1};\rho_{2}\rho_{3}} = c~
(W^{\mu;\rho_{1};\{\nu\rho_{2}\rho_{3}\}}_{4} 
+ {1\over 3}~ W^{\mu;\nu;\{\rho_{1}\rho_{2}\rho_{3}\}}_{4}) 
$; 
moreover if we take 
$
c = {9 i \over 16}
$
we get a new expression (\ref{A4-2}) for which
$
k \leq 3
$. 
As a result we have the expression (3.71) from \cite{cohomology} and we can perform the succession of finite renormalizations from there. In the end the expression (\ref{A3-2}) will have the form from the statement. The Wess-Zumino equation (\ref{WZ2}) is equivalent to:
\bea
d_{Q}W^{\mu}_{2} = i~\partial_{\nu}W_{3}^{[\mu\nu]}
\nonumber \\
W^{\mu;\nu}_{4} = - W^{[\mu\nu]}_{3}.
\eea

(v) From (\ref{genericA}) we have:
\be
A^{[\mu\nu\rho]}_{7}(x_{1},x_{2})
= \sum_{k \leq 3} 
\partial_{\sigma_{1}} \dots \partial_{\sigma_{k}} \delta(x_{2} - x_{1})
W^{[\mu\nu\rho];\{\sigma_{1},\dots,\sigma_{k}\}}_{7}(x_{1})
\label{A7-2}
\ee
and we have the restrictions 
\be
\omega(W^{[\mu\nu\rho];\{\sigma_{1},\dots,\sigma_{k}\}}_{4}) \leq 7 - k, \quad
gh(W^{[\mu\nu\rho];\{\sigma_{1},\dots,\sigma_{k}\}}_{4}) = 4
\ee
for all 
$
k = 0,\dots,3.
$
We perform the finite renormalization
\be
T(T^{[\mu\nu\rho]}(x_{1}),T^{\sigma}(x_{2})) \rightarrow T(T^{[\mu\nu\rho]}(x_{1}),T^{\sigma}(x_{2})) 
+ \partial_{\lambda_{1}}\partial_{\lambda_{2}}\delta(x_{2} - x_{1})~
U_{8}^{[\mu\nu\rho];\sigma;\{\lambda_{1}\lambda_{2}\}}(x_{1})
\label{R8a}
\ee
with
$
U_{8}^{[\mu\nu\rho];\sigma;\{\lambda_{1}\lambda_{2}\}} = 
- i~W^{[\mu\nu\rho];\{\sigma\lambda_{1}\lambda_{2}\}}_{7}
$
and we eliminate the contributions corresponding to 
$k = 3$
from (\ref{A7-2}). Now we consider the finite renormalization
\be
T(T^{[\mu\nu\rho]}(x_{1}),T^{\sigma}(x_{2})) \rightarrow T(T^{[\mu\nu\rho]}(x_{1}),T^{\sigma}(x_{2})) 
+ \partial_{\lambda}\delta(x_{2} - x_{1})~U_{8}^{[\mu\nu\rho];\sigma;\lambda}(x_{1})
\label{R8b}
\ee
with
$
U_{8}^{[\mu\nu\rho];\sigma;\lambda} = - i~W_{7}^{[\mu\nu\rho];\{\sigma\lambda\}} 
$
and we get a new expressions (\ref{A7-2}) with
$
k \leq 1
$.
Finally we consider the finite renormalization
\be
T(T^{[\mu\nu\rho]}(x_{1}),T^{\sigma}(x_{2})) \rightarrow T(T^{[\mu\nu\rho]}(x_{1}),T^{\sigma}(x_{2})) 
+ \delta(x_{2} - x_{1})~U_{8}^{[\mu\nu\rho];\sigma}(x_{1})
\label{R8c}
\ee
with
$
U_{8}^{[\mu\nu\rho];\sigma} = - i~W_{7}^{[\mu\nu\rho];\sigma} 
$
and we get the expression for 
$
A_{7}
$
from the statement.

(vi) From (\ref{genericA}) we have:
\be
A^{[\mu\nu];\rho}_{5}(x_{1},x_{2})
= \sum_{k \leq 3} 
\partial_{\sigma_{1}} \dots \partial_{\sigma_{k}} \delta(x_{2} - x_{1})
W^{[\mu\nu];\rho;\{\sigma_{1},\dots,\sigma_{k}\}}_{5}(x_{1})
\label{A5-2}
\ee
and we have the restrictions 
\be
\omega(W^{[\mu\nu];\rho;\{\sigma_{1},\dots,\sigma_{k}\}}_{4}) \leq 7 - k, \quad
gh(W^{[\mu\nu];\rho;\{\sigma_{1},\dots,\sigma_{k}\}}_{4}) = 4
\ee

We consider the Wess-Zumino consistency conditions (\ref{WZ3}). From the terms with four derivatives on delta we obtain:
\be
{\cal S}_{\rho,\sigma_{1}\sigma_{2}\sigma_{3}}~
W^{[\mu\nu];\rho;\{\sigma_{1}\sigma_{2}\sigma_{3}\}}_{5} = 0.
\ee
This equation can be solved explicitly: if we denote:
\be
\tilde{W}^{[\mu\nu];[\rho\sigma_{1}];\{\sigma_{2}\sigma_{3}\}}_{5} =
{3\over 4}~W^{[\mu\nu];\rho;\{\sigma_{1}\sigma_{2}\sigma_{3}\}}_{5} 
- (\rho \leftrightarrow \sigma_{1})
\ee 
we have:
\be
W^{[\mu\nu];\rho;\{\sigma_{1}\sigma_{2}\sigma_{3}\}}_{5} 
= {\cal S}_{\sigma_{1}\sigma_{2}\sigma_{3}}~
\tilde{W}^{[\mu\nu];[\rho\sigma_{1}];\{\sigma_{2}\sigma_{3}\}}_{5}
\ee
and we can make in (\ref{A5-2})
$
W^{[\mu\nu];\rho;\{\sigma_{1}\sigma_{2}\sigma_{3}\}}_{5} 
\rightarrow
\tilde{W}^{[\mu\nu];[\rho\sigma_{1}];\{\sigma_{2}\sigma_{3}\}}_{5}.
$

From the Wess-Zumino consistency conditions (\ref{WZ4}) we consider again the terms with four derivatives on delta and we obtain after some computations:
\be
{\cal S}_{\rho,\sigma_{1}\sigma_{2}\sigma_{3}}~
(\tilde{W}^{[\mu\rho];[\nu\sigma_{1}];\{\sigma_{2}\sigma_{3}\}}_{5}
- \tilde{W}^{[\nu\rho];[\mu\sigma_{1}];\{\sigma_{2}\sigma_{3}\}}_{5}) = 0. 
\ee
It is convenient to split 
$
\tilde{W}^{[\mu\nu];[\rho\sigma_{1}];\{\sigma_{2}\sigma_{3}\}}_{5}
$
as follows
\be
\tilde{W}^{[\mu\nu];[\rho\sigma];\{\lambda_{1}\lambda_{2}\}}_{5}
= \tilde{W}^{[\mu\nu];[\rho\sigma];\{\lambda_{1}\lambda_{2}\}}_{5,+}
+ \tilde{W}^{[\mu\nu];[\rho\sigma];\{\lambda_{1}\lambda_{2}\}}_{5,-}
\ee
where
\be
\tilde{W}^{[\mu\nu];[\rho\sigma];\{\lambda_{1}\lambda_{2}\}}_{5,\epsilon}
= {1\over 2}~(\tilde{W}^{[\mu\nu];[\rho\sigma];\{\lambda_{1}\lambda_{2}\}}_{5}
+ \epsilon~\tilde{W}^{[\rho\sigma];[\mu\nu];\{\lambda_{1}\lambda_{2}\}}_{5}).
\ee

We now make the finite renormalization
\be
T(T^{[\mu\nu]}(x_{1}),T^{[\rho\sigma]}(x_{2})) \rightarrow T(T^{[\mu\nu]}(x_{1}),T^{[\rho\sigma]}(x_{2})) 
+ \partial_{\lambda_{1}}\partial_{\lambda_{2}}\delta(x_{1} - x_{2})~
U_{6}^{[\mu\nu];[\rho\sigma];\{\lambda_{1}\lambda_{2}\}}(x_{1})
\label{R6a}
\ee
with
$
U_{6}^{[\mu\nu];[\rho\sigma];\{\lambda_{1}\lambda_{2}\}} = 
i~\tilde{W}^{[\mu\nu];[\rho\sigma];\{\lambda_{1}\lambda_{2}\}}_{5,+}
$
such that all symmetry properties of the chronological products are preserved. As a result 
we get a new expression (\ref{A5-2}) with: 
$
\tilde{W}^{[\mu\nu];[\rho\sigma];\{\lambda_{1}\lambda_{2}\}}_{5}
\rightarrow
\tilde{W}^{[\mu\nu];[\rho\sigma];\{\lambda_{1}\lambda_{2}\}}_{5,-}
\equiv W^{[\mu\nu];[\rho\sigma];\{\lambda_{1}\lambda_{2}\}}.
$

The Wess-Zumino consistency conditions (\ref{WZ4}) with four derivatives on delta from above reduces to:
\be
{\cal S}_{\rho\sigma_{1}\sigma_{2}\sigma_{3}}~
W^{[\mu\rho];[\nu\sigma_{1}];\{\sigma_{2}\sigma_{3}\}}_{5} = 0. 
\ee

We note that we still can use the finite renormalization (\ref{R8a}) if we require:
\be
{\cal S}_{\rho\lambda_{1}\lambda_{2}}~
U_{8}^{[\mu\nu\rho];\sigma;\{\lambda_{1};\lambda_{2}\}} = 0 
\ee
i.e. such that we do not spoil the form of
$
A_{7}
$
from the statement. One can write a generic form for such an expression
$
U_{8}^{[\mu\nu\rho];\sigma;\{\lambda_{1};\lambda_{2}\}}
$ 
in terms of 
$
W^{[\mu\nu];[\rho\sigma];\{\lambda_{1}\lambda_{2}\}}.
$
There are five possible combinations meeting the symmetry properties:
\bea
U_{8}^{[\mu\nu\rho];\sigma_{1};\{\sigma_{2};\sigma_{3}\}} =
{\cal A}_{\mu\nu\rho}~{\cal S}_{\sigma_{2}\sigma_{3}}~
(c_{1}~W^{[\mu\rho];[\nu\sigma_{1}];\{\sigma_{2}\sigma_{3}\}}
+ c_{2}~W^{[\mu\rho];[\nu\sigma_{2}];\{\sigma_{1}\sigma_{2}\}}
\nonumber \\
+ c_{3}~W^{[\mu\sigma_{1}];[\nu\sigma_{2}];\{\rho\sigma_{3}\}}
+ c_{4}~W^{[\mu\sigma_{2}];[\nu\sigma_{3}];\{\rho\sigma_{1}\}}
+ c_{5}~W^{[\mu\rho];[\sigma_{1}\sigma_{2}];\{\nu\sigma_{3}\}})
\eea
where we apply the corresponding (anti -)symmetrization operators. It comes after some work that one can fix the coefficients such that we get a new expression (\ref{A5-2}) for which
$
k \leq 2
$; 
we have found the possible values 
$
c_{1} = 3 i,~c_{2} = i,~c_{3} = 0,~c_{4} = 2 i,~c_{5} = 0.
$

We consider again the Wess-Zumino consistency conditions (\ref{WZ3}); from the terms with three derivatives on delta we obtain:
\be
{\cal S}_{\rho\sigma_{1}\sigma_{2}}~W^{[\mu\nu];\rho;\{\sigma_{1}\sigma_{2}\}}_{5} = 0.
\ee
This equation can also be solved explicitly: if we denote:
\be
\tilde{W}^{[\mu\nu];[\rho\sigma_{1}];\{\sigma_{2}\sigma_{3}\}}_{5} =
{2\over 3}~W^{[\mu\nu];\rho;\{\sigma_{1}\sigma_{2}\}}_{5} 
- (\rho \leftrightarrow \sigma_{1})
\ee 
we have:
\be
W^{[\mu\nu];\rho;\{\sigma_{1}\sigma_{2}\}}_{5} 
= {\cal S}_{\sigma_{1}\sigma_{2}}~\tilde{W}^{[\mu\nu];[\rho\sigma_{1}];\sigma_{2}}_{5}
\ee
and we can make in (\ref{A5-2})
$
W^{[\mu\nu];\rho;\{\sigma_{1}\sigma_{2}\}}_{5} 
\rightarrow
\tilde{W}^{[\mu\nu];[\rho\sigma_{1}];\{\sigma_{2}\}}_{5}.
$

From the Wess-Zumino consistency conditions (\ref{WZ4}) we consider the terms with three derivatives on $\delta$ and we obtain:
\be
{\cal S}_{\rho\sigma_{1}\sigma_{2}}~
(\tilde{W}^{[\mu\rho];[\nu\sigma_{1}];\sigma_{2}}_{5}
- \tilde{W}^{[\nu\rho];[\mu\sigma_{1}];\sigma_{2}}_{5}) = 0. 
\ee
It is convenient to split 
$
\tilde{W}^{[\mu\nu];[\rho\sigma];\lambda}_{5}
$
as before
\be
\tilde{W}^{[\mu\nu];[\rho\sigma];\lambda}_{5}
= \tilde{W}^{[\mu\nu];[\rho\sigma];\lambda}_{5,+}
+ \tilde{W}^{[\mu\nu];[\rho\sigma];\lambda}_{5,-}
\ee
where
\be
\tilde{W}^{[\mu\nu];[\rho\sigma];\lambda}_{5,\epsilon}
= {1\over 2}~(\tilde{W}^{[\mu\nu];[\rho\sigma];\lambda}_{5}
+ \epsilon~\tilde{W}^{[\rho\sigma];[\mu\nu];\lambda}_{5}).
\ee

We now make the finite renormalization
\be
T(T^{[\mu\nu]}(x_{1}),T^{[\rho\sigma]}(x_{2})) \rightarrow T(T^{[\mu\nu]}(x_{1}),T^{[\rho\sigma]}(x_{2})) 
+ \partial_{\lambda}\delta(x_{1} - x_{2})~U_{6}^{[\mu\nu];[\rho\sigma];\lambda}(x_{1})
\label{R6b}
\ee
with
$
U_{6}^{[\mu\nu];[\rho\sigma];\lambda} = 
i~\tilde{W}^{[\mu\nu];[\rho\sigma];\lambda}_{5,+}
$
such that all symmetry properties of the chronological products are preserved. As a result 
we get a new expression (\ref{A5-2}) with: 
$
\tilde{W}^{[\mu\nu];[\rho\sigma];\lambda}_{5}
\rightarrow
\tilde{W}^{[\mu\nu];[\rho\sigma];\lambda}_{5,-}
\equiv W^{[\mu\nu];[\rho\sigma];\lambda}.
$

The Wess-Zumino consistency conditions (\ref{WZ4}) with three derivatives on $\delta$ from above reduces to:
\be
{\cal S}_{\rho\sigma_{1}\sigma_{2}}~W^{[\mu\rho];[\nu\sigma_{1}];\sigma_{2}}_{5} = 0. 
\ee

We note that we still can use the finite renormalization (\ref{R8b}) if we require:
\be
U_{8}^{[\mu\nu\rho];\sigma;\lambda} = - (\sigma \leftrightarrow \lambda) 
\ee
such that we do not spoil the form of
$
A_{7}
$
from the statement. One can write a generic form for such an expression
$
U_{8}^{[\mu\nu\rho];[\sigma\lambda]}
$
and the possible combinations meeting the symmetry properties are:
\be
U_{8}^{[\mu\nu\rho];[\sigma\lambda]} =
{\cal A}_{\mu\nu\rho}~{\cal A}_{\sigma\lambda}~
(c_{1}~W^{[\mu\rho];[\nu\sigma];\lambda}
+ c_{2}~W^{[\mu\nu];[\sigma\lambda];\rho})
\ee
where we apply the corresponding antisymmetrization operators. If we take 
$
c_{1} = {2 i \over 3},~c_{2} = {8 i \over 3}
$
we get a new expression (\ref{A5-2}) for which
$
k \leq 1
$.
As a result we have the expression (3.77) from \cite{cohomology} and we can perform the succession of finite renormalizations from there. In the end the expression (\ref{A5-2}) will have the form from the statement.

The Wess-Zumino equation (\ref{WZ3}) becomes equivalent to
\bea
d_{Q}W_{3}^{[\mu\nu]} = - i~\partial_{\rho}W_{7}^{[\mu\nu\rho]}
\nonumber \\
W_{5}^{[\mu\nu];\rho} = W_{7}^{[\mu\nu\rho]}.
\eea

The Wess-Zumino consistency conditions (\ref{WZ4}) is equivalent to
\be
d_{Q}W_{4}^{\mu;\nu} = i~\partial_{\rho}W_{7}^{[\mu\nu\rho]}
\ee
which follows from the preceding relation if we remember the connection between 
$
W_{3}^{[\mu\nu]}
$
and
$
W_{4}^{\mu;\nu}
$
obtained at (iv).

(vii) From (\ref{genericA}) we have:
\be
A^{[\mu\nu\rho];\sigma}_{8}(x_{1},x_{2})
= \sum_{k \leq 2} 
\partial_{\lambda_{1}} \dots \partial_{\lambda_{k}}\delta(x_{2} - x_{1})
W^{[\mu\nu\rho];\sigma;\{\lambda_{1},\dots,\lambda_{k}\}}_{8}(x_{1})
\label{A8-2}
\ee
and we have the restrictions 
\be
\omega(W^{[\mu\nu\rho];\sigma;\{\lambda_{1},\dots,\lambda_{k}\}}_{8}) \leq 7 - k, \quad
gh(W^{[\mu\nu\rho];\sigma;\{\lambda_{1},\dots,\lambda_{k}\}}_{8}) = 5
\ee
for all 
$
k = 0,1,2.
$

We consider the Wess-Zumino consistency condition (\ref{WZ7}). From the terms with three derivatives on $\delta$ we obtain
\be
{\cal S}_{\sigma\lambda_{1}\lambda_{2}}~
W^{[\mu\nu\rho];\sigma;\{\lambda_{1}\lambda_{2}\}}_{8} = 0.
\ee
This equation can be solved explicitly: if we denote:
\be
\tilde{W}^{[\mu\nu\rho];[\sigma\lambda_{1}];\lambda_{2}}_{8} =
{2\over 3}~W^{[\mu\nu\rho];\sigma;\{\lambda_{1}];\lambda_{2}\}}_{8} 
- (\sigma \leftrightarrow \lambda_{1})
\ee 
we have:
\be
W^{[\mu\nu\rho];\sigma;\{\lambda_{1}];\lambda_{2}\}}_{8}
= {\cal S}_{\lambda_{1}\lambda_{2}}~
\tilde{W}^{[\mu\nu\rho];[\sigma\lambda_{1}];\lambda_{2}}_{8}
\ee
and we can make in (\ref{A8-2})
$
W^{[\mu\nu\rho];\sigma;\{\lambda_{1}\lambda_{2}\}}_{8} 
\rightarrow
\tilde{W}^{[\mu\nu\rho];[\sigma\lambda_{1}];\lambda_{2}}_{8}.
$

We make the finite renormalization
\be
T(T^{[\mu\nu\rho]}(x_{1}),T^{[\sigma\lambda]}(x_{2})) \rightarrow T(T^{[\mu\nu\rho]}(x_{1}),T^{[\sigma\lambda]}(x_{2})) 
+ \partial_{\alpha}\delta(x_{1} - x_{2})~
U_{9}^{[\mu\nu\rho];[\sigma\lambda];\alpha}(x_{1})
\label{R9a}
\ee
with
$
U_{9}^{[\mu\nu\rho];\sigma;\{\lambda_{1}\lambda_{2}\}} = 
- i~\tilde{W}^{[\mu\nu\rho];\sigma];\{\lambda_{1}\lambda_{2}\}}_{8}
$ 
and we get a new expression (\ref{A8-2}) with
$
k \leq 1
$.

We consider again the Wess-Zumino consistency condition (\ref{WZ7}); from the terms with 
two derivatives on $\delta$ we obtain: 
\be
W^{[\mu\nu\rho];\sigma;\lambda}_{8} = - (\sigma \leftrightarrow \lambda)
\ee
i.e. we have the writing
$
W^{[\mu\nu\rho];\sigma;\lambda}_{8} = W^{[\mu\nu\rho];[\sigma\lambda]}_{8}
$.

We now make the finite renormalization
\be
T(T^{[\mu\nu\rho]}(x_{1}),T^{[\sigma\lambda]}(x_{2})) \rightarrow T(T^{[\mu\nu\rho]}(x_{1}),T^{[\sigma\lambda]}(x_{2})) 
+ \delta(x_{1} - x_{2})~U_{9}^{[\mu\nu\rho];[\sigma\lambda]}(x_{1})
\label{R9b}
\ee
with
$
U_{9}^{[\mu\nu\rho];[\sigma\lambda]} = 
- i~\tilde{W}^{[\mu\nu\rho];[\sigma\lambda]}_{8}
$ 
we get a new expression
\be
A^{[\mu\nu\rho];\sigma}_{8}(x_{1},x_{2})
= \delta(x_{1} - x_{2}) W^{[\mu\nu\rho;\sigma}_{8}(x_{1}).
\label{A8a}
\ee
But the Wess-Zumino consistency condition (\ref{WZ7}) is in this case equivalent to
\bea
d_{Q}W_{7}^{[\mu\nu\rho]} = 0
\nonumber \\
W^{[\mu\nu\rho;\sigma}_{8} = 0
\eea
so we have in fact:
\be
A^{[\mu\nu\rho];\sigma}_{8}(x_{1},x_{2}) = 0.
\label{A8-2'}
\ee

(viii) From (\ref{genericA}) we have:
\be
A^{[\mu\nu];[\rho\sigma]}_{6}(x_{1},x_{2})
= \sum_{k \leq 2} 
\partial_{\lambda_{1}} \dots \partial_{\lambda_{k}}\delta(x_{2} - x_{1})
W^{[\mu\nu];[\rho\sigma];\{\lambda_{1},\dots,\lambda_{k}\}}_{6}(x_{1})
\label{A6-2}
\ee
and we have the restrictions 
\be
\omega(W^{[\mu\nu];[\rho\sigma];\{\lambda_{1},\dots,\lambda_{k}\}}_{6}) \leq 7 - k
\qquad
gh(W^{[\mu\nu];[\rho\sigma];\{\lambda_{1},\dots,\lambda_{k}\}}_{6}) = 5.
\ee

From the symmetry property (\ref{s6'}) we consider the terms with two derivatives on the $\delta$ function and obtain:
\be
W^{[\mu\nu];[\rho\sigma];\{\lambda_{1}\lambda_{2}\}}_{6} 
= W^{[\rho\sigma];[\mu\nu];\{\lambda_{1}\lambda_{2}\}}_{6}.
\ee

Now the Wess-Zumino consistency condition (\ref{WZ5}) gives:
\be
{\cal S}_{\sigma\lambda_{1}\lambda_{2}}~
W^{[\mu\nu];[\rho\sigma];\{\lambda_{1}\lambda_{2}\}}_{6} = 0.
\ee

We observe that in the renormalization (\ref{R9a}) we have used only the piece
$
{\cal S}_{\lambda\alpha} U_{9}^{[\mu\nu\rho];[\sigma\lambda];\alpha}
$
so we still can use 
$
{\cal A}_{\lambda\alpha} U_{9}^{[\mu\nu\rho];[\sigma\lambda];\alpha}.
$
We make the following ansatz for 
$
U_{9}^{[\mu\nu\rho];\sigma\lambda;\alpha}
$
\bea
U_{9}^{[\mu\nu\lambda_{1}];[\rho\sigma];\lambda_{2}} =
{\cal A}_{\mu\nu\lambda_{1}}~{\cal A}_{\rho\sigma}~
(c_{1}~W^{[\mu\nu];[\rho\sigma];\{\lambda_{1}\lambda_{2}\}}
+ c_{2}~W^{[\mu\nu];[\sigma\lambda_{2}];\{\lambda_{1}\rho\}}
\nonumber \\
+ c_{3}~W^{[\lambda_{1}\lambda_{2}];[\mu\rho];\{\nu\sigma\}}
+ c_{4}~W^{[\mu\rho];[\nu\sigma];\{\lambda_{1}\lambda_{2}\}})
\eea
which is compatible with the symmetry properties. A long computation shows that one can fix these coefficients such that the renormalization (\ref{R9a}) leaves the expression
$
A_{8}
$
unchanged but the expression(\ref{A6-2}) gets modified: we have the restriction
$
k \leq 1.
$
Now from the symmetry property (\ref{s6'}) with one derivatives on the $\delta$ function we obtain:
\be
W^{[\mu\nu];[\rho\sigma];\lambda}_{6} = - W^{[\rho\sigma];[\mu\nu];\lambda}_{6}
\ee
and from the Wess-Zumino equation (\ref{WZ5}):
\be
W^{[\mu\nu];[\rho\sigma];\lambda}_{6} =  - (\sigma \leftrightarrow \lambda).
\ee
If we combine these two equations we arrive at the conclusion that
$
W^{[\mu\nu];[\rho\sigma];\lambda}_{6}
$
is completely antisymmetric in all indices so it must be null (because we are in $4$ dimensions). As a consequence
\be
A^{[\mu\nu];[\rho\sigma]}_{6}(x_{1} - x_{2}) = 
\delta(x_{1} - x_{2})~W^{[\mu\nu];[\rho\sigma]}_{6}(x_{1}).
\ee
Now the Wess-Zumino equation (\ref{WZ5}) is equivalent to
\be
W^{[\mu\nu];[\rho\sigma]}_{6} = 0
\ee
so in fact:
\be
A^{[\mu\nu];[\rho\sigma]}_{6} = 0.
\ee

(ix) From (\ref{genericA}) we have:
\be
A^{[\mu\nu\rho];[\sigma\lambda]}_{9}(x_{1},x_{2})
= \delta(x_{2} - x_{1})~W^{[\mu\nu\rho];[\sigma\lambda]}_{9}(x_{1})
+ \partial_{\omega}\delta(x_{2} - x_{1})
W^{[\mu\nu\rho];[\sigma\lambda];\omega}_{6}(x_{1})
\label{A9-2}
\ee
and we have the restrictions 
\be
\omega(W^{[\mu\nu\rho];[\sigma\lambda]}_{9}) \leq 7
\qquad
\omega(W^{[\mu\nu\rho];[\sigma\lambda];\omega}_{9}) \leq 6
\qquad
gh(W^{[\mu\nu\rho];[\sigma\lambda]}_{9}) = (W^{[\mu\nu\rho];[\sigma\lambda];\omega}_{9}) = 6.
\ee

Now the Wess-Zumino consistency condition (\ref{WZ8}) gives:
\be
W^{[\mu\nu\rho];[\sigma\lambda];\omega}_{9} = - (\lambda \leftrightarrow \omega)
\ee
so we can write 
$
W^{[\mu\nu\rho];[\sigma\lambda];\omega}_{9} = W^{[\mu\nu\rho];[\sigma\lambda\omega]}_{9}. 
$
We perform the finite renormalization
\be
T(T^{[\mu\nu\rho]}(x_{1}),T^{[\sigma\lambda\omega]}(x_{2})) \rightarrow T(T^{[\mu\nu\rho]}(x_{1}),T^{[\sigma\lambda\omega]}(x_{2})) 
+ \delta(x_{2} - x_{1})~U_{10}^{[\mu\nu\rho];[\sigma\lambda\omega]}(x_{1})
\label{R10}
\ee
with
$
U_{10}^{[\mu\nu\rho];[\sigma\lambda\omega]} = i~W_{9}^{[\mu\nu\rho];[\sigma\lambda\omega]} 
$
As a consequence the formula (\ref{A9-2}) becomes
\be
A^{[\mu\nu\rho];[\sigma\lambda]}_{9}(x_{1} - x_{2}) = 
\delta(x_{1} - x_{2})~W^{[\mu\nu\rho];[\sigma\lambda]}_{9}(x_{1}).
\ee
Now the Wess-Zumino equation (\ref{WZ8}) is equivalent to
\be
W^{[\mu\nu\rho];[\sigma\lambda]}_{9} = 0
\ee
so in fact:
\be
A^{[\mu\nu\rho];[\sigma\lambda]}_{9} = 0.
\ee

(x) From (\ref{genericA}) we have:
\be
A^{[\mu\nu\rho];[\sigma\lambda\omega]}_{10}(x_{1},x_{2})
= \delta(x_{2} - x_{1})~W^{[\mu\nu\rho];[\sigma\lambda\omega]}_{9}(x_{1})
W^{[\mu\nu\rho];[\sigma\lambda\omega]}_{10}(x_{1})
\label{A10-2}
\ee
and we have the restrictions 
\be
\omega(W^{[\mu\nu\rho];[\sigma\lambda\omega]}_{10}) \leq 7
\qquad
gh(W^{[\mu\nu\rho];[\sigma\lambda\omega]}_{10}) = 7.
\ee

The Wess-Zumino equation (\ref{WZ9}) is equivalent to
$
W^{[\mu\nu\rho];[\sigma\lambda\omega]}_{10} = 0
$
so in fact 
\be
A^{[\mu\nu\rho];[\sigma\lambda\omega]}_{10}(x_{1},x_{2}) = 0.
\ee

(xi) Finally we observe that we can make some redefinitions of the chronological products without changing the structure of the anomalies. Indeed we have
\be
T(T(x_{1}),T(x_{2})) \rightarrow T(T(x_{1}),T(x_{2})) + \delta(x_{1} - x_{2})~B(x_{1})
\ee
which makes
\be
W \rightarrow W + d_{Q}B 
\ee
and
\be
T(T^{\mu}(x_{1}),T(x_{2})) \rightarrow T(T^{\mu}(x_{1}),T(x_{2})) 
+ \delta(x_{1} - x_{2})~B^{\mu}(x_{1})
\ee
which makes
\be
W \rightarrow W + i~\partial_{\mu}B^{\mu}, \qquad 
W^{\mu} \rightarrow W^{\mu} + d_{Q}B^{\mu}.
\ee
We also observe that we can consider the finite renormalizations
\be
T(T^{[\mu\nu]}(x_{1}),T(x_{2})) \rightarrow T(T^{[\mu\nu]}(x_{1}),T(x_{2})) 
+ \delta(x_{2} - x_{1})~U_{3}^{[\mu\nu]}(x_{1})
\label{R3c}
\ee
and
\be
T(T^{\mu}(x_{1}),T^{\nu}(x_{2})) \rightarrow T(T^{\mu}(x_{1}),T^{\nu}(x_{2})) 
+ \delta(x_{2} - x_{1})~U_{4}^{[\mu\nu]}(x_{1})
\label{R4b}
\ee
with
\be
U_{3}^{[\mu\nu]} = B^{[\mu\nu]}, \qquad U_{4}^{[\mu\nu]} = - B^{[\mu\nu]}
\ee
and they produce the redefinitions
\be
W^{\mu} \rightarrow W^{\mu} + i~\partial_{\nu}B^{[\mu\nu]}, \qquad 
W^{[\mu\nu]} \rightarrow W^{[\mu\nu]} + d_{Q}B^{[\mu\nu]}.
\ee
Finally we have the finite renormalizations
\be
T(T^{[\mu\nu]}(x_{1}),T^{\rho}(x_{2})) \rightarrow T(T^{[\mu\nu]}(x_{1}),T^{\rho}(x_{2})) 
+ \delta(x_{2} - x_{1})~U_{5}^{[\mu\nu];\rho}(x_{1})
\label{R5}
\ee
and
\be
T(T^{[\mu\nu\rho]}(x_{1}),T(x_{2})) \rightarrow T(T^{[\mu\nu\rho]}(x_{1}),T(x_{2})) 
+ \delta(x_{2} - x_{1})~U_{7}^{[\mu\nu\rho]}(x_{1})
\label{R7}
\ee
with
\be
U_{5}^{[\mu\nu];\rho} = U_{7}^{[\mu\nu\rho]} = B^{[\mu\nu\rho]}
\ee
and they produce the redefinitions
\be
W^{[\mu\nu]} \rightarrow W^{[\mu\nu]} + i~\partial_{\rho}B^{[\mu\nu\rho]}, \qquad 
W^{[\mu\nu\rho]} \rightarrow W^{[\mu\nu\rho]} + d_{Q}B^{[\mu\nu\rho]}.
\ee

All these redefinitions do not modify the form of the anomalies from the statement and we have obtained the last assertion of the theorem.
$\qed$

As we can see one can simplify considerably the form of the anomalies in the second order of the perturbation theory if one makes convenient redefinitions of the chronological products. Moreover, the result is of purely cohomological nature i.e. we did not use the explicit form of the expressions
$
T, T^{\mu}, T^{[\mu\nu]}, T^{[\mu\nu\rho]}.
$
The main difficulty of the proof is to find a convenient way of using Wess-Zumino equations, the (anti)symmetry properties and a succession of finite renormalizations. It will be a remarkable fact to extend the preceding result for arbitrary order of the perturbation theory. 

We have proved that renormalization of gauge theories leads to some descent equations. 
We have the expressions 
$
T^{I}
$
and 
$
R^{I}
$
(with ghost numbers 
$
gh(T^{I}) = gh(R^{I}) = |I|
$ 
and canonical dimension $\leq 5$ and $\leq 6$ respectively) for the interaction Lagrangian and the finite renormalizations compatible with gauge invariance; we also have the expressions 
$
W^{I}
$
(with ghost numbers 
$
gh(W^{I}) = |I| + 1
$ 
and canonical dimension $\leq 7$) for the anomalies. In the next Sections we give the most simplest way to solve in general such type of problems.
\section{The Cohomology of the Gauge Charge Operator\label{q}}

We consider the vector space 
$
{\cal H}
$
of Fock type generated (in the sense of Borchers theorem) by the symmetric tensor field 
$
h_{\mu\nu}
$ 
(with Bose statistics) and the vector fields 
$
u^{\rho}, \tilde{u}^{\sigma}
$
(with Fermi statistics). The Fermi fields are usually called {\it ghost fields}. We suppose that all these (quantum) fields are of null mass. Let $\Omega$ be the vacuum state in
$
{\cal H}.
$
In this vector space we can define a sesquilinear form 
$<\cdot,\cdot>$
in the following way: the (non-zero) $2$-point functions are by definition:
\bea
<\Omega, h_{\mu\nu}(x_{1}) h_{\rho\sigma}(x_{2})\Omega> = - {i\over 2}~
(\eta_{\mu\rho}~\eta_{\nu\sigma} + \eta_{\nu\rho}~\eta_{\mu\sigma}
- \eta_{\mu\nu}~\eta_{\rho\sigma})~D_{0}^{(+)}(x_{1} - x_{2}),
\nonumber \\
<\Omega, u_{\mu}(x_{1}) \tilde{u}_{\nu}(x_{2})\Omega> = i~\eta_{\mu\nu}~
D_{0}^{(+)}(x_{1} - x_{2}),
\nonumber \\
<\Omega, \tilde{u}_{\mu}(x_{1}) u_{\nu}(x_{2})\Omega> = - i~\eta_{\mu\nu}~
D_{0}^{(+)}(x_{1} - x_{2})
\eea
and the $n$-point functions are generated according to Wick theorem. Here
$
\eta_{\mu\nu}
$
is the Minkowski metrics (with diagonal $1, -1, -1, -1$) and 
$
D_{0}^{(+)}
$
is the positive frequency part of the Pauli-Villars distribution
$
D_{0}
$
of null mass. To extend the sesquilinear form to
$
{\cal H}
$
we define the conjugation by
\be
h_{\mu\nu}^{\dagger} = h_{\mu\nu}, \qquad 
u_{\rho}^{\dagger} = u_{\rho}, \qquad
\tilde{u}_{\sigma}^{\dagger} = - \tilde{u}_{\sigma}.
\ee

Now we can define in 
$
{\cal H}
$
the operator $Q$ according to the following formulas:
\bea
~[Q, h_{\mu\nu}] = - {i\over 2}~(\partial_{\mu}u_{\nu} + \partial_{\nu}u_{\mu}
- \eta_{\mu\nu} \partial_{\rho}u^{\rho}),\qquad
[Q, u_{\mu}] = 0,\qquad
[Q, \tilde{u}_{\mu}] = i~\partial^{\nu}h_{\mu\nu}
\nonumber \\
Q\Omega = 0
\label{Q-0}
\eea
where by 
$
[\cdot,\cdot]
$
we mean the graded commutator. One can prove that $Q$ is well defined. Indeed, we have the causal commutation relations 
\bea
~[h_{\mu\nu}(x_{1}), h_{\rho\sigma}(x_{2}) ] = - {i\over 2}~
(\eta_{\mu\rho}~\eta_{\nu\sigma} + \eta_{\nu\rho}~\eta_{\mu\sigma}
- \eta_{\mu\nu}~\eta_{\rho\sigma})~D_{0}(x_{1} - x_{2})~\cdot I,
\nonumber \\
~[u(x_{1}), \tilde{u}(x_{2})] = i~\eta_{\mu\nu}~D_{0}(x_{1} - x_{2})~\cdot I
\eea
and the other commutators are null. The operator $Q$ should leave invariant these relations, in particular 
\be
[Q, [ h_{\mu\nu}(x_{1}),\tilde{u}_{\sigma}(x_{2})]] + {\rm cyclic~permutations} = 0
\ee
which is true according to (\ref{Q-0}). It is useful to introduce a grading in 
$
{\cal H}
$
as follows: every state which is generated by an even (odd) number of ghost fields and an arbitrary number of vector fields is even (resp. odd). We denote by 
$
|f|
$
the ghost number of the state $f$. We notice that the operator $Q$ raises the ghost number of a state (of fixed ghost number) by an unit. The usefullness of this construction follows from:
\begin{thm}
The operator $Q$ verifies
$
Q^{2} = 0.
$ 
The factor space
$
Ker(Q)/Ran(Q)
$
is isomorphic to the Fock space of particles of zero mass and helicity $2$ (gravitons). 
\label{fock-0}
\end{thm}
{\bf Proof:} (i) The fact that $Q$ squares to zero follows easily from (\ref{Q-0}): the operator 
$
Q^{2} = 0
$
commutes with all field operators and gives zero when acting on the vacuum. 

(ii) The generic form of a state 
$
\Psi \in {\cal H}^{(1)} \subset {\cal H}
$
from the one-particle Hilbert subspace is
\be
\Psi = \left[ \int f_{\mu\nu}(x) h^{\mu\nu}(x) + \int g^{(1)}_{\mu}(x) u^{\mu}(x) 
+ \int g^{(2)}_{\mu}(x) \tilde{u}^{\mu}(x) \right] \Omega
\ee
with test functions
$
f_{\mu\nu}, g^{(1)}_{\mu}, g^{(2)}_{\mu}
$
verifying the wave equation equation; we can also suppose that
$
f_{\mu\nu}
$
is symmetric. The condition 
$
\Psi \in Ker(Q) \quad \Longleftrightarrow \quad Q\Psi = 0;
$
leads to
$
\partial^{\nu}f_{\mu\nu} = {1\over 2}~\partial_{\mu}f
$
(where
$
f = \eta^{\mu\nu}f_{\mu\nu}
$
is the trace of
$
f_{\mu\nu}
$
and
$
g^{(2)}_{\mu} = 0
$
i.e. the generic element
$
\Psi \in {\cal H}^{(1)} \cap Ker(Q)
$
is
\be
\Psi = \left[ \int f_{\mu\nu}(x) h^{\mu\nu}(x) + \int g_{\mu}(x) u^{\mu}(x) \right] \Omega
\label{kerQ-0}
\ee
with $g_{\mu}$ arbitrary and 
$
f_{\mu\nu}
$
constrained by the transversality condition 
$
\partial^{\nu}f_{\mu\nu} = {1\over 2}~\partial_{\mu}f;
$
so the elements of
$
{\cal H}^{(1)} \cap Ker(Q)
$
are in one-one correspondence with couples of test functions
$
[f_{\mu\nu}, g_{\rho}]
$
with the transversality condition on the first entry. Now, a generic element
$
\Psi^{\prime} \in {\cal H}^{(1)} \cap Ran(Q)
$
has the form 
\be
\Psi^{\prime} = Q\Phi = \left[
- {1\over 2} \int (\partial_{\mu}g^{\prime}_{\nu} + \partial_{\nu}g^{\prime}_{\mu})(x) h^{\mu\nu}(x) 
+ \int \left(\partial^{\nu}g^{\prime}_{\mu\nu} 
- {1\over 2}~\partial_{\mu}g^{\prime}\right)(x) u(x) \right] \Omega
\label{ranQ-0}
\ee
with 
$
g^{\prime} = \eta^{\mu\nu}g^{\prime}_{\mu\nu}
$
so if
$
\Psi \in {\cal H}^{(1)} \cap Ker(Q)
$
is indexed by the couple 
$
[f_{\mu\nu}, g_{\rho}]
$
then 
$
\Psi + \Psi^{\prime}
$
is indexed by the couple
$
\left[
f_{\mu\nu} - {1\over 2}~(\partial_{\mu}g^{\prime}_{\nu} + \partial_{\nu}g^{\prime}_{\mu}), 
g_{\mu} + \left( \partial^{\nu}g^{\prime}_{\mu\nu} 
- {1\over 2}~\partial_{\mu}g^{\prime}\right)\right].
$
If we take 
$
g^{\prime}_{\mu\nu}
$
conveniently we can make 
$
g_{\mu} = 0
$
and if we take 
$
g^{\prime}_{\mu}
$
convenient we can make 
$
f = 0;
$
in this case the transversality condition becomes
$
\partial^{\nu}f_{\mu\nu} = 0.
$
It follows that the equivalence classes from
$
({\cal H}^{(1)} \cap Ker(Q))/({\cal H}^{(1)} \cap Ran(Q))
$ 
are indexed by wave functions
$
f_{\mu\nu}
$
verifying the conditions of transversality and tracelessness
$
\partial^{\nu}f_{\mu\nu} = 0,~f = 0.
$
We still have the freedom to change 
$
f_{\mu\nu} \rightarrow f_{\mu\nu}- {1\over 2}~(\partial_{\mu}g^{\prime}_{\nu} + \partial_{\nu}g^{\prime}_{\mu})
$
with 
$
\partial^{\mu}g^{\prime}_{\mu} = 0
$
without affecting the properties
$
\partial^{\nu}f_{\mu\nu} = 0,~f = 0.
$
It remains to prove that the sesquilinear form 
$<\cdot,\cdot>$ 
induces a positively defined form on
$
({\cal H}^{(1)} \cap Ker(Q))/({\cal H}^{(1)} \cap Ran(Q))
$ 
and we have obtained the usual one-particle Hilbert space for the graviton (i.e. a particle of zero mass and helicity $2$).

(iii) The extension of this argument to the $n$th-particle space is done as in \cite{cohomology} using K\"unneth formula \cite{Dr}. 
$\qed$

We see that the condition 
$
[Q, T] = i~\partial_{\mu}T^{\mu}
$
means that the expression $T$ leaves invariant the physical Hilbert space (at least in the adiabatic limit).

Now we have the physical justification for solving another cohomology problem namely to determine the cohomology of the operator 
$
d_{Q} = [Q,\cdot]
$
induced by $Q$ in the space of Wick polynomials. To solve this problem it is convenient to use the same geometric formalism \cite{jet} used in \cite{cohomology}. We consider that the (classical) fields are
$
h_{\mu\nu}, u_{\rho}, \tilde{u}_{\sigma}
$
of null mass and we consider the set 
$
{\cal P}
$
of polynomials in these fields and their formal derivatives (in the sense of jet bundle theory). The formal derivatives operators
$
d_{\mu}
$
are given by 
\be
d_{\mu}y^{\alpha}_{\nu_{1}\cdots\nu_{n}} \equiv y^{\alpha}_{\mu\nu_{1}\cdots\nu_{n}}
\ee
where 
$
y^{A}
$
are the basic variables
$
y^{\alpha} = (h_{\mu\nu}, u_{\rho}, \tilde{u}_{\sigma})
$
and 
$
y^{\alpha}_{\nu_{1}\cdots\nu_{n}} 
$
are the jet bundle coordinates (see \cite{cohomology} for details). We note that on
$
{\cal P}
$
we have a natural grading. We introduce by convenience the notation:
\be
B_{\mu} \equiv d^{\nu}h_{\mu\nu}
\ee
and define the graded derivation 
$
d_{Q}
$
on
$
{\cal P}
$
according to
\bea
d_{Q}h_{\mu\nu} = - {i\over 2}~(d_{\mu}u_{\nu} + d_{\nu}u_{\mu} 
- \eta_{\mu\nu}~d_{\rho}u^{\rho}), 
\qquad 
d_{Q}u_{\mu} = 0, 
\qquad d_{Q}\tilde{u}_{\mu} = i~B_{\mu}
\nonumber \\
~[d_{Q}, d_{\mu} ] = 0.
\eea
Then one can easily prove that 
$
d_{Q}^{2} = 0
$
and the cohomology of this operator is isomorphic to the cohomology of the preceding operator (denoted also by $d_{Q}$) and acting in the space of Wick polynomials. The operator 
$
d_{Q}
$
raises the grading and the canonical dimension by an unit. To determine the cohomology of 
$
d_{Q}
$
it is convenient to introduce some notations: first
\be
h \equiv \eta^{\mu\nu}h_{\mu\nu} \qquad
\hat{h}_{\mu\nu} \equiv h_{\mu\nu} - {1\over 2}~\eta_{\mu\nu}~h
\ee
and the we define the {\it Christoffel symbols} according to:
\be
\Gamma_{\mu;\nu\rho} \equiv d_{\rho}\hat{h}_{\mu\nu} + d_{\nu}\hat{h}_{\mu\rho} - d_{\mu}\hat{h}_{\nu\rho}.
\ee 
We observe that 
\be
d_{Q}\Gamma_{\mu;\nu\rho} = - i~d_{\nu}d_{\rho}u_{\mu}. 
\ee
and we can express the first order derivatives through the Christoffel symbols 
\be
d_{\rho}\hat{h}_{\mu\nu} = {1\over 2}~(\Gamma_{\mu;\nu\rho} + \Gamma_{\nu;\mu\rho}).
\ee

The expression
\be
R_{\mu\nu;\rho\sigma} \equiv d_{\rho}\Gamma_{\mu;\nu\sigma} - (\rho \leftrightarrow \sigma)
\ee
is called the {\it Riemann tensor}; we can easily prove 
\bea
R_{\mu\nu;\rho\sigma} = - R_{\nu\mu;\rho\sigma} = - R_{\mu\nu;\sigma\rho} = R_{\rho\sigma;\mu\nu}, 
\nonumber \\
d_{Q}R_{\mu\nu;\rho\sigma} = 0,
\nonumber \\
R_{\mu\nu;\rho\sigma} + R_{\mu\rho;\nu\sigma} + R_{\mu\sigma;\nu\rho} = 0;
\nonumber \\
d_{\lambda}R_{\mu\nu;\rho\sigma} + d_{\rho}R_{\mu\nu;\sigma\lambda} + d_{\sigma}R_{\mu\nu;\lambda\rho} = 0
\eea
the last two relations are called {\it Bianchi identities}. 

Next we consider, as in the case of the Yang-Mills fields, more convenient variables: (i) first one can expresses the derivatives of the Christoffel symbols in terms of the completely symmetric derivatives
\be
\Gamma_{\mu;\rho_{1},\dots,\rho_{n}} \equiv {\cal S}_{\rho_{1},\dots,\rho_{n}}~ (d_{\rho_{3}}\dots d_{\rho_{n}}~\Gamma_{\mu;\rho_{1}\rho_{2}})
\ee 
and derivatives of the Riemann tensor;
(ii) next, one expresses the variables
$
\Gamma_{\mu;\rho_{1},\dots,\rho_{n}} 
$
in terms of the expressions 
$
\Gamma^{(0)}_{\mu;\rho_{1},\dots,\rho_{n}} 
$
(which is, by definition, the traceless part in
$
\rho_{1},\dots,\rho_{n}
$)
and
$
B_{\mu;\rho_{1},\dots,\rho_{n-2}}; 
$
(iii) finally one expresses the derivatives of the Riemann tensor
$
d_{\lambda_{1}}\dots d_{\lambda_{n}}~R_{\mu\nu;\rho\sigma}
$
in terms of the traceless part in all indices
$
R^{(0)}_{\mu\nu;\rho\sigma;\lambda_{1},\dots,\lambda_{n}} 
$
and
$
B_{\mu;\rho_{1},\dots,\rho_{n+1}}.
$

We will use the K\"unneth theorem:
\begin{thm}
Let 
$
{\cal P}
$
be a graded space of polynomials and $d$ an operator verifying
$
d^{2} = 0
$ 
and raising the grading by an unit. Let us suppose that
$
{\cal P}
$
is generated by two subspaces
$
{\cal P}_{1}, {\cal P}_{2}
$
such that
$
{\cal P}_{1} \cap {\cal P}_{2} = \{0\}
$
and
$
d{\cal P}_{j} \subset {\cal P}_{j}, j = 1,2.
$
We define by 
$
d_{j}
$
the restriction of $d$ to
$
{\cal P}_{j}.
$
Then there exists the canonical isomorphism
$
H(d) \cong H(d_{1}) \times H(d_{2})
$ 
of the associated cohomology spaces. 
\label{kunneth}
\end{thm}
(see \cite{Dr}). Now we can give a generic description for the co-cycles of  
$
d_{Q};
$
we denote by
$
Z_{Q}
$
and 
$
B_{Q}
$
the co-cycles and the co-boundaries of this operator. First we define
\bea
u_{\mu\nu} = u_{[\mu\nu]} \equiv {1\over 2}~(d_{\mu}u_{\nu} - d_{\nu}u_{\mu})
\nonumber \\
u_{\{\mu\nu\}} \equiv {1\over 2}~(d_{\mu}u_{\nu} + d_{\nu}u_{\mu})
\eea 
such that we have:
\be
d_{\mu}u_{\nu} = u_{\mu\nu} + u_{\{\mu\nu\}}.
\ee
Now we have:
\begin{thm}
Let 
$
p \in Z_{Q}.
$
Then $p$ is cohomologous to a polynomial in 
$u_{\mu}, u_{\mu\nu}$ 
and
$
R^{(0)}_{\mu\nu;\rho\sigma;\lambda_{1},\dots,\lambda_{n}}.
$
\label{m=0}
\end{thm}
{\bf Proof:} (i) The idea is to define conveniently two subspaces 
$
{\cal P}_{1}, {\cal P}_{2}
$
and apply K\"unneth theorem. We will take
$
{\cal P}_{1} = {\cal P}_{0}
$
from the statement and
$
{\cal P}_{2}
$
the subspace generated by the variables
$
B_{\mu;\nu_{1},\dots,\nu_{n}}~(n \geq 0),~
\Gamma^{(0)}_{\mu;\nu_{1},\dots,\nu_{n}}~(n \geq 2),~
\tilde{u}_{\mu;\nu_{1},\dots,\nu_{n}}~(n \geq 0),~
u_{\mu;\nu_{1},\dots,\nu_{n}}(n \geq 2),
u_{\{\mu\nu\}}
$
and
$
\hat{h}_{\mu\nu}.
$
We have
$
d_{Q}{\cal P}_{1} = \{0\}
$
and
\bea 
d_{Q}u_{\{\mu;\nu\}} = 0,\qquad
d_{Q}u_{\mu;\nu_{1},\dots,\nu_{n}} = 0~(n \geq 2)
\nonumber \\
d_{Q}\Gamma^{(0)}_{\mu;\nu_{1},\dots,\nu_{n}} = - i~u_{\mu;\nu_{1},\dots,\nu_{n}}~
(n \geq 2)
\nonumber \\
d_{Q}\tilde{u}_{\mu;\nu_{1},\dots,\nu_{n}} = i~B_{\mu;\nu_{1},\dots,\nu_{n}}~
(n \geq 0)
\nonumber \\
d_{Q}B_{\mu;\nu_{1},\dots,\nu_{n}} = 0~(n \geq 0)
\nonumber \\
d_{Q}\hat{h}_{\mu\nu} = - i~u_{\{\mu\nu\}}
\eea
so we meet the conditions of K\"unneth theorem. Let us define in
$
{\cal P}_{2}
$
the graded derivation ${\mathfrak h}$ by:
\bea
{\mathfrak h}u_{\{\mu\nu\}} = i~\hat{h}_{\mu\nu} 
\nonumber \\
{\mathfrak h}u_{\mu;\nu_{1},\dots,\nu_{n}} = i~\Gamma^{(0)}_{\mu;\nu_{1},\dots,\nu_{n}}~
(n \geq 2)
\nonumber \\
{\mathfrak h}B_{\mu;\nu_{1},\dots,\nu_{n}} = - i~\tilde{u}_{\mu;\nu_{1},\dots,\nu_{n}}~
(n \geq 0)
\eea
and zero on the other variables from 
$
{\cal P}_{2}.
$
It is easy to prove that ${\mathfrak h}$ is well defined: the condition of tracelessness is essential to avoid conflict with the equations of motion. Then one can prove that
\be
[d_{Q},{\mathfrak h}] = Id
\ee
on polynomials of degree one in the fields and because the left hand side is a derivation operator we have
\be
[d_{Q},{\mathfrak h}] = n \cdot Id
\ee 
on polynomials of degree $n$ in the fields. It means that ${\mathfrak h}$ is a homotopy for 
$
d_{Q}
$
restricted to 
$
{\cal P}_{2}
$
so the the corresponding cohomology is trivial: indeed, if 
$
p \in {\cal P}_{2}
$
is a co-cycle of degree $n$ in the fields then it is a co-boundary
$
p = {1\over n} d_{Q}{\mathfrak h}p.
$

According to K\"unneth formula if $p$ is an arbitrary co cycle from 
$
{\cal P}
$
it can be replaced by a cohomologous polynomial from
$
{\cal P}_{0}
$
and this proves the theorem.
$\qed$

\begin{rem}
There is an important difference with respect to the Yang-Mills case, namely the space
$
{\cal P}_{0}
$
is not isomorphic to the cohomology group
$
H_{Q}
$
and this follows from the fact that
$
{\cal P}_{0}~\cap B_{Q} \not= 0.
$
We provide an example of an expression belonging to this intersection. We start with the expression
\be
B^{\mu\nu\rho\sigma\lambda} \equiv u^{\mu}~u^{\nu}~u^{\rho}~u^{\sigma}~u^{\lambda};
\ee
because of the complete antisymmetry we have in fact
\be
B^{\mu\nu\rho\sigma\lambda} = 0.
\ee
On the other hand we have
\be
d_{\lambda}B^{\mu\nu\rho\sigma\lambda} = p^{\mu\nu\rho\sigma} + d_{Q}(\cdots)
\ee
where
\be
p^{\mu\nu\rho\sigma} \equiv u_{\lambda}~
(u^{\mu}~u^{\nu}~u^{\rho}~u^{\sigma\lambda}
+ u^{\nu}~u^{\rho}~u^{\sigma}~u^{\mu\lambda} 
+ u^{\rho}~u^{\sigma}~u^{\mu}~u^{\nu\lambda} 
+ u^{\sigma}~u^{\mu}~u^{\nu}~u^{\rho\lambda}) 
\ee
so we have
$
p^{\mu\nu\rho\sigma} \in {\cal P}_{0}~\cap B_{Q}.
$
\end{rem}

We repeat the whole argument for the case of massive graviton i.e. particles of spin $1$ and positive mass. 

We consider a vector space 
$
{\cal H}
$
of Fock type generated (in the sense of Borchers theorem) by the tensor field 
$
h_{\mu\nu},
$ 
the vector field 
$
v_{\mu}
$
(with Bose statistics) and the vector fields 
$
u_{\mu}, \tilde{u}_{\mu}
$
(with Fermi statistics). We suppose that all these (quantum) fields are of mass
$
m > 0.
$
In this vector space we can define a sesquilinear form 
$<\cdot,\cdot>$
in the following way: the (non-zero) $2$-point functions are by definition:
\bea
<\Omega, h_{\mu\nu}(x_{1}) h_{\rho\sigma}(x_{2})\Omega> = - {i\over 2}~
(\eta_{\mu\rho}~\eta_{\nu\sigma} + \eta_{\nu\rho}~\eta_{\mu\sigma}
- \eta_{\mu\nu}~\eta_{\rho\sigma})~D_{m}^{(+)}(x_{1} - x_{2}),
\nonumber \\
<\Omega, u_{\mu}(x_{1}) \tilde{u}_{\nu}(x_{2})\Omega> = i~\eta_{\mu\nu}~
D_{m}^{(+)}(x_{1} - x_{2}),
\nonumber \\
<\Omega, \tilde{u}_{\mu}(x_{1}) u_{\nu}(x_{2})\Omega> = - i~\eta_{\mu\nu}~
D_{m}^{(+)}(x_{1} - x_{2}),
\nonumber \\
<\Omega, v_{\mu}(x_{1}) v_{\mu}(x_{2})\Omega> =i~\eta_{\mu\nu}~D_{m}^{(+)}(x_{1} - x_{2})
\eea
and the $n$-point functions are generated according to Wick theorem. Here
$
D_{m}^{(+)}
$
is the positive frequency part of the Pauli-Villars distribution
$
D_{m}
$
of mass $m$. To extend the sesquilinear form to
$
{\cal H}
$
we define the conjugation by
\bea
h_{\mu\nu}^{\dagger} = h_{\mu\nu}, \qquad 
u_{\rho}^{\dagger} = u_{\rho}, \qquad
\tilde{u}_{\sigma}^{\dagger} = - \tilde{u}_{\sigma}, \qquad
v_{\mu}^{\dagger} = v_{\mu}.
\eea

Now we can define in 
$
{\cal H}
$
the operator $Q$ according to the following formulas:
\bea
~[Q, h_{\mu\nu}] = - {i\over 2}~(\partial_{\mu}u_{\nu} + \partial_{\nu}u_{\mu}
- \eta_{\mu\nu} \partial_{\rho}u^{\rho}),
\nonumber \\
~[Q, u_{\mu}] = 0,\qquad
[Q, \tilde{u}_{\mu}] = i~(\partial^{\nu}h_{\mu\nu} + m v_{\mu}),
\nonumber \\
~[Q, v_{\mu}] = - {i~m\over 2}~u_{\mu}
\nonumber \\
Q\Omega = 0.
\label{Q-m}
\eea
One can prove that $Q$ is well defined. Indeed, we have the causal commutation relations 
\bea
~[h_{\mu\nu}(x_{1}), h_{\rho\sigma}(x_{2}) ] = - {i\over 2}~
(\eta_{\mu\rho}~\eta_{\nu\sigma} + \eta_{\nu\rho}~\eta_{\mu\sigma}
- \eta_{\mu\nu}~\eta_{\rho\sigma})~D_{m}(x_{1} - x_{2})~\cdot I,
\nonumber \\
~[u(x_{1}), \tilde{u}(x_{2})] = i~\eta_{\mu\nu}~D_{m}(x_{1} - x_{2})~\cdot I
\nonumber \\
~[v_{\mu}(x_{1}) v_{\mu}(x_{2})] = i~\eta_{\mu\nu}~D_{m}(x_{1} - x_{2})~\cdot I
\eea
and the other commutators are null. The operator $Q$ should leave invariant these relations, in particular 
\bea
[Q, [ h_{\mu\nu}(x_{1}),\tilde{u}_{\sigma}(x_{2})]] + {\rm cyclic~permutations} = 0,
\nonumber \\
~[Q, [ v_{\mu}(x_{1}),\tilde{u}_{\sigma}(x_{2})]] + {\rm cyclic~permutations} = 0.
\eea

We have a result similar to the first theorem of this Section:
\begin{thm}
The operator $Q$ verifies
$
Q^{2} = 0.
$ 
The factor space
$
Ker(Q)/Ran(Q)
$
is isomorphic to the Fock space of particles of mass $m$ and spin $2$ (massive gravitons). 
\end{thm}
{\bf Proof:} (i) The fact that $Q$ squares to zero follows easily from (\ref{Q-m}).

(ii) The generic form of a state 
$
\Psi \in {\cal H}^{(1)} \subset {\cal H}
$
from the one-particle Hilbert subspace is
\be
\Psi = \left[ \int f_{\mu\nu}(x) h^{\mu\nu}(x) + \int g^{(1)}_{\mu}(x) u^{\mu}(x) 
+ \int g^{(2)}_{\mu}(x) \tilde{u}^{\mu}(x) + \int h_{\mu}(x) v^{\mu}(x) \right] \Omega
\ee
with test functions
$
f_{\mu\nu}, g^{(1)}_{\mu}, g^{(2)}_{\mu}, h_{\mu}
$
verifying the wave equation equation; we can also suppose that
$
f_{\mu\nu}
$
is symmetric. The condition 
$
\Psi \in Ker(Q)~\Longleftrightarrow~Q\Psi = 0
$
leads to
$
h_{\mu} = {2\over m}~
\left(\partial^{\nu}f_{\mu\nu} - {1\over 2}~\partial_{\mu}f\right)
$
(where
$
f = \eta^{\mu\nu}f_{\mu\nu}
$
is the trace of
$
f_{\mu\nu})
$
and
$
g^{(2)}_{\mu} = 0
$
i.e. the generic element
$
\Psi \in {\cal H}^{(1)} \cap Ker(Q)
$
is
\be
\Psi = \left[ \int f_{\mu\nu}(x) h^{\mu\nu}(x) + \int g_{\mu}(x) u^{\mu}(x) 
+ {2\over m}~\int
\left(\partial^{\nu}f_{\mu\nu} - {1\over 2}~\partial_{\mu}f\right)(x) v^{\mu}(x)\right] \Omega
\label{kerQ-m}
\ee
with 
$g_{\mu}$ 
and 
$
f_{\mu\nu}
$
arbitrary so
$
\Psi \in {\cal H}^{(1)} \cap Ker(Q)
$
is indexed by couples of test functions
$
[f_{\mu\nu},g_{\mu}].
$
Now, a generic element
$
\Psi^{\prime} \in {\cal H}^{(1)} \cap Ran(Q)
$
has the form 
\be
\Psi^{\prime} = Q\Phi = \left[
- {1\over 2} \int (\partial_{\mu}g^{\prime}_{\nu} + \partial_{\nu}g^{\prime}_{\mu})(x) h^{\mu\nu}(x) 
+ \int \left(\partial^{\nu}g^{\prime}_{\mu\nu} 
- {1\over 2}~\partial_{\mu}g^{\prime} - {m\over 2} h^{\prime}_{\mu}\right)(x) u^{\mu}(x) \right] \Omega
\label{ranQ-m}
\ee
with 
$
g^{\prime} = \eta^{\mu\nu}g^{\prime}_{\mu\nu}
$
so if
$
\Psi \in {\cal H}^{(1)} \cap Ker(Q)
$
is indexed by the couple 
$
[f_{\mu\nu}, g_{\rho}]
$
then 
$
\Psi + \Psi^{\prime}
$
is indexed by the couple
$
\left[
f_{\mu} - {1\over 2}~(\partial_{\mu}g^{\prime}_{\nu} + \partial_{\nu}g^{\prime}_{\mu}), 
g_{\mu} + \left( \partial^{\nu}g^{\prime}_{\mu\nu} 
- {1\over 2}~\partial_{\mu}g^{\prime} - {m\over 2} h^{\prime}_{\mu}\right)\right].
$
If we take 
$
h^{\prime}_{\mu}
$
conveniently we can make 
$
g_{\mu} = 0
$
and if we take 
$
g^{\prime}_{\mu\nu}
$
convenient we can make 
$
\partial^{\nu}f_{\mu\nu} = 0.
$
We still have the freedom to change 
$
f_{\mu\nu} \rightarrow f_{\mu\nu}- {1\over 2}~(\partial_{\mu}g^{\prime}_{\nu} + \partial_{\nu}g^{\prime}_{\mu})
$
with transverse functions
$
\partial^{\mu}g^{\prime}_{\mu} = 0
$
without affecting the property
$
\partial^{\nu}f_{\mu\nu} = 0.
$
It remains to prove that the sesquilinear form 
$<\cdot,\cdot>$ 
induces a positively defined form on
$
({\cal H}^{(1)} \cap Ker(Q))/({\cal H}^{(1)} \cap Ran(Q))
$ 
and we have obtained a direct sum of the one-particle Hilbert space for the  graviton of mass $m$ (i.e. a particle of mass $m$ and helicity $2$) and a scalar particle of the same mass $m$.

(iii) The extension of this argument to the $n$th-particle space is done as in \cite{cohomology} using K\"unneth formula \cite{Dr}. 
$\qed$

Now we determine the cohomology of the operator 
$
d_{Q} = [Q,\cdot]
$
induced by $Q$ in the space of Wick polynomials. As before, it is convenient to use the formalism from the preceding Section. We consider that the (classical) fields
$
y^{\alpha}
$
are
$
h_{\mu\nu}, u_{\mu}, \tilde{u}_{\mu}, v_{\mu}
$
of mass $m$ and we consider the set 
$
{\cal P}
$
of polynomials in these fields and their derivatives. We introduce by convenience the notation:
\be
C_{\mu} \equiv d^{\nu}h_{\mu\nu} + m v_{\mu}
\ee
and define the graded derivation 
$
d_{Q}
$
on
$
{\cal P}
$
according to
\bea
d_{Q}h_{\mu\nu} = - {i\over 2}~(d_{\mu}u_{\nu} + d_{\nu}u_{\mu} 
- \eta_{\mu\nu}~d_{\rho}u^{\rho}), 
\nonumber \\
d_{Q}u_{\mu} = 0,\qquad 
d_{Q}\tilde{u}_{\mu} = i~B_{\mu}, \qquad
d_{Q}v_{\mu} = - {i~m\over 2}~u_{\mu}
\nonumber \\
~[d_{Q}, d_{\mu} ] = 0.
\eea
Then one can prove that 
$
d_{Q}^{2} = 0
$
and the cohomology of this operator is isomorphic to the cohomology of the preceding operator (denoted also by $d_{Q}$) and acting in the space of Wick monomials. To determine the cohomology of 
$
d_{Q}
$
it is convenient to introduce the Riemann tensor
$
R_{\mu\nu;\rho\sigma}
$ 
as before and also
\bea
\phi_{\mu\nu} \equiv 
d_{\mu}v_{\nu} + d_{\nu}v_{\mu} - \eta_{\mu\nu} d_{\rho}v^{\rho} - m~h_{\mu\nu}
\nonumber \\
\phi \equiv \eta^{\mu\nu}~\phi_{\mu\nu}
\eea
and observe that we also have
\be
d_{Q}\phi_{\mu\nu} = 0.
\ee

Then we construct new variables as in the massless case: (i) we express the variables
$
v_{\{\mu\nu\}} \equiv {1\over 2} (d_{\mu}v_{\nu} +d_{\nu}v_{\mu}),
$
and
$
d_{\nu_{1}}\dots d_{\nu_{n}}v_{\mu}~(n \geq 2)
$
through
$
\phi_{\mu\nu}, h_{\mu\nu} 
$
and their derivatives; (ii) next, we express the derivatives
$
\phi_{\mu\nu;\rho_{1}\dots\rho_{n}}
$
through the traceless parts
$
\phi^{(0)}_{\mu\nu;\rho_{1}\dots\rho_{n}}, \phi^{(0)}_{;\rho_{1}\dots\rho_{n}}
$
and
$
C_{\mu;\rho_{1},\dots,\rho_{n}}.
$
(iii) Finally we express the variables:
$
\Gamma_{\mu;\nu_{1},\dots,\nu_{n}} 
$
and 
$
d_{\lambda_{1}}\dots d_{\lambda_{n}}R_{\mu\nu;\rho\sigma} 
$
in terms of the traceless parts
$
\Gamma^{(0)}_{\mu;\nu_{1},\dots,\nu_{n}},
R^{(0)}_{\mu\nu;\rho\sigma;\lambda_{1},\dots,\lambda_{n}},
\phi^{(0)}_{\mu\nu;\rho_{1}\dots\rho_{n}}, 
\phi^{(0)}_{;\rho_{1}\dots\rho_{n}} 
C_{\mu;\nu_{1},\dots,\nu_{n}}
$
and
$
h_{\mu\nu},~v_{[\mu\nu]}.
$
Now we can describe the cohomology of the operator 
$
d_{Q}
$
in the massive case.
\begin{thm}
Let 
$
p \in Z_{Q}.
$
Then $p$ is cohomologous to a polynomial in the traceless variables:
$
R^{(0)}_{\mu\nu;\rho\sigma;\lambda_{1},\dots,\lambda_{n}}
$
and
$
\phi^{(0)}_{\mu\nu;\rho_{1}\dots\rho_{n}}, 
\phi^{(0)}_{;\rho_{1}\dots\rho_{n}} 
$ 
\label{m>0}
\end{thm}

{\bf Proof:} (i) Is similar to the proof of theorem \ref{m=0}. We take
$
{\cal P}_{1} = {\cal P}_{0}
$
as in the statement of the theorem and
$
{\cal P}_{2}
$
generated by the other variables. The graded derivation ${\mathfrak h}$ is defined in this case by:
\bea
{\mathfrak h}u_{\mu} = {2 i\over m} ~v_{\mu},\qquad
{\mathfrak h}u_{\{\mu\nu\}} = i~\hat{h}_{\mu\nu},\qquad
{\mathfrak h}u_{[\mu\nu]} = {2 i\over m} ~v_{[\mu\nu]},
\nonumber \\
{\mathfrak h}u^{(0)}_{\mu;\nu_{1},\dots,\nu_{n}} = i~\Gamma^{(0)}_{\mu;\nu_{1},\dots,\nu_{n}}~
(n \geq 2)
\nonumber \\
{\mathfrak h}C^{(0)}_{\mu;\nu_{1},\dots,\nu_{n}} = 
- i~\tilde{u}^{(0)}_{\mu;\nu_{1},\dots,\nu_{n}}~
(n \geq 0)
\eea
and it follows that ${\mathfrak h}$ is a homotopy for 
$
d_{Q}
$
restricted to 
$
{\cal P}_{2}
$
so the the corresponding cohomology is trivial.

According to K\"unneth formula if $p$ is an arbitrary co-cycle from 
$
{\cal P}
$
it can be replaced by a cohomologous polynomial from
$
{\cal P}_{0}
$
and this proves the theorem.
$\qed$

We note that in the case of null mass the operator 
$
d_{Q}
$
raises the canonical dimension by one unit and this fact is not true anymore in the massive case. We are lead to another cohomology group. Let us take as the space of co-chains the space 
$
{\cal P}^{(n)}
$
of polynomials of canonical dimension
$
\omega \leq n;
$
then 
$
Z_{Q}^{(n)} \subset {\cal P}^{(n)}
$
and
$
B_{Q}^{(n)} \equiv d_{Q}{\cal P}^{(n-1)}
$
are the co-cycles and the co-boundaries respectively. It is possible that a polynomial is a co-boundary as an element of
$
{\cal P}
$
but not as an element of
$
{\cal P}^{(n)}.
$
The situation is described by the following generalization of the preceding theorem.
\begin{thm}
Let 
$
p \in Z^{(n)}_{Q}.
$
Then $p$ is cohomologous to a polynomial of the form  
$
p_{1} + d_{Q}p_{2}
$
where
$
p_{1} \in {\cal P}_{0}
$ 
and
$
p_{2} \in {\cal P}^{(n)}.
$
\label{Q-cohomology}
\end{thm}

We will call the co-cycles of the type
$
p_{1}
$
(resp.
$
d_{Q}p_{2})
$
{\it primary} (resp. {\it secondary}).

\section{The Relative Cohomology of the Operator $d_{Q}$\label{relative}}
A polynomial 
$
p \in {\cal P}
$
verifying the relation
\be
d_{Q}p = i~d_{\mu}p^{\mu}
\label{rel-co}
\ee
for some polynomials
$
p^{\mu}
$
is called a {\it relative co-cycle} for 
$
d_{Q}.
$
The expressions of the type
\be
p = d_{Q}b + i~d_{\mu}b^{\mu}, \qquad (b, b^{\mu} \in {\cal P})
\ee
are relative co-cycles and are called {\it relative co-boundaries}. We denote by
$
Z_{Q}^{\rm rel}, B_{Q}^{\rm rel} 
$
and
$
H_{Q}^{\rm rel}
$
the corresponding cohomological spaces. In (\ref{rel-co}) the expressions
$
p_{\mu}
$
are not unique. It is possible to choose them Lorentz covariant.

Now we consider the framework and notations of the preceding Section in the case
$
m = 0
$. Then we have the following result which describes the most general form of the self-interaction of the gravitons. Summation over the dummy indices is used everywhere. 
\begin{thm}
Let $T$ be a relative co-cycle for 
$
d_{Q}
$
which is as least tri-linear in the fields and is of canonical dimension
$
\omega(T) \leq 5
$
and ghost number
$
gh(T) = 0.
$
Then:
(i) $T$ is (relatively) cohomologous to a non-trivial co-cycle of the form:
\bea
t = \kappa ( 2~h_{\mu\rho}~d^{\mu}h^{\nu\lambda}~d^{\rho}h_{\nu\lambda}
+ 4~h_{\nu\rho}~d^{\lambda}h^{\mu\nu}~d_{\mu}{h_{\nu}}^{\lambda}
- 4~h_{\rho\lambda}~d^{\mu}h^{\nu\rho}~d_{\mu}{h_{\nu}}^{\lambda}
\nonumber \\
+ 2~h^{\rho\lambda}~d_{\mu}h_{\rho\lambda}~d^{\mu}h
- h_{\mu\rho}~d^{\mu}h~d^{\rho}h 
- 4~u^{\rho}~d^{\nu}\tilde{u}^{\lambda}~d_{\rho}h_{\nu\lambda}
\nonumber \\
+ 4~d^{\rho}u^{\nu}~d_{\nu}\tilde{u}^{\lambda}~h_{\rho\lambda}
+ 4~d^{\rho}u_{\nu}~d^{\lambda}\tilde{u}_{\nu}~h_{\rho\lambda}
- 4~d^{\nu}u_{\nu}~d^{\rho}\tilde{u}^{\lambda}~h_{\rho\lambda})
\eea
where
$
\kappa \in \R.
$

(ii) The relation 
$
d_{Q}t = i~d_{\mu}t^{\mu}
$
is verified by:
\bea
t^{\mu} = \kappa ( - 2 u^{\mu}~d_{\nu}h_{\rho\lambda}~d^{\rho}h^{\nu\lambda} 
+ u^{\mu}~d_{\rho}h_{\nu\lambda}~d^{\rho}h^{\nu\lambda} 
- {1\over 2} u^{\mu}~d_{\rho}h~d^{\rho}h
\nonumber \\
+ 4~u^{\rho}~d^{\nu}h^{\mu\lambda}~d_{\rho}h_{\nu\lambda}
- 2~u^{\rho}~d^{\mu}h^{\nu\lambda}~d_{\rho}h_{\nu\lambda}
+ u^{\rho}~d^{\mu}h~d_{\rho}h
\nonumber \\
- 4~d^{\rho}u^{\nu}~d_{\nu}h^{\mu\lambda}~h_{\rho\lambda}
- 4~d^{\rho}u_{\nu}~d^{\lambda}h^{\mu\nu}~h_{\rho\lambda}
+ 4~d^{\lambda}u_{\rho}~d^{\mu}h^{\nu\rho}~h_{\nu\lambda}
\nonumber \\
+ 4~d_{\nu}u^{\nu}~d^{\rho}h^{\mu\lambda}~h_{\rho\lambda}
- 2~d_{\nu}u^{\nu}~d^{\mu}h^{\rho\lambda}~h_{\rho\lambda}
- 2~d^{\rho}u^{\lambda}~h_{\rho\lambda}~d^{\mu}h
+ d^{\nu}u_{\nu}~h~d^{\mu}h
\nonumber \\
- 2~u^{\mu}~d_{\nu}d_{\rho}u^{\rho}~\tilde{u}^{\nu}
+ 2~u_{\rho}~d^{\rho}d^{\sigma}u_{\sigma}~\tilde{u}^{\mu}
- 2~u^{\mu}~d_{\lambda}u_{\rho}~d^{\rho}\tilde{u}^{\nu}
\nonumber \\
+ 2~u_{\rho}~d_{\lambda}u^{\mu}~d^{\rho}\tilde{u}^{\lambda}
+ 2~d^{\rho}u_{\rho}~d_{\lambda}u^{\mu}~\tilde{u}^{\lambda}
- 2~u_{\rho}~d^{\rho}u_{\lambda}~d^{\mu}\tilde{u}^{\lambda})
\label{Tmu}
\eea

(iii) The relation 
$
d_{Q}t^{\mu} = i~d_{\nu}t^{\mu\nu}
$
is verified by:
\bea
t^{\mu\nu} \equiv \kappa [ 2 ( - u^{\mu}~d_{\lambda}u_{\rho}~d^{\rho}h^{\nu\lambda}
+ u_{\rho}~d_{\lambda}u^{\mu}~d^{\rho}h^{\nu\lambda}
+ u_{\rho}~d^{\rho}u_{\lambda}~d^{\nu}h^{\mu\lambda}
+ d_{\rho}u^{\rho}~d_{\lambda}u^{\mu}~h^{\nu\lambda})
\nonumber \\
- (\mu \leftrightarrow \nu)
+ 4~d^{\lambda}u^{\mu}~d^{\rho}u^{\nu}~h_{\rho\lambda} ].
\label{Tmunu}
\eea

(iv) The relation 
$
d_{Q}t^{\mu\nu} = i~d_{\rho}t^{\mu\nu\rho}
$
is verified by:
\bea
t^{\mu\nu\rho} \equiv \kappa [ 2 u_{\lambda}~d^{\lambda}u^{\rho}~u^{\mu\nu}
- u_{\rho}~(d^{\mu}u^{\lambda}~d_{\lambda}u^{\nu}
- d^{\nu}u^{\lambda}~d_{\lambda}u^{\mu}) 
+ {\rm circular~perm.}]
\label{Tmunurho}
\eea
and we have
$
d_{Q}t^{\mu\nu\rho} = 0.
$

(v) The co-cycles
$
t, t^{\mu}, t^{\mu\nu}
$
and
$
t^{\mu\nu\rho}
$
are non-trivial and invariant with respect to parity.
\label{T1}
\end{thm}
{\bf Proof:}
(i) By hypothesis we have
\be
d_{Q}T = i~d_{\mu}T^{\mu}.
\label{descent-t0}
\ee
If we apply 
$
d_{Q}
$
we obtain
$
d_{\mu}d_{Q}~T^{\mu} = 0
$
so with the Poincar\'e lemma there must exist the polynomials
$
T^{\mu\nu}
$
antisymmetric in $\mu, \nu$ such that
\be
d_{Q}T^{\mu} = i~d_{\nu}T^{\mu\nu}.
\label{descent-t}
\ee
Continuing in the same way we find
$
T^{\mu\nu\rho},~T^{\mu\nu\rho\sigma}
$
which are completely antisymmetric and we also have
\bea
d_{Q}T^{\mu\nu} = i~d_{\rho}T^{\mu\nu\rho}
\nonumber \\
d_{Q}T^{\mu\nu\rho} = i~d_{\sigma}T^{\mu\nu\rho\sigma}
\nonumber \\
d_{Q}T^{\mu\nu\rho\sigma} = 0.
\label{descent-T}
\eea
According to a theorem proved in \cite{cohomology} can choose the expressions
$
T^{I}
$
to be Lorentz covariant; we also have
\be
gh(T^{I}) = |I|.
\ee 

From the last relation we find, using Theorem \ref{m=0} that
\be
T^{\mu\nu\rho\sigma} = d_{Q}B^{\mu\nu\rho\sigma} + T_{0}^{\mu\nu\rho\sigma}
\ee
with
$
T_{0}^{\mu\nu\rho\sigma} \in {\cal P}_{0}^{(5)}
$
and we can choose the expressions
$
B^{\mu\nu\rho\sigma}
$
and
$
T_{0}^{\mu\nu\rho\sigma}
$
completely antisymmetric. The generic form of 
$
T_{0}^{\mu\nu\rho\sigma}
$
is:
\be
T_{0}^{\mu\nu\rho\sigma} = a~u^{\mu}~u^{\nu}~u^{\rho}~u^{\sigma}
\ee
with $a$ a constant. If we substitute the expression obtained for
$
T^{\mu\nu\rho\sigma}
$
in the second relation (\ref{descent-T}) we find out
\be
d_{Q}(T^{\mu\nu\rho} - i~d_{\sigma}B^{\mu\nu\rho\sigma}) = i~d_{\sigma}T_{0}^{\mu\nu\rho\sigma}
\ee 
so the expression in the right hand side must be a co-boundary: we use systematically
\be
d_{\sigma}u_{\mu} = u_{\sigma\mu} + u_{\{\sigma\mu\}}
= u_{\sigma\mu} + i~d_{Q}\hat{h}_{\sigma\mu}
\ee
and find out
\be
d_{\sigma}T_{0}^{\mu\nu\rho\sigma} = a (u^{\sigma\mu}~u^{\nu}~u^{\rho}
+ u^{\sigma\nu}~u^{\rho}~u^{\mu} + u^{\sigma\rho}~u^{\mu}~u^{\nu})~u_{\sigma} + d_{Q}(\cdots)
\ee
and we obtain
$
a = 0.
$
It follows that
\be
T^{\mu\nu\rho\sigma} = d_{Q}B^{\mu\nu\rho\sigma}
\ee
and
\be
d_{Q}(T^{\mu\nu\rho} - i~d_{\sigma}B^{\mu\nu\rho\sigma}) = 0
\ee 
We apply again Theorem \ref{m=0} and obtain
\be
T^{\mu\nu\rho} = d_{Q}B^{\mu\nu\rho} + i~d_{\sigma}B^{\mu\nu\rho\sigma} 
+ T^{\mu\nu\rho}_{0}
\ee 
where
$
T_{0}^{\mu\nu\rho} \in {\cal P}_{0}^{(5)}
$
and we can choose the expressions
$
B^{\mu\nu\rho}
$
and
$
T_{0}^{\mu\nu\rho}
$
completely antisymmetric. The generic form of
$
T_{0}^{\mu\nu\rho}
$
is:
\bea
T_{0}^{\mu\nu\rho} = a_{0}~u^{\mu}~u^{\nu}~u^{\rho}
+ a_{1}~( u^{\rho}~u^{\mu\lambda}~{u^{\nu}}_{\lambda} 
+ u^{\mu}~u^{\nu\lambda}~{u^{\rho}}_{\lambda}
+ u^{\nu}~u^{\rho\lambda}~{u^{\mu}}_{\lambda})
\nonumber \\
+ a_{2}~u_{\lambda} ( u^{\lambda\rho}~u^{\mu\nu} + u^{\lambda\mu}~u^{\nu\rho}
+ u^{\lambda\nu}~u^{\rho\mu})
+ a^{\prime}~\epsilon^{\mu\nu\rho\sigma}~u_{\sigma\alpha}~u^{\alpha\beta}~u_{\beta}.
\label{t0mnr}
\eea

We substitute the expression
$
T^{\mu\nu\rho}
$
into the first relation (\ref{descent-T}) and obtain
\be
d_{Q}(T^{\mu\nu} - i~d_{\rho}B^{\mu\nu\rho}) = i~d_{\rho}T_{0}^{\mu\nu\rho}.
\ee 
The right hand side must be a co-boundary. If we compute the divergence
$
d_{\rho}T_{0}^{\mu\nu\rho}
$
and impose that it is a co-boundary we obtain
$
a_{0} = a^{\prime} = 0
$
and no constraints on
$
a_{j}~(j = 1,2)
$
so apparently we have two possible solutions, namely the corresponding polynomials 
$
T_{0j}^{\mu\nu\rho}~(j = 1,2)
$
from the expression of
$
T^{\mu\nu\rho}.
$
However, let us define
\bea
b^{\mu\nu\rho\sigma} \equiv {\cal A}~u^{\mu\nu}~u^{\rho}~u^{\sigma}
\nonumber
\eea
where
$
{\cal A}
$
performs antisymmetrization in all indices. Then it is not hard to obtain that
\bea
d_{\sigma}b^{\mu\nu\rho\sigma} = - {1\over 3}~T_{01}^{\mu\nu\rho}
- {1\over 6}~T_{02}^{\mu\nu\rho} + d_{Q}b^{\mu\nu\rho}
\nonumber 
\eea
where we can choose the expression
$
b^{\mu\nu\rho}
$
completely antisymmetric. It follows that if we modify conveniently the expressions
$
B^{\mu\nu\rho\sigma}
$
and
$
B^{\mu\nu\rho}
$
we make
$
a_{1} \rightarrow a_{1} + 2~c,\quad 
a_{2} \rightarrow a_{2} + c
$
with $c$ arbitrary. In particular we can arrange such that
$
a_{1} = a_{2} \equiv 2~\kappa
$
(this is the choice made in \cite{descent}). In this case one can prove rather easily that
$
T_{0}^{\mu\nu\rho} = t^{\mu\nu\rho} + d_{Q}(\cdots)
$ 
where
$
t^{\mu\nu\rho}
$
is the expression from the statement. It follows that one can exhibit
$
T^{\mu\nu\rho}
$
in the following form:
\be
T^{\mu\nu\rho} = t^{\mu\nu\rho}+ d_{Q}B^{\mu\nu\rho} + i~d_{\sigma}B^{\mu\nu\rho\sigma}
\ee
Now one proves by direct computation that
\be
d_{\rho}t^{\mu\nu\rho} = - i~d_{Q}~t^{\mu\nu}
\ee
where
$
t^{\mu\nu}
$
is the expression from the statement so we obtain
\be
d_{Q}(T^{\mu\nu} - t^{\mu\nu} - i~d_{\rho}B^{\mu\nu\rho}) = 0.
\ee 

(ii) We use again Theorem \ref{m=0} and obtain
\be
T^{\mu\nu} = t^{\mu\nu} + d_{Q}B^{\mu\nu} + i~d_{\rho}B^{\mu\nu\rho} 
+ T^{\mu\nu}_{0}
\ee 
where
$
T_{0}^{\mu\nu} \in {\cal P}_{0}^{(5)}
$
and we can choose the expressions
$
B^{\mu\nu}
$
and
$
T_{0}^{\mu\nu}
$
antisymmetric. The generic form of the expression 
$
T_{0}^{\mu\nu}
$
is:
\be
T_{0}^{\mu\nu} = b~u_{\rho}~u_{\sigma}~R^{(0)\mu\nu;\rho\sigma}
+ b^{\prime}~\epsilon^{\mu\nu\rho\sigma}~
u^{\alpha}~u^{\beta}~R^{(0)}_{\rho\sigma;\alpha\beta};
\ee
the monomials
$
u_{\rho}~u_{\sigma}~R^{(0)\mu\rho;\nu\sigma}
$
and
$
\epsilon^{\mu\nu\rho\sigma}~u^{\alpha}~u^{\beta}~R^{(0)}_{\rho\alpha;\sigma\beta}
$
can be eliminated if we use the following consequence of the Bianchi identity:
\be
R^{(0)}_{\mu\nu;\rho\sigma} + R^{(0)}_{\mu\rho;\nu\sigma} 
+ R^{(0)}_{\mu\sigma;\nu\rho} = d_{Q}(\cdots)
\ee
and redefine the expression
$
B^{\mu\nu}.
$
We substitute the expression of
$
T^{\mu\nu}
$
in (\ref{descent-t}) and get:
\be
d_{Q}(T^{\mu} - i~d_{\nu}B^{\mu\nu}) = i~d_{\nu}(T_{0}^{\mu\nu} + t^{\mu\nu}).
\ee

But one proves by direct computation that we have
\be
d_{\rho}t^{\mu\nu} = - i~d_{Q}~t^{\mu}
\ee
where
$
t^{\mu}
$
is the expression from the statement so the preceding relation becomes
\be
d_{Q}(T^{\mu} - t^{\mu}- i~d_{\nu}B^{\mu\nu}) = i~d_{\nu}T_{0}^{\mu\nu}.
\label{descent-t'}
\ee
The right hand side must be a co-boundary and one easily obtains that
$
b = b^{\prime} = 0
$
so we have:
\be
d_{Q}(T^{\mu} - t^{\mu}- i~d_{\nu}B^{\mu\nu}) = 0.
\ee

(iii) Now it is again time we use Theorem \ref{m=0} and obtain
\be
T^{\mu} = t^{\mu} + d_{Q}B^{\mu} + i~d_{\nu}B^{\mu\nu} + T^{\mu}_{0}
\ee 
where
$
T_{0}^{\mu} \in {\cal P}_{0}^{(5)}.
$
But there are no such expression i.e.
$
T^{\mu}_{0} = 0
$
and we have
\be
T^{\mu} = t^{\mu} + d_{Q}B^{\mu} + i~d_{\nu}B^{\mu\nu}.
\ee 
Now we get from (\ref{descent-t0}) 
\be
d_{Q}(T - i~d_{\mu}B^{\mu}) = i~d_{\mu}t^{\mu}.
\label{descent-0t'}
\ee
But we obtain by direct computation that we have
\be
d_{\rho}t^{\mu} = - i~d_{Q}~t
\ee
where $t$ is the expression from the statement so the preceding relation becomes
\be
d_{Q}(T - t - i~d_{\mu}B^{\mu}) = 0
\ee
so a last use of Theorem \ref{m=0} gives
\be
T = t + d_{Q}B + i~d_{\mu}B^{\mu} + T_{0}
\ee 
where
$
T_{0} \in {\cal P}_{0}^{(5)}.
$
But there are no such expression i.e.
$
T_{0} = 0
$
and we have 
\be
T = t + d_{Q}B + i~d_{\mu}B^{\mu}
\ee 
i.e. we have obtained the first four assertions from the statement.

(v) We prove now that $t$ from the statement is not a trivial (relative) co-cycle. Indeed, if this would be true i.e.
$
t = d_{Q}B + i~d_{\mu}B^{\mu}
$
then we get 
$
d_{\mu}(t^{\mu} - d_{Q}B^{\mu}) = 0
$
so with Poincar\'e lemma we have
$
t^{\mu} = d_{Q}B^{\mu} + i~d_{\nu}B^{[mu\nu}.
$
In the same way we obtain from here:
$
t^{\mu\nu} = d_{Q}B^{\mu\nu} + i~d_{\rho}B^{\mu\nu\rho}
$
and
$
t^{\mu\nu\rho} = d_{Q}B^{\mu\nu\rho} + i~d_{\sigma}B^{\mu\nu\rho\sigma}
$
But it is easy to see that there is no such an expression
$
B^{\mu\nu\rho\sigma}
$
with the desired antisymmetry property in ghost number $4$ so we have in fact
$
t^{\mu\nu\sigma} = d_{Q}B^{\mu\nu\sigma}.
$
This relation contradicts the fact that
$
t^{\mu\nu\sigma}
$
is a non-trivial co-cycle for
$
d_{Q}
$
as it follows from Theorem \ref{m=0}. The invariance with respect to parity invariance is obvious.
$\qed$

If $T$ is bi-linear in the fields we cannot use the Poincar\'e lemma but we can make a direct analysis. The result is the following.
\begin{thm}
Let $T$ be a relative co cycle for 
$
d_{Q}
$
which is bi-linear in the fields, of canonical dimension
$
\omega(T) \leq 5
$
and ghost number
$
gh(T) = 0.
$
Then:
(i) $T$ is (relatively) cohomologous to an expression of the form:
\be
t = \kappa^{\prime}~
( - 2~h_{\mu\nu}~h^{\mu\nu} + h^{2} - 4 u_{\mu}~\tilde{u}^{\mu}) 
\ee

(ii) The relation 
$
d_{Q}t = i~d_{\mu}t^{\mu}
$
is verified with 
\be
t^{\mu} = 4~\kappa^{\prime}~u_{\nu} h^{\mu\nu}
\ee
and we also have
\be
d_{Q}t^{\mu} = 2 i \kappa^{\prime}~[u_{\nu}~d^{\mu}u^{\nu} 
+ d_{\nu}~(u^{\mu}u^{\mu})].
\ee
\end{thm}
{\bf Proof:} The cases
$
\omega(T) = 3, 5
$
are not possible on grounds of Lorentz covariance. The cases
$
\omega(T) = 2, 4
$
must be investigated starting from a general ansatz and the solution from the statement emerges.
$\qed$

All linear solutions of this problem are trivial. Now we extend this result to the case
$
m > 0.
$

\begin{thm}
Let 
$
T_{m}
$ 
be a relative co cycle for 
$
d_{Q}
$
which is as least tri-linear in the fields and is of canonical dimension
$
\omega(T_{m}) \leq 5
$
and ghost number
$
gh(T_{m}) = 0.
$
Then:
(i) 
$
T_{m}
$ 
is (relatively) cohomologous to a non-trivial co-cycle of the form:
\bea
t_{m} = t + \kappa \Bigl[ m^{2}~\left( {4\over 3}~h^{\mu\nu}~h_{\nu\rho}~{h_{\mu}}^{\rho}
- h^{\mu\nu}~h_{\mu\nu}~h + {1\over 6} h^{3}\right) 
\nonumber \\
+ 4~m~u_{\rho}~d^{\rho}v^{\lambda}~\tilde{u}_{\lambda}
- 4~d^{\rho}v^{\sigma}~d^{\lambda}v_{\sigma}~h_{\rho\lambda}\Bigl]
\eea
where $t$ is the expression from the preceding theorem.

(ii) The relation 
$
d_{Q}t_{m} = i~d_{\mu}t^{\mu}_{m}
$
is verified by:
\bea
t_{m}^{\mu} = t_{\mu} + \kappa \Bigl[ 4 u_{\rho}~d^{\rho}v_{\lambda}~d^{\mu}v^{\lambda} 
- 2 u^{\mu}~d^{\rho}v^{\lambda}~d_{\rho}v_{\lambda} 
+ 4 m (u^{\mu}~d_{\rho}v_{\lambda}~h^{\rho\lambda}
- u^{\rho}~d_{\rho}v_{\lambda}~h^{\mu\lambda})
\nonumber \\
- m^{2}~u^{\mu}~\left(h^{\rho\sigma}~h_{\rho\sigma} - {1\over 2}h^{2}\right)\Bigl]
\label{Tmu-m}
\eea
where 
$
t_{\mu}
$
is the expression from the preceding theorem.

(iii) The relation 
$
d_{Q}t_{m}^{\mu} = i~d_{\nu}t_{m}^{\mu\nu}
$
is verified by:
\be
t_{m}^{\mu\nu} \equiv t^{\mu\nu} + 2~\kappa m (v^{\mu}~u^{\nu} - v^{\nu}~u^{\mu})
~d^{\rho}u_{\rho}
\label{Tmunu-m}
\ee
where 
$
t^{\mu\nu}
$
is the expression from the preceding theorem.

(iv) The relation 
$
d_{Q}t_{m}^{\mu\nu} = i~d_{\rho}t_{m}^{\mu\nu\rho}
$
is verified by:
\be
t_{m}^{\mu\nu\rho} \equiv t^{\mu\nu\rho} - 2~\kappa m^{2} u^{\mu}~u^{\nu}~u^{\rho}
\label{Tmunurho-m}
\ee
where
$
t^{\mu\nu\rho}
$
is the expression from the preceding theorem. We also have
$
d_{Q}t_{m}^{\mu\nu\rho} = 0.
$

(v) The co-cycles
$
t_{m}, t_{m}^{\mu}, t_{m}^{\mu\nu}
$
and
$
t_{m}^{\mu\nu\rho}
$
are non-trivial, parity invariant and have smooth limit for
$
m \searrow 0.
$
\label{T1-m}
\end{thm}
{\bf Proof:}
(i) As in the preceding theorem we can prove that we must have
\be
d_{Q}T_{m} = i~d_{\mu}T_{m}^{\mu}.
\label{descent-t-m}
\ee
and
\bea
d_{Q}T_{m}^{\mu} = i~d_{\nu}T_{m}^{\mu\nu},
\nonumber \\
d_{Q}T_{m}^{\mu\nu} = i~d_{\rho}T_{m}^{\mu\nu\rho}
\nonumber \\
d_{Q}T_{m}^{\mu\nu\rho} = i~d_{\sigma}T_{m}^{\mu\nu\rho\sigma}
\nonumber \\
d_{Q}T_{m}^{\mu\nu\rho\sigma} = 0.
\label{descent-T-m}
\eea
According to a theorem proved in \cite{cohomology} can choose the expressions
$
T^{I}_{m}
$
to be Lorentz covariant; we also have
\be
gh(T_{m}^{I}) = |I|.
\ee 

From the last relation we find, using Theorem \ref{Q-cohomology} that
\be
T_{m}^{\mu\nu\rho\sigma} = d_{Q}B^{\mu\nu\rho\sigma} + T_{0,m}^{\mu\nu\rho\sigma}
\ee
with
$
T_{0,m}^{\mu\nu\rho\sigma} \in {\cal P}_{0}^{(5)}
$
and we can choose the expressions
$
B^{\mu\nu\rho\sigma}
$
and
$
T_{0,m}^{\mu\nu\rho\sigma}
$
completely antisymmetric. The generic form of 
$
T_{0,m}^{\mu\nu\rho\sigma}
$
is the same as in the preceding theorem and if we substitute the expression obtained for
$
T^{\mu\nu\rho\sigma}
$
in the third relation (\ref{descent-T-m}) we find out as there that
$
a = 0.
$
It follows that
\be
T_{m}^{\mu\nu\rho\sigma} = d_{Q}B^{\mu\nu\rho\sigma}
\ee
and
\be
d_{Q}(T_{m}^{\mu\nu\rho} - i~d_{\sigma}B^{\mu\nu\rho\sigma}) = 0
\ee 
We apply again Theorem \ref{Q-cohomology} and obtain
\be
T_{m}^{\mu\nu\rho} = B^{\mu\nu\rho} + i~d_{\sigma}B^{\mu\nu\rho\sigma} + T^{\mu\nu\rho}_{0,m}
\ee 
where
$
T_{0,m}^{\mu\nu\rho} \in {\cal P}_{0}^{(5)}
$
and we can choose the expressions
$
B^{\mu\nu\rho}
$
and
$
T_{0,m}^{\mu\nu\rho}
$
completely antisymmetric. The generic form of
$
T_{0,m}^{\mu\nu\rho}
$
is:
\bea
T_{0,m}^{\mu\nu\rho} = T_{0}^{\mu\nu\rho} 
+ c_{1}~u^{\mu}~u^{\nu}~u^{\rho}~\phi
+ c_{2}~( u^{\mu}~u^{\nu}~\phi^{\rho\lambda} 
+ u^{\nu}~u^{\rho}~\phi^{\mu\lambda}
+ u^{\rho}~u^{\mu}~\phi^{\nu\lambda})~u_{\lambda}
\eea
where 
$
T_{0}^{\mu\nu\rho}
$
is the expression (\ref{t0mnr}) from the massless case with
$
a = 0
$
(the corresponding term is a secondary co-cycles). We substitute the expression
$
T^{\mu\nu\rho}
$
into the first relation (\ref{descent-T-m}) and obtain
\be
d_{Q}(T^{\mu\nu} - i~d_{\rho}B^{\mu\nu\rho}) = i~d_{\rho}T_{0}^{\mu\nu\rho}.
\ee 
The right hand side must be a co-boundary. If we compute the divergence
$
d_{\rho}T_{0,m}^{\mu\nu\rho}
$
and impose that it is a co-boundary we obtain immediately
$
c_{j} = 0~(j = 1,2)
$
and 
$
a^{\prime} = 0
$
so
$
T_{0}^{\mu\nu\rho}
$
is given by the same expression as in the massless case and we also can take
$
a_{j} = 2~\kappa~(j = 1,2)
$
as we have argued there. 

In this case one can prove rather easily that
$
T_{0,m}^{\mu\nu\rho} = t_{m}^{\mu\nu\rho} + d_{Q}(\cdots)
$ 
where
$
t_{m}^{\mu\nu\rho}
$
is the expression from the statement. It follows that one can exhibit
$
T_{m}^{\mu\nu\rho}
$
in the following form:
\be
T_{m}^{\mu\nu\rho} = d_{Q}B^{\mu\nu\rho} + i~d_{\sigma}B^{\mu\nu\rho\sigma}
+ t_{m}^{\mu\nu\rho}
\ee
Now one proves by direct computation that
\be
d_{\rho}t_{m}^{\mu\nu\rho} = - i~d_{Q}~t_{m}^{\mu\nu}
\ee
where
$
t_{m}^{\mu\nu}
$
is the expression from the statement. We substitute this expression in the second relation (\ref{descent-T-m}) and obtain
\be
d_{Q}(T_{m}^{\mu\nu} - t_{m}^{\mu\nu} - i~d_{\rho}B^{\mu\nu\rho}) = 0.
\ee 

(ii) We use again Theorem \ref{m>0} and obtain
\be
T_{m}^{\mu\nu} = t_{m}^{\mu\nu} + d_{Q}B^{\mu\nu} + i~d_{\rho}B^{\mu\nu\rho} 
+ T^{\mu\nu}_{0,m}
\ee 
where
$
T_{0,m}^{\mu\nu} \in {\cal P}_{0}^{(5)}
$
and we can choose the expressions
$
B^{\mu\nu}
$
and
$
T_{0,m}^{\mu\nu}
$
antisymmetric. The generic form of the expression 
$
T_{0,m}^{\mu\nu}
$
is the same as in the massless case
$
T_{0,m}^{\mu\nu} = T_{0}^{\mu\nu}
$
so we obtain from the first relation in (\ref{descent-T-m}):
\be
d_{Q}(T_{m}^{\mu} - i~d_{\nu}B^{\mu\nu}) = i~d_{\nu}(T_{0}^{\mu\nu} + t_{m}^{\mu\nu}).
\ee

But one proves by direct computation that we have
\be
d_{\rho}t_{m}^{\mu\nu} = - i~d_{Q}~t_{m}^{\mu}
\ee
where
$
t_{m}^{\mu}
$
is the expression from the statement so the preceding relation becomes
\be
d_{Q}(T_{m}^{\mu} - t_{m}^{\mu}- i~d_{\nu}B^{\mu\nu}) = i~d_{\nu}T_{0}^{\mu\nu}.
\ee
The right hand side must be a co-boundary and one obtains as in the massless case
$
T_{0}^{\mu\nu} = 0
$
so we have:
\be
T_{m}^{\mu\nu} = t_{m}^{\mu\nu} + d_{Q}B^{\mu\nu} + i~d_{\rho}B^{\mu\nu\rho} 
\ee 
and
\be
d_{Q}(T_{m}^{\mu} - t_{m}^{\mu}- i~d_{\nu}B^{\mu\nu}) = 0.
\ee

(iii) Now it is again time we use Theorem \ref{m>0} and obtain
\be
T_{m}^{\mu} = t_{m}^{\mu} + d_{Q}B^{\mu} + i~d_{\nu}B^{\mu\nu} + T^{\mu}_{0,m}
\ee 
where
$
T_{0,}^{\mu} \in {\cal P}_{0}^{(5)}.
$
The generic form of such expression is
\be
T^{\mu}_{0,m} = d_{1}~u^{\mu}~\phi^{2} 
+ d_{2}~u^{\mu}~\phi^{\rho\sigma}~\phi_{\rho\sigma}
+ d_{3}~u_{\nu}~\phi^{\mu\nu}~\phi
+ d_{4}~u^{\sigma}~\phi^{\mu\rho}~\phi_{\rho\sigma}
\ee
and we have from the relation (\ref{descent-t-m})
\be
d_{Q}(T_{m} - i~d_{\nu}B^{\mu\nu}) = i~(d_{\mu}t_{m}^{\mu} + d_{\mu}T^{\mu}_{0,m})
\ee
so the right hand side must be a co-boundary. But one proves by direct computation that
\be
d_{\rho}t_{m}^{\mu} = - i~d_{Q}~t_{m}
\ee
where
$
t_{m}
$
is the expression from the statement so the preceding relation becomes
\be
d_{Q}(T_{m} - t_{m} - i~d_{\mu}B^{\mu}) = i~d_{\mu}T^{\mu}_{0,m}
\ee
so the expression
$
d_{\mu}T^{\mu}_{0,m}
$
must be a co-boundary. By direct computation we obtain from this condition
$
d_{j} = 0~(j = 1,\dots,4)
$
i.e.
$
T^{\mu}_{0,m} = 0.
$
It follows that 
\be
T_{m}^{\mu} = t_{m}^{\mu} + d_{Q}B^{\mu} + i~d_{\nu}B^{\mu\nu}
\ee 
and
\be
d_{Q}(T_{m} - t_{m} - i~d_{\mu}B^{\mu}) = 0
\ee
so a last use of Theorem \ref{m>0} gives
\be
T_{m} = t_{m} + d_{Q}B + i~d_{\mu}B^{\mu} + T_{0,m}
\ee 
where
$
T_{0,m} \in {\cal P}_{0}^{(5)}.
$
But there are no such expression i.e.
$
T_{0,m} = 0
$
and we have 
\be
T_{m} = t_{m} + d_{Q}B + i~d_{\mu}B^{\mu}
\ee 
i.e. we have obtained the first four assertions from the statement.

(v) We prove now that 
$
t_{m}
$ from the statement is not a trivial (relative) co-cycle as in the massless case.
Parity invariance and the existence of a smooth limit
$
m \searrow 0
$
are obvious.
$\qed$

If 
$
T_{m}
$ 
is bi-linear in the fields we cannot use the Poincar\'e lemma but we can make a direct analysis as in the massless case. The result is the following.
\begin{thm}
Let 
$
T_{m}
$ 
be a relative co-cycle for 
$
d_{Q}
$
which is bi-linear in the fields, of canonical dimension
$
\omega(T_{m}) \leq 5
$
and ghost number
$
gh(T_{m}) = 0.
$
Then:
(i) 
$
T_{m}
$ 
is (relatively) cohomologous to an expression of the form:
\be
t_{m} = \kappa^{\prime}~
( - 2~h_{\mu\nu}~h^{\mu\nu} + h^{2} - 4 u_{\mu}~\tilde{u}^{\mu} 
+ 4 v_{\mu}~v^{\mu}) 
\ee

(ii) The expression 
$
t_{m}^{\mu}
$
coincides with the expression
$
t^{\mu}
$
from the massless case.
\end{thm}

\section{Gauge Invariance and Renormalization in the Second Order of Perturbation Theory\label{int}}

In the same way one can analyze the descent equations (\ref{W}) and study the form of the anomalies in the second order of perturbation theory.

\begin{thm}
In the massless case the second order chronological products can be chosen such that
the expression
$
W^{I}
$
from theorem \ref{ano-2} are
\be
W = 0,~W^{\mu} = 0~,W^{\mu\nu} = 0
\ee
and
\be
W^{\mu\nu\rho} = d_{Q}B^{\mu\nu\rho}
\ee
with the expression
$
B^{\mu\nu\rho}
$
completely antisymmetric and constrained by the conditions
$
\omega(B^{\mu\nu\rho}) \leq 6
$ 
and
$
gh(B^{\mu\nu\rho}) = 4.
$
\end{thm}
{\bf Proof:} We will need the relations (\ref{W}) in which we prefer to change some signs
$
W^{\mu} \rightarrow - W^{\mu},~~
W^{\mu\nu} \rightarrow - W^{\mu\nu}
$
i.e.
\be
d_{Q}W = i~\partial_{\mu}W^{\mu},\qquad
d_{Q}W^{\mu} = i~\partial_{\nu}W^{\mu\nu},\qquad
d_{Q}W^{\mu\nu} = i\partial_{\rho}W^{\mu\nu\rho},\qquad
d_{Q}W^{\mu\nu\rho} = 0
\label{W1}
\ee
and we also have from (\ref{power4}) the bound
$
\omega(W^{I}) \leq 7.
$
Moreover, the parity invariance obtained in theorem \ref{m=0} can be used to prove that the polynomials
$
W^{I}
$
are also parity invariant.

(i) From the last relation (\ref{W1}) and theorem \ref{m=0} we obtain:
\be
W^{\mu\nu\rho} = d_{Q}B^{\mu\nu\rho} + W_{0}^{\mu\nu\rho}
\label{w3}
\ee
with
$
W_{0}^{\mu\nu\rho} \in {\cal P}_{0}^{(7)}
$
and we can choose the expressions
$
B^{\mu\nu\rho}
$
and
$
W_{0}^{\mu\nu\rho}
$
completely antisymmetric. The generic form of 
$
W_{0}^{\mu\nu\rho}
$
is:
\bea
W_{0}^{\mu\nu\rho} = 
a_{1}~( u^{\mu\lambda}~u^{\nu}~u^{\rho} + u^{\nu\lambda}~u^{\rho}~u^{\mu} 
+ u^{\rho\lambda}~u^{\mu}~u^{\nu})~u_{\lambda}
\nonumber \\
+ a_{2}~( u^{\mu\nu}~u^{\rho\lambda} + u^{\nu\rho}~u^{\mu\lambda} 
+ u^{\rho\mu}~u^{\nu\lambda})~u_{\lambda\sigma}~u^{\sigma}
\nonumber \\
+ a_{3}~( u^{\mu\lambda}~{u^{\nu}}_{\lambda}~u^{\rho\sigma} 
+ u^{\nu\lambda}~{u^{\rho}}_{\lambda}~u^{\mu\sigma} 
+ u^{\rho\lambda}~{u^{\mu}}_{\lambda}~u^{\nu\sigma})~u_{\sigma}
\nonumber \\
+ a_{4}~( u^{\mu\lambda}~u^{\nu\sigma}~u^{\rho} 
+ u^{\nu\lambda}~u^{\rho\sigma}~u^{\mu} 
+ u^{\rho\lambda}~u^{\mu\sigma}~u^{\nu})~u_{\lambda\sigma}
\eea
with
$
a_{j} \in \R~(j = 1,\dots,4).
$
We denote by
$
T_{j}~(j = 1,\dots,4)
$
the polynomials multiplied by
$
a_{j}~(j = 1,\dots,4)
$
respectively. If we define the completely antisymmetric expressions
\bea
b_{1}^{\mu\nu\rho\sigma} \equiv u^{\mu}~u^{\nu}~u^{\rho}~u^{\sigma}
\nonumber \\
b_{2}^{\mu\nu\rho\sigma} \equiv (u^{\mu\lambda}~{u^{\nu}}_{\lambda}~u^{\rho} 
+ u^{\nu\lambda}~{u^{\rho}}_{\lambda}~u^{\mu} 
+ u^{\rho\lambda}~{u^{\mu}}_{\lambda}~u_{\nu})~u^{\sigma}
\nonumber \\
+ u^{\sigma\lambda}~({u^{\mu}}_{\lambda}~u^{\rho}~u^{\nu} 
+ {u^{\nu}}_{\lambda}~u^{\mu}~u^{\rho} + {u^{\rho}}_{\lambda}~u^{\nu}~u^{\mu})
\nonumber \\
b_{3}^{\mu\nu\rho\sigma} \equiv (u^{\mu\nu}~u^{\rho\lambda} 
+ u^{\nu\rho}~u^{\mu\lambda} + u^{\rho\mu}~u^{\nu\lambda})~u^{\sigma}~u_{\lambda}
\nonumber \\
+ (u^{\mu\sigma}~u^{\nu\lambda}~u^{\rho} 
+ u^{\nu\sigma}~u^{\rho\lambda}~u^{\mu} + u^{\rho\sigma}~u^{\mu\lambda}~u^{\nu})~
u_{\lambda}
\eea
then it is not hard to obtain that
\bea
d_{\sigma}b_{1}^{\mu\nu\rho\sigma} = - T_{1}^{\mu\nu\rho} + d_{Q}b_{1}^{\mu\nu\rho}
\nonumber \\
d_{\sigma}b_{2}^{\mu\nu\rho\sigma} = - T_{3}^{\mu\nu\rho} + d_{Q}b_{2}^{\mu\nu\rho}
\nonumber \\
d_{\sigma}b_{3}^{\mu\nu\rho\sigma} = T_{2}^{\mu\nu\rho}  - T_{3}^{\mu\nu\rho}
+ T_{4}^{\mu\nu\rho} + d_{Q}b_{3}^{\mu\nu\rho}
\eea
where we can choose the expressions
$
b_{j}^{\mu\nu\rho}~(j = 1,\dots,3)
$
completely antisymmetric. It follows that we can rewrite (\ref{w3}) in the form
\be
W^{\mu\nu\rho} = d_{Q}B^{\mu\nu\rho} + i~d_{\sigma}B^{\mu\nu\rho\sigma} 
+ W_{0}^{\mu\nu\rho}
\ee
where in the expression of
$
W_{0}^{\mu\nu\rho}
$
we can make
$
a_{1} = a_{3} = a_{4} = 0. 
$
We substitute the preceding expression in the third relation (\ref{W1}) and get
\be
d_{Q}(W^{\mu\nu} - i~d_{\rho}B^{\mu\nu\rho}) = i~d_{\rho}W_{0}^{\mu\nu\rho}
\ee 
so the right hand side must be a co-boundary. From this condition we easily find
$
a_{2} = 0
$
so in fact we can take
$
T_{0}^{\mu\nu\rho} = 0
$ 
It follows that one can exhibit
$
W^{\mu\nu\rho}
$
in the following form:
\be
W^{\mu\nu\rho} = d_{Q}B^{\mu\nu\rho} + i~d_{\sigma}B^{\mu\nu\rho\sigma}
\ee
and we also have
\be
d_{Q}(W^{\mu\nu} - i~d_{\rho}B^{\mu\nu\rho}) = 0.
\ee 

(ii) We use again Theorem \ref{m=0} and obtain
\be
W^{\mu\nu} = d_{Q}B^{\mu\nu} + i~d_{\rho}B^{\mu\nu\rho} + W^{\mu\nu}_{0}
\label{w2}
\ee 
where
$
W_{0}^{\mu\nu} \in {\cal P}_{0}^{(7)}
$
and we can choose the expressions
$
B^{\mu\nu}
$
and
$
W_{0}^{\mu\nu}
$
antisymmetric. The generic form of the expression 
$
W_{0}^{\mu\nu}
$
is:
\bea
W_{0}^{\mu\nu} = b_{1}~(u^{\mu\rho}~u^{\nu} - u^{\nu\rho}~u^{\mu})~u_{\rho}
+ b_{2}~u^{\mu\rho}~u^{\nu\lambda}~u_{\rho\lambda}
+ b_{3}~R^{(0)\mu\nu;\alpha\beta}~u_{\alpha\rho}~u_{\beta}~u^{\rho}
\nonumber \\
+ b_{4}~(R^{(0)\mu\rho;\alpha\beta}~u^{\nu} - R^{(0)\nu\rho;\alpha\beta}~u^{\mu})
~u_{\alpha\beta}~u_{\rho}
+ b_{5}~(R^{(0)\mu\rho;\alpha\beta}~{u^{\nu}}_{\rho} 
- R^{(0)\nu\rho;\alpha\beta}~{u^{\mu}}_{\rho})~u_{\alpha}~u_{\beta};
\eea
many other possible expressions can be eliminated, or reduced to these above if we use Bianchi identity. We denote by
$
T_{j}~(j = 1,\dots,5)
$
the polynomials multiplied by
$
b_{j}~(j = 1,\dots,5)
$
respectively. If we define the completely antisymmetric expressions
\bea
b_{1}^{\mu\nu\rho} \equiv u^{\mu}~u^{\nu}~u^{\rho}
\nonumber \\
b_{2}^{\mu\nu\rho} \equiv (R^{(0)\mu\nu;\alpha\beta}~u^{\rho}
+ R^{(0)\nu\rho;\alpha\beta}~u^{\mu} + R^{(0)\rho\mu;\alpha\beta}~u^{\nu})
~u_{\alpha}~u_{\beta}
\eea
we easily obtain that 
\bea
d_{\rho}b_{1}^{\mu\nu\rho} = - T_{1}^{\mu\nu} + d_{Q}b_{1}^{\mu\nu}
\nonumber \\
d_{\rho}b_{2}^{\mu\nu\rho} = T_{5}^{\mu\nu} - 2 T_{3}^{\mu\nu} - T_{4}^{\mu\nu} 
+ d_{Q}b_{2}^{\mu\nu}
\eea
where we can choose the expressions
$
b_{j}^{\mu\nu}~(j = 1,\dots,3)
$
antisymmetric. It follows that if we redefine the expressions
$
B^{\mu\nu\rho}
$
and
$
B^{\mu\nu}
$
from (\ref{w2}) we can make 
$
b_{1} = b_{4} = 0
$
in
$
W_{0}^{\mu\nu}.
$
We substitute the expression of
$
W^{\mu\nu}
$
in the second relation (\ref{W1}) and get:
\be
d_{Q}(W^{\mu} - i~d_{\nu}B^{\mu\nu}) = i~d_{\nu}W_{0}^{\mu\nu}
\ee
so the right hand side must be a co-boundary. One easily obtains that
$
b_{3} = b_{5} = 0
$
so we are left with one nontrivial co-cycle corresponding to
$
b \equiv b_{2}:
$
\be
W_{0}^{\mu\nu} = b~u^{\mu\rho}~u^{\nu\lambda}~u_{\rho\lambda}.
\ee

Now one proves immediately that
\be
d_{\nu}W_{0}^{\mu\nu} = - i~d_{Q}U^{\mu}
\ee
where
\be
U^{\mu} \equiv b \left(d_{\rho}\hat{h}^{\mu\nu}~u_{\nu\lambda}~u^{\rho\lambda}
+ {1\over 2}~d^{\lambda}h~u^{\mu\rho}~u_{\rho\lambda}
+ d_{\lambda}\hat{h}_{\nu\rho}~u^{\mu\rho}~u^{\nu\lambda}\right) 
\ee
and we have 
\be
d_{Q}(W^{\mu} - U^{\mu}- i~d_{\nu}B^{\mu\nu}) = 0.
\ee

(iii) Now it is again time we use Theorem \ref{m=0} and obtain
\be
W^{\mu} = U^{\mu} + d_{Q}B^{\mu} + i~d_{\nu}B^{\mu\nu} + W^{\mu}_{0}
\ee 
where
$
W_{0}^{\mu} \in {\cal P}_{0}^{(7)}.
$
The generic form of such an expression is:
\be
W^{\mu}_{0} = c_{1}~u^{\mu\nu}~u_{\nu}
+ c_{2}~R^{(0)\mu\nu;\alpha\beta}~u_{\alpha\beta}~u_{\nu};
\label{w1}
\ee
we denote by
$
T_{j}~(j = 1,2)
$
the polynomials multiplied by
$
c_{j}~(j = 1,2).
$
However let us consider the antisymmetric expressions
\bea
b_{1}^{\mu\nu} \equiv u^{\mu}~u^{\nu}
\nonumber \\
b_{2}^{\mu\nu} \equiv R^{(0)\mu\nu;\alpha\beta}~~u_{\alpha}~u_{\beta}
\eea
and we have
\bea
d_{\nu}b_{1}^{\mu\nu} = - T_{1}^{\mu} + d_{Q}b_{1}^{\mu}
\nonumber \\
d_{\nu}b_{2}^{\mu\nu} = T_{2}^{\mu} + d_{Q}b_{2}^{\mu}
\eea
so if we redefine the expressions
$
B^{\mu\nu}
$
and
$
B^{\mu}
$
from (\ref{w1}) we can make 
$
c_{1} = c_{2} = 0
$
i.e.
$
W_{0}^{\mu} = 0.
$
As a consequence we have:
\be
W^{\mu} = U^{\mu} + d_{Q}B^{\mu} + i~d_{\nu}B^{\mu\nu}.
\ee 
Now we get from the first equation (\ref{W1}) 
\be
d_{Q}(W - i~d_{\mu}B^{\mu}) = i~d_{\mu}U^{\mu}
\ee
so the right hand side must be a co-boundary. One can prove that this is not possible so we must have
$
b = 0
$
i.e.
$
U^{\mu} = 0.
$
As a consequence we have
\be
W^{\mu} = d_{Q}B^{\mu} + i~d_{\nu}B^{\mu\nu}
\ee 
and
\be
d_{Q}(W - i~d_{\mu}B^{\mu}) = 0
\ee
so a last use of Theorem \ref{m=0} gives
\be
W = d_{Q}B + i~d_{\mu}B^{\mu} + W_{0}
\ee 
where
$
W_{0} \in {\cal P}_{0}^{(7)}.
$
But there are no such expression i.e.
$
W_{0} = 0
$
and we have 
\be
W = d_{Q}B + i~d_{\mu}B^{\mu}.
\ee 

Now we use finite renormalizations to eliminate the expressions
$
B^{I}~(|I| \leq 3)
$
as in the end of theorem \ref{ano-2} and end up with the expression from the statement.
$\qed$

\begin{rem}

(i) One can extend the preceding result to the massive case also. The complications are only of technical nature: more terms can appear in the generic expressions of the expressions
$
W^{I}_{0}
$
but they eventually are eliminated.

(ii) The preceding proof stays true if we do not use parity invariance: as in the preceding remark, more terms can appear in the expressions
$
W^{I}_{0}.
$ 

(iii) We cannot eliminate
$
B^{\mu\nu\rho\sigma}
$
by finite renormalizations.

(iv) In higher orders of perturbation theory some expressions of the type
$
W^{I}_{0}
$
will survive the algebraic machinery we have used and new ideas are needed to eliminate them.
\end{rem}
We have reduced the gauge invariance problem in the second order to a much simple computation namely of the anomaly
$
A_{7}
$
where the expression
$
W^{\mu\nu\rho}
$
do appear. But we have
\begin{thm}
The anomaly
$
A_{7}
$
can be eliminated by finite renormalizations.
\label{a7}
\end{thm}
{\bf Proof:} We consider the massless case. The standard procedure is to show by direct computation that
\bea
[t^{\mu\nu\rho}(x_{1}), t^{\sigma}(x_{2}) ] =
A^{\mu\nu\rho}(x_{1},x_{2})~\partial^{\sigma}D_{0}(x_{1} - x_{2}) 
+ A^{\mu\nu\rho;\alpha}(x_{1},x_{2})~
\partial^{\sigma}\partial_{\alpha}D_{0}(x_{1} - x_{2}) + \cdots
\eea 
where by
$
\cdots
$
we mean terms for which the index $\sigma$ does not act on the Pauli-Jordan distribution. If we transform this distribution in the corresponding Feynman propagator
$
D_{0}^{F}
$
we obtain in the left hand side of the relation (\ref{g7}) an anomaly of the form:
\be
A^{[\mu\nu\rho]}_{7}(x_{1},x_{2})
= \delta(x_{2} - x_{1})~A^{\mu\nu\rho}(x_{1},x_{2})
+ [\partial_{\alpha}\delta(x_{2} - x_{1})]~A^{\mu\nu\rho;\alpha}(x_{1},x_{2}).
\ee
Now we make simple computation to put the preceding expression in the standard form (\ref{A7-2}):
\be
A^{[\mu\nu\rho]}_{7}(x_{1},x_{2})
= \delta(x_{2} - x_{1})~W^{\mu\nu\rho}(x_{1})
+ \partial_{\alpha}\delta(x_{2} - x_{1})~W^{\mu\nu\rho;\alpha}(x_{1}).
\ee
The second term can be eliminated by a finite renormalization of the chronological products 
$
T(T^{[\mu\nu\rho]}(x_{1}),T^{\sigma}(x_{2})) 
$
of the type (\ref{R8c}). Now the only thing to prove is that the first term is a co-boundary:
\be
W^{\mu\nu\rho}_{7} = d_{Q}B^{\mu\nu\rho}
\ee
and we can eliminate it by a redefinition of the chronological products
$
T(T^{[\mu\nu\rho]}(x_{1}),T(x_{2})).
$

Finally one can prove that in the massive case no new contributions to the anomaly
$
A_{7}
$
do appear.
$\qed$

We now turn to the renormalization problem for the second order of the perturbation theory. We have the following result:
\begin{thm}
In the massless case the finite renormalizations for the second order chronological products are of the form
\be
R^{I} = t^{\prime} + d_{Q}B^{I} + i~d_{\mu}B^{I\mu}
\ee
where
$
t^{\prime}
$
has the same form as the interaction Lagrangian $t$ from theorem \ref{T1} (but with a different overall constant) and can be eliminated by a redefinition of the gravitational coupling $\kappa$. The rest of the terms can be eliminated by finite renormalizations of the chronological products. In the massive case the contribution
\be
R_{0} = r_{1}~\phi^{3} + r_{2}~\phi~\phi^{(0)}_{\alpha\beta}~\phi^{(0)\alpha\beta}
+ r_{3}~\phi^{(0)}_{\alpha\gamma}~\phi^{(0)}_{\alpha\beta}~{\phi^{(0)\mu}}_{\beta}
\ee
of
$
R^{\emptyset}
$
cannot be eliminated. 
\end{thm}
{\bf Proof:} According to (\ref{gauge5}) we have the following descent procedure:
\bea
d_{Q}R = i~d_{\mu}R^{\mu}.
\nonumber \\
d_{Q}R^{\mu} = i~d_{\nu}R^{\mu\nu}.
\nonumber \\
d_{Q}R^{\mu\nu} = i~d_{\rho}R^{\mu\nu\rho}
\nonumber \\
d_{Q}R^{\mu\nu\rho} = i~d_{\sigma}R^{\mu\nu\rho\sigma}
\nonumber \\
d_{Q}R^{\mu\nu\rho\sigma} = 0
\label{R1}
\eea
and we have the limitations
$
\omega(R^{I}) \leq 6,~gh(R^{I}) = |I|
$
and also the expressions
$
T^{I}
$
are Lorentz covariant. We consider only the case
$
\omega(R^{I}) = 6
$
because the case
$
\omega(R^{I}) \leq 5
$
has been covered by theorem \ref{T1}.

From the last relation we find, using Theorem \ref{m=0} that
\be
R^{\mu\nu\rho\sigma} = d_{Q}B^{\mu\nu\rho\sigma} + R_{0}^{\mu\nu\rho\sigma}
\ee
with
$
R_{0}^{\mu\nu\rho\sigma} \in {\cal P}_{0}^{(6)}
$
and we can choose the expressions
$
B^{\mu\nu\rho\sigma}
$
and
$
R_{0}^{\mu\nu\rho\sigma}
$
completely antisymmetric. The generic form of 
$
R_{0}^{\mu\nu\rho\sigma}
$
is:
\be
R_{0}^{\mu\nu\rho\sigma} = 
a_{1}~(u^{\mu\nu}~u^{\rho\lambda}~u^{\sigma} + \cdots)
+ a_{2}~(u^{\mu\lambda}~{u^{\nu}}_{\lambda}~u^{\rho}~u^{\sigma} + \cdots)
+ a^{\prime}~\epsilon^{\mu\nu\rho\sigma}~
u^{\alpha\beta}~u_{\alpha\gamma}~u_{\beta}~u^{\gamma}
\ee
where by
$
\cdots
$
we mean the rest of the terms needed to make the expression completely antisymmetric. We denote by
$
R_{j}~(j = 1,2)
$
the polynomials multiplied by
$
a_{j}~(j = 1,2)
$
respectively. We define the completely antisymmetric expression
\be
b^{\mu\nu\rho\sigma\lambda} \equiv u^{\mu\nu}~u^{\rho}~u^{\sigma}~u^{\lambda}
+ \cdots
\ee
which must be obviously null. On the other hand we have
\bea
d_{\lambda}b^{\mu\nu\rho\sigma\lambda} = - R_{1}^{\mu\nu\rho\sigma} 
+ 2 R_{2}^{\mu\nu\rho\sigma} + d_{Q}b^{\mu\nu\rho\sigma}
\nonumber
\eea
where we can take
$
b^{\mu\nu\rho\sigma}
$
completely antisymmetric. In other words we have proved that
\bea
R_{1}^{\mu\nu\rho\sigma} - 2 R_{2}^{\mu\nu\rho\sigma} = d_{Q}b^{\mu\nu\rho\sigma}
\label{intersection}
\eea
and this relation can be used to make 
$
a_{1} = 0
$
in the expression of
$
R_{0}^{\mu\nu\rho\sigma}.
$
We substitute the expression of
$
R^{\mu\nu\rho\sigma}
$
in the fourth relation (\ref{W1}) and get
\be
d_{Q}(R^{\mu\nu\rho} - i~d_{\sigma}B^{\mu\nu\rho\sigma}) = i~d_{\sigma}R_{0}^{\mu\nu\rho\sigma}
\ee 
so the right hand side must be a co-boundary. From this condition we easily find
$
a_{2} = a^{\prime} = 0
$
so in fact we can take
$
R_{0}^{\mu\nu\rho\sigma} = 0.
$ 
It follows that one can exhibit
$
R^{\mu\nu\rho\sigma}
$
in the following form:
\be
R^{\mu\nu\rho} = d_{Q}B^{\mu\nu\rho\sigma}
\ee
and we also have
\be
d_{Q}(R^{\mu\nu\rho} - i~d_{\sigma}B^{\mu\nu\rho\sigma}) = 0.
\ee 

(ii) We use again Theorem \ref{m=0} and obtain
\be
R^{\mu\nu\rho} = d_{Q}B^{\mu\nu\rho} + i~d_{\sigma}B^{\mu\nu\rho\sigma} 
+ R^{\mu\nu\rho}_{0}
\ee 
where
$
R_{0}^{\mu\nu\sigma} \in {\cal P}_{0}^{(6)}
$
and we can choose the expressions
$
B^{\mu\nu\rho}
$
and
$
R_{0}^{\mu\nu\rho}
$
completely antisymmetric. The generic form of the expression 
$
R_{0}^{\mu\nu\rho}
$
is:
\bea
R_{0}^{\mu\nu\sigma} = b~(u^{\mu}~R^{(0)\nu\rho;\alpha\beta} 
+ u^{\nu}~R^{(0)\rho\mu;\alpha\beta} + u^{\rho}~R^{(0)\mu\nu;\alpha\beta})
~u_{\alpha}~u_{\beta};
\eea
other possible expressions can be eliminated, or reduced to this above if we use Bianchi identity. We substitute the expression of
$
R^{\mu\nu\rho}
$
in the third relation (\ref{R1}) and get:
\be
d_{Q}(R^{\mu\nu} - i~d_{\rho}B^{\mu\nu\rho}) = i~d_{\rho}R_{0}^{\mu\nu\rho}
\ee
so the right hand side must be a co-boundary. One easily obtains that
$
b = 0
$
so we have
$
R_{0}^{\mu\nu\rho} = 0.
$
This means that
\be
R^{\mu\nu\rho} = d_{Q}B^{\mu\nu\rho} + i~d_{\sigma}B^{\mu\nu\rho\sigma} 
\ee 
and we have
\be
d_{Q}(R^{\mu\nu} - i~d_{\rho}B^{\mu\nu\rho}) = 0.
\ee

(iii) Now it is again time to use Theorem \ref{m=0} and obtain
\be
R^{\mu\nu} = d_{Q}B^{\mu\nu} + i~d_{\rho}B^{\mu\nu\rho} + R^{\mu\nu}_{0}
\ee 
where
$
R_{0}^{\mu} \in {\cal P}_{0}^{(6)}
$
and we can take the expressions
$
B^{\mu\nu}
$
and 
$
R^{\mu\nu}_{0}
$
antisymmetric. But there is no such an expression
$
R^{\mu\nu}_{0}
$
i.e. we have
$
R^{\mu\nu}_{0} = 0
$
and it follows that
\be
R^{\mu\nu} = d_{Q}B^{\mu\nu} + i~d_{\rho}B^{\mu\nu\rho}.
\ee

We substitute this in the second relation (\ref{R1}) and we obtain
\be
d_{Q}(R^{\mu} - i~d_{\nu}B^{\mu\nu}) = 0.
\ee

(iv) Once more we use theorem \ref{m=0} and get
\be
R^{\mu} = d_{Q}B^{\mu} + i~d_{\nu}B^{\mu\nu} + R^{\mu}_{0}.
\ee
with
$
R^{\mu}_{0} \in {\cal P}_{0}^{(6)};
$
but there is no such an expression i.e. we have
$
R^{\mu}_{0} = 0
$
so in fact:
\be
R^{\mu} = d_{Q}B^{\mu} + i~d_{\nu}B^{\mu\nu}.
\ee

We substitute this in the first relation (\ref{R1}) and we obtain
\be
d_{Q}(R - i~d_{\mu}B^{\mu}) = 0.
\ee

(v) A last use of Theorem \ref{m=0} gives
\be
R = d_{Q}B + i~d_{\mu}B^{\mu} + R_{0}
\ee 
where
$
R_{0} \in {\cal P}_{0}^{(7)}.
$
But there are no such expression i.e.
$
R_{0} = 0
$
and we have 
\be
R = d_{Q}B + i~d_{\mu}B^{\mu}
\ee 
and this finishes the massless case. The massive case brings some new terms in
$
R_{0}
$
which survive and are given in the statement.
$\qed$
\begin{rem}
(i) The expression (\ref{intersection}) is another example of a non-trivial element from
$
{\cal P}_{0} \cap B_{Q}.
$

(ii) In higher orders we can have, even in the massless case, expressions
$
R_{0}^{I}
$ 
which cannot be eliminated by the descent procedure. 
\end{rem}

\section{The Interaction of Gravity with other Quantum Fields\label{int1}}

In \cite{cohomology} we have given the generic structure of the interaction between a system of Yang-Mills fields (particles of spin $1$ and mass 
$m \geq 0$)
with ``matter" fields i.e scalar fields of spin $0$ and Dirac fields of spin $1/2$.
In this Section we add the interaction with massless gravitons. 

First we remind the results from \cite{cohomology}. We take 
$
I = I_{1} \cup I_{2} \cup I_{3}
$
a set of indices and for any index we take a quadruple
$
(v^{\mu}_{a}, u_{a}, \tilde{u}_{a},\Phi_{a}), a \in I
$
of fields with the following conventions:
(a) the first entry are vector fields and the last three ones are scalar fields;
(b) the fields
$
v^{\mu}_{a},~\Phi_{a}
$
are obeying Bose statistics and the fields
$
u_{a},~\tilde{u}_{a}
$
are obeying Fermi statistics;
(c) For
$
a \in I_{1}
$
we impose 
$
\Phi_{a} = 0
$
and we take the masses to be null
$
m_{a} = 0;
$
(d) For
$
a \in I_{2}
$
we take the all the masses strictly positive:
$
m_{a} > 0;
$
(e) For 
$
a \in I_{3}
$
we take 
$
v_{a}^{\mu}, u_{a}, \tilde{u}_{a}
$
to be null and the fields
$
\Phi_{a} \equiv \phi^{H}_{a} 
$
of mass 
$
m^{H}_{a} \geq 0;
$
The fields
$
u_{a},~\tilde{u}_{a},~~a \in I_{1} \cup I_{2}
$
and
$
\Phi_{a}~~a \in I_{2}
$
are called {\it ghost fields} and the fields
$
\phi^{H}_{a},~~a \in I_{3} 
$
are called {\it Higgs fields};
(f) we include matter fields also i.e some set of Dirac fields with Fermi statistics:
$
\Psi_{A}, A \in I_{4};
$ 
(g) we consider that the Hilbert space is generated by all these fields applied on the vacuum and  define in 
$
{\cal H}
$
the operator $Q$ according to the following formulas for all indices
$
a \in I:
$
\bea
~[Q, v^{\mu}_{a}] = i~\partial^{\mu}u_{a},\qquad
[Q, u_{a}] = 0,
\nonumber \\
~[Q, \tilde{u}_{a}] = - i~(\partial_{\mu}v^{\mu}_{a} + m_{a}~\Phi_{a})
\qquad
[Q,\Phi_{a}] = i~m_{a}~u_{a},
\label{Q-general}
\eea
\be
[Q,\Psi_{A}] = 0,
\ee
and
\be
Q\Omega = 0.
\ee
Here 
$
[\cdot,\cdot]
$
is the graded commutator. In \cite{descent} we have determined the most general interaction between these fields in theorem 6.1: 
\begin{thm}
Let $T$ be a relative co-cycle for 
$
d_{Q}
$
which is as least tri-linear in the fields and is of canonical dimension
$
\omega(T) \leq 4
$
and ghost number
$
gh(T) = 0.
$
Then:
(i) $T$ is (relatively) cohomologous to a non-trivial co-cycle of the form:
\bea
t_{YM} = f_{abc} \left( {1\over 2}~v_{a\mu}~v_{b\nu}~F_{c}^{\nu\mu}
+ u_{a}~v_{b}^{\mu}~d_{\mu}\tilde{u}_{c}\right)
\nonumber \\
+ f^{\prime}_{abc} (\Phi_{a}~\phi_{b}^{\mu}~v_{c\mu} + m_{b}~\Phi_{a}~\tilde{u}_{b}~u_{c})
\nonumber \\
+ {1\over 3!}~f^{\prime\prime}_{abc}~\Phi_{a}~\Phi_{b}~\Phi_{c}
+ {1\over 4!}~\sum_{a,b,c,d \in I_{3}}~g_{abcd}~\Phi_{a}~\Phi_{b}~\Phi_{c}~\Phi_{d}
+ j^{\mu}_{a}~v_{a\mu} + j_{a}~\Phi_{a}
\eea
where we make the following conventions:
$
f_{abc} = 0
$
if one of the indices is in
$
I_{3};
$
$
f^{\prime}_{abc} = 0
$
if 
$
c \in I_{3}
$
or one of the indices $a$ and $b$ are from
$
I_{1};
$
$
j^{\mu}_{a} = 0
$
if
$
a \in I_{3};
$
$
j_{a} = 0
$
if
$
a \in I_{1}.
$
Moreover we have:

(a) The constants
$
f_{abc}
$
are completely antisymmetric
\be
f_{abc} = f_{[abc]}.
\label{anti-f}
\ee

(b) The expressions
$
f^{\prime}_{abc}
$
are antisymmetric  in the indices $a$ and $b$:
\be
f^{\prime}_{abc} = - f^{\prime}_{bac}
\label{anti-f'}
\ee
and are connected to 
$f_{abc}$
by:
\be
f_{abc}~m_{c} = f^{\prime}_{cab} m_{a} - f^{\prime}_{cba} m_{b}.
\label{f-f'}
\ee

(c) The (completely symmetric) expressions 
$f^{\prime\prime}_{abc} = f^{\prime\prime}_{\{abc\}}$
verify
\be
f^{\prime\prime}_{abc}~m_{c} = \left\{\begin{array}{rcl} 
{1 \over m_{c}}~f'_{abc}~(m_{a}^{2} - m_{b}^{2}) & \mbox{for} & a, b \in I_{3}, c \in I_{2} \\
- {1 \over m_{c}}~f'_{abc}~m_{b}^{2} & \mbox{for} & a, c \in I_{2}, b \in I_{3}.
\end{array}\right.
\label{f"}
\ee

(d) the expressions
$
j^{\mu}_{a}
$
and
$
j_{a}
$
are bi-linear in the Fermi matter fields: in tensor notations;
\bea
j_{a}^{\mu} = \overline{\psi} t^{\epsilon}_{a} \otimes \gamma^{\mu}\gamma_{\epsilon} \psi \qquad
j_{a} = \overline{\psi} s^{\epsilon}_{a} \otimes \gamma_{\epsilon} \psi
\label{current}
\eea
where  for every
$
\epsilon = \pm
$
we have defined the chiral projectors of the algebra of Dirac matrices
$
\gamma_{\epsilon} \equiv {1\over 2}~(I + \epsilon~\gamma_{5})
$
and
$
t^{\epsilon}_{a},~s^{\epsilon}_{a}
$
are 
$
|I_{4}| \times |I_{4}|
$
matrices; the summation over 
$
\epsilon = \pm
$
is assumed. If $M$ is the mass matrix
$
M_{AB} = \delta_{AB}~M_{A}
$
then we must have
\be
d_{\mu}j^{\mu}_{a} = m_{a}~j_{a} 
\qquad \Leftrightarrow \qquad
m_{a}~s_{a}^{\epsilon} = i(M~t^{\epsilon}_{a} - t^{-\epsilon}_{a}~M)~~(\forall~a \in I_{1} \cup I_{2}.)
\label{conserved-current}
\ee

(ii) The relation 
$
d_{Q}t_{YM} = i~d_{\mu}t_{YM}^{\mu}
$
is verified by:
\be
t_{YM}^{\mu} = f_{abc} \left( u_{a}~v_{b\nu}~F^{\nu\mu}_{c} -
{1\over 2} u_{a}~u_{b}~d^{\mu}\tilde{u}_{c} \right)
+ f^{\prime}_{abc}~\Phi_{a}~\phi_{b}^{\mu}~u_{c}
+ j^{\mu}_{a}~u_{a}
\ee

(iii) The relation 
$
d_{Q}t_{YM}^{\mu} = i~d_{\nu}t_{YM}^{\mu\nu}
$
is verified by:
\be
t_{YM}^{\mu\nu} \equiv {1\over 2} f_{abc}~u_{a}~u_{b}~F_{c}^{\mu\nu}.
\ee

(iv) The constants
$
f_{abc}~f^{\prime}_{abc},~f^{\prime\prime}_{abc}
$
and
$
g_{abcd}
$
are real and the matrices
$
t_{a}^{\epsilon},~s_{a}^{\epsilon}
$
are Hermitean.
\end{thm}

Now we extend the argument including massless gravitation. We include in the set of fields generating the Hilbert space
$
{\cal H}
$
the fields
$
h_{\mu\nu},~u^{\rho},~\tilde{u}^{\sigma}
$ 
the first one being a tensor fields with Bose statistics and the last are vector fields with Fermi statistics. We also extend the definition of the gauge charge
$Q$ given by (\ref{Q-general}) with 
\bea
~[Q, h_{\mu\nu}] = - {i\over 2}~(\partial_{\mu}u_{\nu} + \partial_{\nu}u_{\mu}
- \eta_{\mu\nu} \partial_{\rho}u^{\rho}),\qquad
[Q, u_{\mu}] = 0,\qquad
[Q, \tilde{u}_{\mu}] = i~\partial^{\nu}h_{\mu\nu}
\eea
and we can easily generalize theorem \ref{fock-0} and the corresponding result for the Yang-Mills system from \cite{cohomology}: the Fock space describes in this case massless gravitons, spin $0$ and spin $1$ particles. By definition the ghost number is the sum of the ghost numbers of the YM and gravity sectors. We want to extend the preceding theorem to this more general case. For simplicity we treat here only the case of massless Yang-Mills fields i.e. we take in the general scheme from above
$
I_{2} = 0.
$
Beside the expression 
$
t_{YM}
$
given above and
$
t_{gh}
$
determined in theorem \ref{m=0} we need the interaction between the two sets of fields (Yang-Mills and gravitational). The result is described in the following:
\begin{thm}
Suppose that the interaction Lagrangian 
$
T_{\rm int}
$
is restricted by Lorentz covariance, is at least tri-linear in the fields and
$
\omega(T_{\rm int}) \leq 5,~~gh(T_{\rm int}) = 0.
$
We also suppose that 
$
I_{2} = 0.
$
Then: (i) 
$
T_{\rm int}
$
is cohomologous to the expression
\bea
t_{\rm int} \equiv \sum_{a,b \in I_{1}}~f_{ab}~
(4 h_{\mu\nu}~F_{a}^{\mu\rho}~{F_{b}^{\nu}}_{\rho} 
- h~F_{a\rho\sigma}~F_{b}^{\rho\sigma} 
+ 4~u_{\mu}~d_{\nu}\tilde{u}_{a}~F_{b}^{\mu\nu})
\nonumber \\
+ \sum_{c,d \in I_{3}}~f^{\prime}_{cd}~
\left(h_{\mu\nu}~d^{\mu}\Phi_{c}~d^{\nu}\Phi_{d} 
- {m^{2}_{a} + m^{2}_{b} \over 4}~h~\Phi_{a}~\Phi_{b}\right)
\nonumber \\
+ h_{\mu\nu}~
(d^{\mu}\bar{\psi}~c^{\epsilon} \otimes \gamma^{\nu}\gamma_{\epsilon}\psi
- \bar{\psi}~c^{\epsilon} \otimes \gamma^{\nu}\gamma_{\epsilon}d^{\mu}\psi).
\label{t-int}
\eea

Moreover we have

(a) the constants
$
f_{ab}
$
are symmetric
$
f_{ab} = f_{ba};
$

(b)
the constants
$
f^{\prime}_{cd}
$
are symmetric
$
f^{\prime}_{cd}= f^{\prime}_{dc};
$
also if we denote by $m$ the mass matrix of the scalar fields
$
m_{cd} \equiv m_{c}~\delta_{cd},~\forall c,d \in I_{3}
$
it commutes with the matrix
$
f^{\prime}_{cd}:
$
\be
[ m, f^{\prime} ] = 0
\ee

(c) the matrices
$
c^{\epsilon}
$
verify
\be
c^{\epsilon}~M = M~c^{- \epsilon}
\ee
where $M$ is the mass matrix of the Dirac fields:
$
M_{AB} \equiv M_{A}~\delta_{AB},~\forall A,B \in I_{4}.
$

(ii) The relation 
$
d_{Q}t_{\rm int} = i~d_{\mu}t_{\rm int}^{\mu}
$
is verified by:
\bea
t_{\rm int}^{\mu} \equiv \sum_{a,b \in I_{1}}~f_{ab}~
( u^{\mu}~F_{a}^{\rho\sigma}~F_{b\rho\sigma} 
+ 4~u^{\rho}~F_{a}^{\mu\nu}~F_{b\nu\rho} )
\nonumber \\
+ \sum_{c,d \in I_{3}}~f^{\prime}_{cd}~
\left({1\over 2}~u^{\mu}~d^{\nu}\Phi_{c}~d^{\nu}\Phi_{d} 
- u_{\nu}~d^{\mu}\Phi_{c}~d^{\nu}\Phi_{d} 
- {m^{2}_{a} + m^{2}_{b} \over 4}~u^{\mu}~\Phi_{a}~\Phi_{b}\right)
\nonumber \\
- {1\over 2}~u_{\nu}~
[ (d^{\mu}\bar{\psi}~c^{\epsilon} \otimes \gamma^{\nu}\gamma_{\epsilon}\psi
- \bar{\psi}~c^{\epsilon} \otimes \gamma^{\nu}\gamma_{\epsilon}d^{\mu}\psi)
+ (\mu \leftrightarrow \nu) ]
\eea
and we also have
\be
d_{Q}t_{\rm int}^{\mu} = 0.
\ee

(iii) the constants
$
f_{ab}
$
and
$
f^{\prime}_{cd}
$
are real and we also have the Hermiticity property
$
(c^{\epsilon})^{\dagger} = c^{- \epsilon}.
$
\label{T-int}
\end{thm}
{\bf Proof:} (i) By hypothesis we have
\be
d_{Q}T_{\rm int} = i~d_{\mu}T_{\rm int}^{\mu}
\label{descent-tint}
\ee
and the descent procedure leads to
\bea
d_{Q}T_{\rm int}^{\mu} = i~d_{\nu}T_{\rm int}^{\mu\nu}.
\nonumber\\
d_{Q}T_{\rm int}^{\mu\nu} = i~d_{\rho}T_{\rm int}^{\mu\nu\rho}
\nonumber \\
d_{Q}T_{\rm int}^{\mu\nu\rho} = i~d_{\sigma}T_{\rm int}^{\mu\nu\rho\sigma}
\nonumber \\
d_{Q}T_{\rm int}^{\mu\nu\rho\sigma} = 0
\label{descent-T-int}
\eea
and can choose the expressions
$
T_{\rm int}^{I}
$
to be Lorentz covariant; we also have
\be
gh(T_{\rm int}^{I}) = |I|, \omega(T_{\rm int}^{I}) \leq 5. 
\ee 

From the last relation we find, using Theorem \ref{m=0} and the corresponding result from \cite{cohomology}, that
\be
T_{\rm int}^{\mu\nu\rho\sigma} = d_{Q}B^{\mu\nu\rho\sigma} 
+ T_{{\rm int},0}^{\mu\nu\rho\sigma}
\ee
with
$
T_{{\rm int},0}^{\mu\nu\rho\sigma} \in {\cal P}_{0}^{(5)}
$
and we can choose the expressions
$
B^{\mu\nu\rho\sigma}
$
and
$
T_{{\rm int},0}^{\mu\nu\rho\sigma}
$
completely antisymmetric. The generic form of 
$
T_{{\rm int},0}^{\mu\nu\rho\sigma}
$
is:
\be
T_{{\rm int},0}^{\mu\nu\rho\sigma} = \sum_{a \in I_{3}}~ f_{a}~u^{\mu}~u^{\nu}~u^{\rho}~u^{\sigma}~\Phi_{a}.
\ee
If we substitute the expression obtained for
$
T_{\rm int}^{\mu\nu\rho\sigma}
$
in the third relation (\ref{descent-T-int}) we find out
\be
d_{Q}(T_{\rm int}^{\mu\nu\rho} - i~d_{\sigma}B^{\mu\nu\rho\sigma}) = i~d_{\sigma}T_{{\rm int},0}^{\mu\nu\rho\sigma}
\ee 
so the expression in the right hand side must be a co-boundary and we immediately obtain
$
f_{a} = 0.
$
It follows that
\be
T_{\rm int}^{\mu\nu\rho\sigma} = d_{Q}B^{\mu\nu\rho\sigma}
\ee
and
\be
d_{Q}(T_{\rm int}^{\mu\nu\rho} - i~d_{\sigma}B^{\mu\nu\rho\sigma}) = 0
\ee 
so we obtain
\be
T_{\rm int}^{\mu\nu\rho} = B^{\mu\nu\rho} + i~d_{\sigma}B^{\mu\nu\rho\sigma} 
+ T^{\mu\nu\rho}_{{\rm int},0}
\ee 
where
$
T_{{\rm int},0}^{\mu\nu\rho} \in {\cal P}_{0}^{(5)};
$
we can choose the expressions
$
B^{\mu\nu\rho}
$
and
$
T_{{\rm int},0}^{\mu\nu\rho}
$
completely antisymmetric. The generic form of
$
T_{0}^{\mu\nu\rho}
$
has a two contributions: even and odd with respect to parity invariance. We do not give the generic form here but we give the result: the second relation (\ref{descent-T-int}) gives
\be
d_{Q}(T_{\rm int}^{\mu\nu} - i~d_{\rho}B^{\mu\nu\rho}) 
= i~d_{\rho}T_{{\rm int},0}^{\mu\nu\rho}
\ee 
so the right hand side must be a co-boundary and a direct computation gives that in fact
$
T_{{\rm int},0}^{\mu\nu\rho} = 0.
$
It follows that
\be
T_{\rm int}^{\mu\nu\rho} = d_{Q}B^{\mu\nu\rho} + i~d_{\sigma}B^{\mu\nu\rho\sigma}
\ee
and
\be
d_{Q}(T_{\rm int}^{\mu\nu} - i~d_{\rho}B^{\mu\nu\rho}) = 0.
\ee 

(ii) We obtain from the preceding relation that
\be
T_{\rm int}^{\mu\nu} = d_{Q}B^{\mu\nu} + i~d_{\rho}B^{\mu\nu\rho} 
+ T^{\mu\nu}_{{\rm int},0}
\ee 
where
$
T_{{\rm int},0}^{\mu\nu} \in {\cal P}_{0}^{(5)}
$
and we can choose the expressions
$
B^{\mu\nu}
$
and
$
T_{{\rm int},0}^{\mu\nu}
$
antisymmetric. Again we do not give the generic form of the expression 
$
T_{{\rm int},0}^{\mu\nu}
$
but we give the final result of this standard computation: by conveniently modifying the expressions
$
B^{I}
$
we can arrange such that
$
T_{{\rm int},0}^{\mu\nu} = 0.
$
We substitute the expression of
$
T_{\rm int}^{\mu\nu}
$
in the first relation (\ref{descent-T-int}) and get:
\be
d_{Q}(T_{\rm int}^{\mu} - i~d_{\nu}B^{\mu\nu}) = 0
\ee

(iii) Now it is again time we use known results and obtain
\be
T_{\rm int}^{\mu} = d_{Q}B^{\mu} + i~d_{\nu}B^{\mu\nu} + T^{\mu}_{{\rm int},0}
\ee 
where
$
T_{{\rm int},0}^{\mu} \in {\cal P}_{0}^{(5)}.
$ 
Now we get from the first relation (\ref{descent-T-int}) 
\be
d_{Q}(T_{\rm int} - i~d_{\mu}B^{\mu}) = i~d_{\mu}T^{\mu}_{{\rm int},0}
\ee
so the right hand side must be a co-boundary. If one writes the generic form of
$
T_{{\rm int},0}^{\mu}
$ 
one gets after tedious computations that by modifying the expressions
$
B^{I}
$
one can take
\be
T_{{\rm int},0}^{\mu} = t^{\mu}_{\rm int}
\ee
with
$
t^{\mu}_{\rm int}
$
the expression from the statement of the theorem. Because we have
$
d_{Q}t_{\rm int} = i~d_{\mu}t_{\rm int}^{\mu}
$
we get
\be
d_{Q}(T_{\rm int} - t_{\rm int} - i~d_{\mu}B^{\mu}) = 0
\ee
so known results leads to
\be
T_{\rm int} = t_{\rm int} + d_{Q}B + i~d_{\mu}B^{\mu} + T_{{\rm int},0}
\ee 
where
$
T_{{\rm int}0} \in {\cal P}_{0}^{(5)}.
$
But there are no such expression i.e.
$
T_{{\rm int},0} = 0
$
and we have 
\be
T_{\rm int} = t_{\rm int} + d_{Q}B + i~d_{\mu}B^{\mu}
\ee 
which is the final result.
$\qed$
\begin{rem}
(i) We note that we have obtained in a natural way the known expression of the energy-momentum tensor
$
T_{\mu\nu}
$
which is, up to a factor, the coefficient of 
$
h^{\mu\nu}
$
from the expression
$
t_{\rm int}.
$
However, there is a supplementary ghost term in the first line of the formula (\ref{t-int}). This is due to the fact already explained in the Introduction: because we cannot impose in the quantum framework the Maxwell equation
\be 
\partial_{\mu}v_{a}^{\mu} = 0
\ee
the energy-momentum tensor
\be
T^{\mu\nu} \equiv \sum_{a,b \in I_{1}}~f_{ab}~
\left(F_{a}^{\mu\rho}~{F_{b}^{\nu}}_{\rho} - 
{1\over 4}\eta^{\mu\nu}~F_{a\rho\sigma}~F_{b}^{\rho\sigma} \right)
\ee
does not verify anymore the divergenceless condition
\be 
\partial_{\mu}T^{\mu\nu}  = 0
\ee
and without the extra ghost term in the first line of formula (\ref{t-int}) we do not have gauge invariance.

We note however that the ghost term from the first line of formula (\ref{t-int}) gives a null contributions between physical states (described as in theorem \ref{fock-0}) and this result propagates to all orders of perturbation theory. So it can be neglected in practical computations.

(ii) There are other approaches to the quantization of the massless vector fields in which one can impose the condition
$
\partial_{\mu}v_{a}^{\mu} = 0
$
namely the so-called Coulomb gauge, but the price to pay is the loss of the manifest Lorentz covariance and the appearence of a non-local interaction term so Epstein-Glaser method cannot be implemented in this approach.

(iii) One can prove that there are no bi-linear solution for the interaction.
\end{rem}

\section{Conclusions}

The cohomological methods presented in a previous paper \cite{cohomology} leads to the a simple understanding of quantum gravity in lower orders of perturbation theory. If we use the consistency Wess-Zumino equations we can give simple proofs for the gauge invariance and renormalization in the second order of perturbation theory for the massless and massive pure gravity. 

The descent technique can be used to give the most general interaction including Yang-Mills fields (massless and massive), matter and massless gravity. In this paper we have considered only massless Yang-Mills fields and the general case will be treated in a forthcoming paper.  Further restrictions follow from the cancellation of the anomalies in the second order of the perturbation theory. The analysis can be extended  to the third order of perturbation theory and it will also be done elsewere. One should expect the appearance of the known gravitational anomaly (see for instance \cite{We}.) 
\vskip 1cm
{\bf Acknowledgment:} The author wishes to thank Professor G. Scharf for the critical reading of the typescript and many valuable suggestions.


\begin{thebibliography}{99}

\bibitem{BS}
N. N. Bogoliubov, D. Shirkov,
``{\it Introduction to the Theory of Quantized Fields}",
John Wiley and Sons, 1976 (3rd edition) 

\bibitem{Dr}
N. Dragon, 
{\it BRS Symmetry and Cohomology}, 
Schladming lectures, \\
hep-th/9602163

\bibitem{DB}
M. D\"utsch, F. M. Boas,
``{\it The Master Ward Identity}", \\
hep-th/0111101, Rev. Math. Phys. {\bf 14} (2002) 977-1049,

\bibitem{DF}
M. D\"utsch, K. Fredenhagen,
``{\it A Local (Perturbative) Construction of Observables in Gauge Theories:
the Example of QED}",\\
hep-th/9807078, Commun. Math. Phys. {\bf 203} (1999) 71-105

\bibitem{EG}
H. Epstein, V. Glaser,
``{\it The R\^ole of Locality in Perturbation Theory}",\\
Ann. Inst. H. Poincar\'e {\bf 19 A} (1973) 211-295

\bibitem{Gl}
V. Glaser,
``{\it Electrodynamique Quantique}",
L'enseignement du 3e cycle de la physique en Suisse Romande (CICP), Semestre
d'hiver 1972/73

\bibitem{GW}
J. Gomis, S. Weinberg,
``{\it Are Nonrenormalizable Gauge Theories Renormalizable?}",\\
hep-th/9510087

\bibitem{jet} D. R. Grigore,
``{\it The Variational Sequence on Finite Jet Bundle Extensions and the Lagrangian 
Formalism}",\\
Differential Geometry and Applications {\bf 10} (1999) 43-77

\bibitem{YM} D. R. Grigore
``{\it On the Uniqueness of the Non-Abelian Gauge Theories in Epstein-Glaser 
Approach to Renormalisation Theory}",\\
hep-th/9806244, Romanian J. Phys. {\bf 44} (1999) 853-913

\bibitem{gravity} D. R. Grigore,
``{\it On the Quantization of the Linearized Gravitational Field}",\\
hep-th/9905190, Class. Quant. Grav. {\bf 17} (2000) 319-344

\bibitem{ano} D. R. Grigore,
``{\it The Structure of the Anomalies of the Non-Abelian Gauge Theories 
in the Causal Approach }",\\
hep-th/0010226, Journ. Phys. {\bf A 35} (2002) 1665-1689

\bibitem{cohomology}
D. R. Grigore, 
``{\it Cohomological Aspects of Gauge Invariance in the Causal Approach}",\\
hep-th/0711.3986

\bibitem{massive}
D. R. Grigore, G. Scharf,
``{\it Massive Gravity as a Quantum Gauge Theory}",\\
hep-th/0404157, General Relativity and Gravitation {\bf 37} (2005) 1075-1096 

\bibitem{descent}
D. R. Grigore, G. Scharf,
``{\it Massive Gravity from Descent Equations}",\\
hep-th/0711.0869, to appear in Classical and Quantum Gravity

\bibitem{Grillo}
N. Grillo,
``{\it Finite One-Loop Calculation in Quantum Gravity: Graviton Self-Energy,
Perturbative Gauge Invariance and Slavnov-Ward Identities}",\\
hep-th/9912097

\bibitem{Kr}
D. Kreimer,
``{\it A Remark on Quantum Gravity}",\\
hep-th/0705.3897

\bibitem{PS}
G. Popineau, R. Stora, ``{\it A Pedagogical Remark on the Main Theorem of
Perturbative Renormalization Theory}", unpublished preprint

\bibitem{Sc2}
G. Scharf,
``{\it Quantum Gauge Theories. A True Ghost Story}",
John Wiley, 2001

\bibitem{Sto1}
R. Stora,
``{\it Lagrangian Field Theory}",
Les Houches lectures, Gordon and Breach, N.Y., 1971, 
C. De Witt, C. Itzykson eds.

\bibitem{St1}
O. Steinmann,
``{\it Perturbation Expansions in Axiomatic Field Theory}",
Lect. Notes in Phys. {\bf 11}, Springer, 1971

\bibitem{We}
S. Weinberg,
``{\it The Quantum Theory of Fields}", vol. 1 and 2,
Cambridge Univ. Press, 1995
\end{thebibliography}
\end{document}